\documentclass[epj]{svjour}
\usepackage{amsfonts,amssymb,amsmath}
\usepackage{graphicx,color}
\usepackage{dcolumn,slashed}
\usepackage{adjustbox}

\makeatletter
\DeclareRobustCommand{\cev}[1]{
  {\mathpalette\do@cev{#1}}
}
\newcommand{\do@cev}[2]{
  \vbox{\offinterlineskip
    \sbox\z@{$\m@th#1 x$}
    \ialign{##\cr
      \hidewidth\reflectbox{$\m@th#1\vec{}\mkern4mu$}\hidewidth\cr
      \noalign{\kern-\ht\z@}
      $\m@th#1#2$\cr
    }
  }
}
\makeatother

\newcommand{\beq}{\begin{equation}}
\newcommand{\eeq}{\end{equation}}
\newcommand{\bea}{\begin{eqnarray}}
\newcommand{\eea}{\end{eqnarray}}
\newcommand{\sst}[1]{{\scriptscriptstyle #1}}
\def\vec#1{{\bf #1}}    
\def\bpp{{b_1^{1/2,\,1/2}}}
\def\bpm{{b_1^{1/2,\,-1/2}}}
\def\bppm{{b_1^{1/2,\,\pm 1/2}}}

\def\bra#1{{\langle#1\vert}}
\def\ket#1{{\vert#1\rangle}}
\def\pbar{{\bar{p}}}
\def\mns{{m^2}}
\def\mks{{M_{\sst{K}}^2}}

\newcommand{\eps}{\epsilon}

\begin{document}

%
%
%
%
%
%
%
%

\title{Dispersion-theoretical analysis of the electromagnetic form factors of the nucleon:
  Past, present and future}

\author{Yong-Hui Lin\inst{1} 
\and Hans-Werner Hammer\inst{2,3}  
\and Ulf-G.~Mei{\ss}ner\inst{1,4,5}
%
}                     
%
%
\institute{
Helmholtz-Institut~f\"{u}r~Strahlen-~und~Kernphysik~and~Bethe~Center~for
Theoretical~Physics, Universit\"{a}t~Bonn, \\ D-53115~Bonn,~Germany
\and
Technische Universit\"at Darmstadt, Department of Physics,
Institut f\"ur Kernphysik,\\ 64289 Darmstadt, Germany
\and
ExtreMe Matter Institute EMMI and Helmholtz Forschungsakademie Hessen f\"ur
FAIR (HFHF),\\ GSI Helmholtzzentrum f\"ur Schwerionenforschung GmbH,
64291 Darmstadt, Germany
\and
Institut~f\"{u}r~Kernphysik,~Institute~for~Advanced~Simulation and
J\"{u}lich~Center~for~Hadron~Physics, \\ 
Forschungszentrum~J\"{u}lich, D-52425~J\"{u}lich,~Germany
\and Tbilisi State University, 0186 Tbilisi, Georgia}

\titlerunning{Dispersion-theoretical analysis of the electromagnetic form factors of the nucleon}
\authorrunning{Lin, Hammer, Mei{\ss}ner}
\date{Received: date / Revised version: date}
%

\abstract{We review the dispersion-theoretical analysis of the electromagnetic
form factors of the nucleon. We emphasize in particular the role of unitarity and analyticity
in the construction of the isoscalar and isovector spectral functions.
We present new results on the
extraction of the nucleon radii, the electric and magnetic form factors and the
extraction of $\omega$-meson couplings. All this is supplemented by a detailed
calculation of the theoretical uncertainties, using bootstrap and Bayesian methods
to pin down the statistical errors, while systematic errors are determined from variations
of the spectral functions. We also discuss the physics of the time-like form factors
and point out further issues to be addressed in this framework. 
} 


\maketitle


\section{Introduction}

Nucleons and electrons are the constituents of everyday matter
with nucleons accounting for essentially all of its mass. The mass of the
nucleon as a bound state of quarks and gluons, on the other hand,
arises from the complicated strong
interaction dynamics of quarks and gluons in
Quantum Chromodynamics (QCD)~\cite{Wilczek:2012ab}.
The electromagnetic form factors of the nucleon describe the
structure of the nucleon as seen by an electromagnetic probe.
As such, they provide a window on strong interaction dynamics
over a large range of momentum, for recent reviews see, e.g. Refs.~\cite{Denig:2012by,Punjabi:2015bba}.
Moreover, they are an important ingredient in the description of
a wide range of observables ranging from the Lamb shift in atomic
physics to the strangeness content of the
nucleon~\cite{Armstrong:2012bi,Maas:2017snj}.
At small momentum transfers, they are sensitive
to the gross properties of the nucleon like the charge and magnetic moment
as well as the radii. At large momentum transfer, in contrast, they
probe the  quark substructure of the nucleon as described by QCD.

A new twist was recently added to this picture by measurements
of the proton charge radius in muonic hydrogen.
The proton charge radius was first indirectly measured in the Nobel
prize winning electron scattering experiments by Hofstadter
et al.~\cite{Hofstadter:1956,Hofstadter:1957wk}.
While electron scattering was the
method of choice to refine the proton radius in the
decades following these pioneering experiments, the
Lamb shift in electronic hydrogen and muonic hydrogen is also sensitive
to the proton radius~\cite{Karplus:1952}.
The electronic Lamb shift measurements as well as
most electron scattering experiments gave the so-called {\bf large} radius,
$r_E^P\simeq 0.88\,$fm, which was
also the value given by CODATA~\cite{Mohr:2008fa}.\footnote{Note, however,
  that dispersion-theoretical analyses of the same data always gave a smaller
  radius as will be discussed in detail below.}
It then came as a true surprise to most researchers
when  the first measurement of the muonic
hydrogen Lamb shift, which has a larger sensitivity to $r_E^P$
because of the much larger muon mass, led to the so-called {\bf small} radius,
$r_E^P= 0.84184(67)\,$fm, differing by 5$\sigma$ from the
CODATA value \cite{Pohl:2010zza}. At about the same time, a
high-precision electron-proton  scattering experiment performed at the Mainz Microtron
(MAMI) reinforced the large radius~\cite{Bernauer:2010wm}.
This glaring discrepancy in such a fundamental quantity, which was believed to
be understood since long, became known as the ``proton radius puzzle''.
It led to much experimental and theoretical activity dedicated to
uncover its cause, either  physics beyond the standard model, or
more mundane reasons, such as an underestimation of the experimental
uncertainties. Recent experiments on the electronic Lamb shift
\cite{Beyer:2017gug,Fleurbaey:2018fih,Bezginov:2019mdi} and
a novel measurement of electron-proton scattering at unprecedented small
momentum transfer~\cite{Xiong:2019umf} now point to the latter reason.
With the exception of Ref.~\cite{Fleurbaey:2018fih}, all of these new
determinations of $r_p$ consistently
give a small proton radius. Consequently, the newest addition of the CODATA
compilation lists the proton charge radius as
$r_E^p = 0.8414(19)\,$fm~\cite{CODATAnew}.  A short review of the current
status is given in Ref.~\cite{Hammer:2019uab}. The important role of
dispersion theory in solving this ``puzzle'' will be discussed below.

This paper is structured as follows: In Sec.~\ref{sec:hist}, we briefly review earlier
dispersion-theoretical analyses of the electromagnetic nucleon from factors. The complete
formalism to perform such type of analyses is given in Sec.~\ref{sec:form}, where all
basic definitions are given and the various contributions 
to the spectral functions, the central objects of the dispersive method, are discussed
in detail. Furthermore, constraints on the nucleon form factors and two-photon corrections
to the electron-proton scattering  cross section are presented. Finally, we display in detail
methods to determine the theoretical  uncertainties, both the  statistical and the systematical ones.
In Sec.~\ref{sec:phys}, we display the results of our new dispersion-theoretical
analysis of the electromagnetic form factors in the space-like region, including novel
determinations of the various radii, form factors as well as the $\omega$-meson couplings.
Then, we consider the extension to the form factors in the time-like region and discuss the
physics encoded in these. We end with a brief summary and an outlook in Sec.~\ref{sec:summ}.
In the appendices,
we give further details on the extraction of neutron form factors
from light nuclei as well as on the construction of the continuum contributions to
the spectral functions.
We also collect the various parameters of our best fit discussed in the main text.

\section{Short history of dispersive analyses of the nucleon form factors}  
\label{sec:hist}

Here, we briefly review earlier work using dispersion theory to analyze
the electromagnetic structure of the nucleon. To be more precise, we
only consider investigations that include explicitly the two-pion continuum,
which generates the $\rho$-meson in the isovector part of the spectral
function in addition to a very important uncorrelated two-pion contribution
as first discussed by Frazer and Fulco~\cite{Frazer:1959gy,Frazer:1960zza,Frazer:1960zzb}.
For other work on dispersion relations applied to the nucleon electromagnetic
from factors, we refer the reader to the review Ref.~\cite{Pacetti:2015iqa}.

The first  groundbreaking work was done by the Karls\-ruhe group in 1976~\cite{Hohler:1976ax}.
Here, electron-proton ($ep$) cross section data supplemented by neutron form factor data
from elastic and quasi-elastic electron-deuteron scattering were fitted.
Besides the two-pion continuum, the spectral functions contained the $\omega$-meson
plus additional isoscalar and isovector poles and
normalization constants for the data sets. It should be noted that the $ep$
data base was pruned in the sense that in case of inconsistencies between
data sets, only one was retained. A dozen of fits with varying number of vector
mesons poles and excluding  various subsets of data were performed.
The best fit (fit~8.2) featured 8 resonance parameters. Theoretical errors were
estimated from the variations in the different fits. The resulting proton radii are
tabulated in Tab.~\ref{tab:protonradii} and the neutron radii in Tab.~\ref{tab:neutronradii}.
The neutron magnetic radius could not be determined very precisely at that time.
Also notable were the sizable $\phi NN$ couplings, where $N$ denotes the
nucleon,  at odd with expectations
from the OZI rule~\cite{Okubo:1963fa,Zweig:1964jf,Iizuka:1966fk}.

In 1995, the Bonn-Mainz group (MMD) rejuvenated the dispersion-theoretical approach
to the nucleon electromagnetic form factors, as many new form factor results had
become available and perturbative QCD had firmly established the behavior of
the form factors at large momentum transfer~\cite{Mergell:1995bf}. Fits
were performed to the existing form factor data basis of the Bochum group
(updated from Refs.~\cite{Gari:1984ia,Gari:1986rj}). The two-pion continuum
was still based on the Karlsruhe-Helsinki pion-nucleon ($\pi N$) partial wave amplitudes $f_\pm^1(t)$,
but the $\rho$-$\omega$ mixing visible in the pion vector form factor was included
for the first time. The best fits where obtained
with three additional isovector poles and and one additional isoscalar one
(besides the $\omega$ and the $\phi$). It was found that
the onset of perturbative QCD was not seen in these data and the radii and vector meson
couplings were consistent with the findings of the Karlsruhe group, see Tab.~\ref{tab:protonradii} and
Tab.~\ref{tab:neutronradii}.
Remarkably, these dispersive fits could not be made consistent with the
then existing best value for $r_E^p$ from $ep$ scattering, $r_E^p = (0.862\pm 0.012)\,$fm
\cite{Simon:1980hu}. Further, the large deviation in the OZI rule of the $\phi$
couplings was confirmed.
One year later, the sparse and not very precise existing  data on the proton and
neutron form factors in the time-like region were included, which revealed some
inconsistencies in the time-like data basis for the neutron~\cite{Hammer:1996kx}.

In view of new data on the proton and neutron form factors, in particular the
first polarization transfer measurements at Jefferson Lab at few GeV$^2$ squared momentum transfer
\cite{Jones:1999rz,Gayou:2001qd}, the MMD work was updated, with a particular
emphasis on the magnetic radius of the proton and the neutron in~\cite{Hammer:2003ai}.
In this work, no error analysis was performed.

A significant improvement of the dispersion relation (DR) analysis was performed
in Ref.~\cite{Belushkin:2006qa} (BHM). Not only was the data basis enlarged, but also
the description of the isoscalar spectral function was improved by including the
$K\bar{K}$~\cite{Hammer:1998rz,Hammer:1999uf} and the $\pi\rho$~\cite{Meissner:1997qt}
continua. Furthermore, the $2\pi$ continuum was updated in view of new data for the
pion vector form factor~\cite{Belushkin:2005ds}. All data from the space-like and the
time-like regions were included in the fit. In these fits, besides the mentioned continua,
the isoscalar spectral function featured the $\omega$, the $\phi$ and two 
poles, where as the $2\pi$ continuum was supplemented by  5 effective poles.
The uncertainties were calculated by large scale Monte Carlo samplings of all solution with
a $\chi^2/$dof  in the range $[\chi^2_{\rm min},\chi^2_{\rm min} +1.04]$,  where $\chi^2_{\rm min}$
refers to the best fit, corresponding to the $1\sigma$ coincidence
in the $p$-value. Different to all earlier fits, the neutron charge radius squared
was not included as a constraint. Nonetheless, the extracted value of $(r_E^n)^2$
came out consistent with determinations from low energy atom-neutron scattering, see
Tab.~\ref{tab:neutronradii}. In that paper, it was stated that the then
accepted proton charge radius determined from the Lamb shift in electronic hydrogen,
$r_E^p = 0.88 \ldots 0.90\,$fm, see Ref.~\cite{Melnikov:1999xp} (and references therein),
was inconsistent with the dispersion analysis of the electron scattering data, thus
previewing what was later called the ``proton radius puzzle''. The various radii
came out consistent with earlier DR determinations, see Tab.~\ref{tab:protonradii} and
Tab.~\ref{tab:neutronradii}.  The same spectral functions were also used to extract
the strength of two-photon corrections from the difference of data obtained by
Rosenbluth separation and direct polarization transfer measurements~\cite{Belushkin:2007zv}.
The so determined two-photon corrections came out comparable to direct calculations available in the literature, such as
Refs.~\cite{Blunden:2003sp,Blunden:2005ew,Kondratyuk:2005kk}.

\begin{table}[t]
\centering  
\begin{tabular}{|c|cc|}
\hline
Ref.                      &   $r_E^p$~[fm]         &  $r_M^p$~[fm] \\
\hline
\cite{Hohler:1976ax}      & $0.836\pm 0.025$       &  $0.843\pm 0.025$ \\
\cite{Mergell:1995bf}     & $0.847\pm 0.008$       &  $0.836\pm 0.008$ \\
\cite{Hammer:2003ai}
                          & $0.848$*                & $0.857$* \\
\cite{Belushkin:2006qa}   & $0.844^{+0.008}_{-0.004}$  &  $0.854\pm 0.005$ \\ 
\cite{Lorenz:2012tm}      & $0.84\pm 0.01$          &  $0.86^{+0.02}_{-0.03}$ \\
\cite{Lorenz:2014yda}     & $0.840^{+0.015}_{-0.012}$  &  $0.848^{+0.06}_{-0.05}$ \\
\cite{Lin:2021umk}        & $0.838^{+0.005}_{-0.004}{}^{+0.004}_{-0.003}$
                          & $0.847\pm{0.004}\pm{0.004}$\\
\hline
\end{tabular}
\caption{Proton electromagnetic radii from various dispersion-theoretical analysis.
The * marks a quantity, where no error  was given.}
\label{tab:protonradii}
\vspace{-3mm}
\end{table}

\begin{table}[t]
\centering  
\begin{tabular}{|c|cc|}
\hline
Ref.                      &   $(r_E^n)^2$~[fm$^2$]         &  $r_M^n$~[fm] \\
\hline
\cite{Hohler:1976ax}      & $-0.117\pm 0.004^\dagger$       &  $0.873\pm 0.087$ \\
\cite{Mergell:1995bf}     & $-0.113\pm 0.004^\dagger$       &  $0.889\pm 0.009$ \\
\cite{Hammer:2003ai}
                          & $-0.113\pm 0.004^\dagger$       & $0.879$* \\
\cite{Belushkin:2006qa}   & $-0.117^{+0.007}_{-0.011}$     &  $0.862^{+0.009}_{-0.008}$ \\ 
\cite{Lorenz:2012tm}      & $-0.127$*               &  $0.88\pm 0.05$ \\
\hline
\end{tabular}
\caption{Neutron electromagnetic radius squared and magnetic radius
  from various dispersion-theoretical analysis.
The ${}^\dagger$  denotes an input quantity.  
The * marks a quantity, where no error  was given.}
\label{tab:neutronradii}
\vspace{-3mm}
\end{table}

The high-precision data with $Q^2 \leq 1\,$GeV$^2$ that emer\-ged from MAMI-C in
2010~\cite{Bernauer:2010wm,Bernauer:2013tpr} called for a further update of the DR analysis.  
A first DR analysis in Ref.~\cite{Lorenz:2012tm}
utilized the same continua as BHM with the  $\omega$, the $\phi$ and three/five effective
isoscalar/isovector poles. The fit was done to the reconstructed MAMI cross section data
in the one-photon approximation
and simultaneously to the neutron form factor data. The uncertainties were obtained varying
the continua within reasonable ranges, namely the $2\pi$ continuum by 5\% and  the $K\bar{K}$ 
and $\pi\rho$ continua by 20\%.
Again, a small proton charge radius, $r_E^p =0.84\,$fm, emerged and the other radii
were also agreeing with early DR determinations, very different to the values quoted
in~\cite{Bernauer:2010wm}.

This work was further improved in various aspects in Ref.~\cite{Lorenz:2014yda}.
Here, only proton data were investigated, but two-photon corrections to the cross section
were calculated and systematically included to the MAMI-C data, overcoming some inconsistencies in older
approaches to this problem.  Furthermore, to extract  the statistical error
due to the fitting procedure, a bootstrap approach was implemented. The spectral function
was the same as in ~\cite{Lorenz:2012tm}, but in addition, normalization constants for the
various data sets were included (in total 31 new parameters) and the $\chi^2$ definition
was augmented by the correlation matrix. This method constituted an improvement about
earlier error determinations. The uncertainties in the radii were somewhat
increased compared to earlier determinations, see Tab.~\ref{tab:protonradii} and
Tab.~\ref{tab:neutronradii}.
The measured proton form factor ratio data for $Q^2 <1\,$GeV$^2$
\cite{Ron:2011rd,Zhan:2011ji} were not included in the fits but well described.

The work of Ref.~\cite{Lorenz:2014yda} was extended by including neutron space-like
form factor data as well as then existing data for the proton and the neutron
in the time-like region in~\cite{Lorenz:2015pba}. The emphasis of this work was to
understand the structures seen by the BaBar collaboration~\cite{Lees:2013ebn}
in the region between threshold up to highest measured momentum transfers.
These structures (and similar but less pronounced ones in $e^+e^-\to n\bar{n}$)
could be explained by including a $\phi(2170)$ meson as well as the $N\bar{\Delta}$
and $\Delta\bar{\Delta}$ thresholds.

A significant improvement of the isovector spectral functions was reported in
Ref.~\cite{Hoferichter:2016duk}, based on the high-precision analysis of
pion-nucleon scattering in the framework of the so-called Roy-Steiner
equations~\cite{Hoferichter:2015hva}. This work also featured a detailed
investigations of the corresponding isospin breaking effects in the pion form factor and
the pion-nucleon P-wave amplitudes. The spectral functions given there serve as
input for any DR analysis.


The most recent DR analysis in~\cite{Lin:2021umk} was triggered by the
PRad data~\cite{Xiong:2019umf}, that measured $ep$ cross sections at
extreme forward angles corresponding to unprecedented small momentum transfers.
In Ref.~\cite{Lin:2021umk}, fits to the PRad as well as the PRad and MAMI-C
data were performed. The best fit to the combined data featured 5 isoscalar
and 5 isovector poles, while the PRad data could be well described with 2+2 poles
only. Again, the low-$Q^2$ data for $\mu_p G_E^p/G_M^p$  were not included in the
fit but could be well described. The error analysis was improved compared to earlier
DR work, the bootstrap method was used to determine the ``statistical'' error,
while the ``systematic'' error was obtained from varying the number of effective
poles, e.g. in the combined analysis the range from (2+2) to (7+7) isoscalar +
isovector poles was covered. This led to the very precise proton radii given in
Tab.~\ref{tab:protonradii}. It was pointed out that the statistical error in the
PRad analysis is underestimated, consistent with the earlier findings of
Ref.~\cite{Paz:2020prs}.
In Ref.~\cite{Lin:2021umk}, no uncertainties on the proton form factors were given
and no neutron data were analyzed. In this review, we will fill this gap and
present detailed results on these topics. Also, a Bayesian approach to calculate
the statistical errors will be presented and compared to the bootstrap method.

It is remarkable how little fluctuations in the extracted values of the nucleon em radii
based on dispersion relations  have appeared with time, despite a dramatic
improvement in the data base 
and a number of theoretical improvements, related in particular to the isoscalar
and the isovector spectral functions and calculation of the theoretical uncertainties.

There has also been some related work in the so-called dispersively improved
chiral perturbation theory,
see~\cite{Alarcon:2017lhg,Alarcon:2018irp,Alarcon:2018zbz,Alarcon:2020kcz}.
The extracted proton charge radius is consistent with our result, but
as noted in  Ref.~\cite{Leupold:2017ngs}, this  approach is subject to
uncertainties
in the $\rho$-region, different from the exact representation used in the
papers discussed above.

We end this section by noting that the so-called strange\-ness form factors of the
nucleon can also be calculated (under certain assumptions) using the DR results
for the isoscalar vector mesons, see e.g. Refs.~\cite{Jaffe:1989mj,Hammer:1995de,Forkel:1995ff,Hammer:1999uf}.
For more details on this interesting topic, see the reviews~\cite{Armstrong:2012bi,Maas:2017snj}.

\section{Formalism}
\label{sec:form}

\subsection{Definitions}

\begin{figure}[t] 
\centerline{\includegraphics*[width=3.95cm,angle=0]{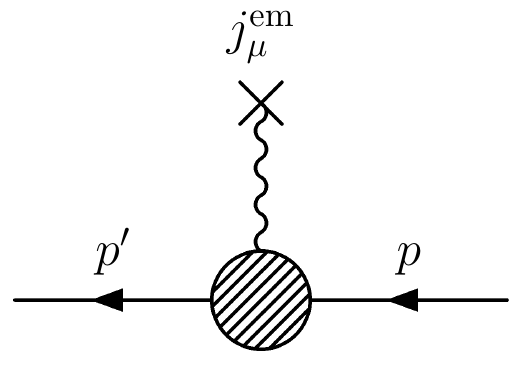}}
\caption{
The nucleon matrix element of the electromagnetic current $j_\mu^{\rm em}$.}
\label{fig:curr}
\vspace{-3mm}
\end{figure}

The electromagnetic (em) structure of the nucleon is determined by
the matrix element of the vector current operator
\begin{equation}
j_\mu^{\rm em} = \bar{q} {\cal Q} \gamma^\mu q~,
\end{equation}
for the light quarks $q=(u,d,s)^T$ with the charges ${\cal Q}=\\ {\rm diag}(2,-1,-1)/3$
(in terms of the elementary charge), sandwiched between nucleon states as depicted
in Fig.~\ref{fig:curr}.
Denoting a nucleon state with four-momentum $p$ as $|p\rangle$ (for ease
of notation, we do not display the corresponding spin or helicity index), with the
help of Lorentz and gauge invariance and assuming CP invariance, this matrix
element can   be expressed in terms of two form factors, 
\bea\label{eq:NME}
\langle p' | j_\mu^{\rm em} | p \rangle &=& \bar{u}(p')
\left[ F_1 (t) \gamma_\mu +i\frac{F_2 (t)}{2 m} \sigma_{\mu\nu}
q^\nu \right] u(p)\,, \nonumber \\
&=& \bar{u}(p')\, \Gamma^\mu(t)\, u(p)~,
\eea
where $m$ is the nucleon mass (which can be either the neutron, the
proton or the isospin averaged mass) and $t=(p'-p)^2$  the four-momentum transfer
squared. For the analysis of data in the space-like region, it is convenient to use the
variable $Q^2=-t>0$. The scalar functions $F_1(t)$ and $F_2(t)$ are the Dirac and 
Pauli form factors, respectively. They are normalized at $t=0$ as
\begin{eqnarray}
\label{norm}
F_1^p(0) &=& 1\,, \quad  \; F_1^n(0) = 0\,,\nonumber\\
F_2^p(0) &=&  \kappa_p\,, \quad  F_2^n(0) = \kappa_n\, ,
\end{eqnarray}
with $\kappa_p=1.793$ and $\kappa_n=-1.913$ the anomalous magnetic moment of
the proton and the neutron, respectively, in units of the nuclear magneton, $\mu_N = e/(2m_p)$.
The magnetic moment of the proton and the neutron is thus given by $\mu_{p} = 1 + \kappa_{p}$
and $\mu_n=\kappa_n$, respectively.

For the theoretical analysis, it is often convenient to work in the isospin basis and to 
decompose the form factors into isoscalar ($s$) and isovector ($v$) parts,
\begin{equation}
F_i^s = \frac{1}{2} (F_i^p + F_i^n) \, , \quad
F_i^v = \frac{1}{2} (F_i^p - F_i^n) \, ,
\end{equation}
where $i = 1,2 \,$. 
The experimental data are usually given in terms of the Sachs form factors
\begin{eqnarray}
G_{E}(t) &=& F_1(t) - \tau F_2(t) \, , \nonumber\\
G_{M}(t) &=& F_1(t) + F_2(t) \, , 
\label{sachs}
\end{eqnarray}
where $\tau = -t/(4 m^2)$.
In the Breit frame, $G_{E}$ and $G_{M}$ may be interpreted as
the Fourier transforms of the charge and magnetization distributions,
respectively.              

The nucleon radii
\begin{equation}
\label{def:r1}
  r \equiv \sqrt{\langle r^2 \rangle}
\end{equation}  
are defined via the low-$t$ expansion of the form factors,
\beq
\label{def:r2}
F(t)=F(0)\left[1+t \frac{\langle r^2 \rangle}{6} +\ldots \right]\,,
\eeq
where $F(t)$ is a generic form factor. In the case of the electric
and Dirac form factors of the neutron, $G_E^n$ and $F_1^n$, the
expansion starts with the term linear in $t$ and the 
normalization factor $F(0)$ is dropped. Note that the slopes of $G_E^n$ and $F_1^n$ are 
related via
\beq
\frac{dG_E^n(Q^2)}{dQ^2}\biggr|_{Q^2=0} = \frac{dF_1^n(Q^2)}{dQ^2}\biggr|_{Q^2=0} 
- \frac{F_2^n(0)}{4m^2_n}~,
\eeq
with $m_n$ the neutron mass. We remark that alternative information on the
proton charge radius can be obtained from Lamb shift measurements in electronic
as well as muonic hydrogen, see e.g. the reviews~\cite{Pohl:2013yb,Karr:2020wgh}.

In the space-like region with $t<0$, the form factors are real valued quantities.
This is different in the time-like region with $t>0$. By their very definition, at the
nucleon-antinucleon threshold, $t_{\rm thr} = 4m^2$, they fulfill the relation
\begin{equation}
G_E(4m^2) = G_M(4m^2)~,
\end{equation} 
for both the proton and the neutron. In the physical region $t>4m^2$, the FFs are
complex valued quantities.

\begin{figure}[t] 
\centerline{\includegraphics*[width=0.9\linewidth,angle=0]{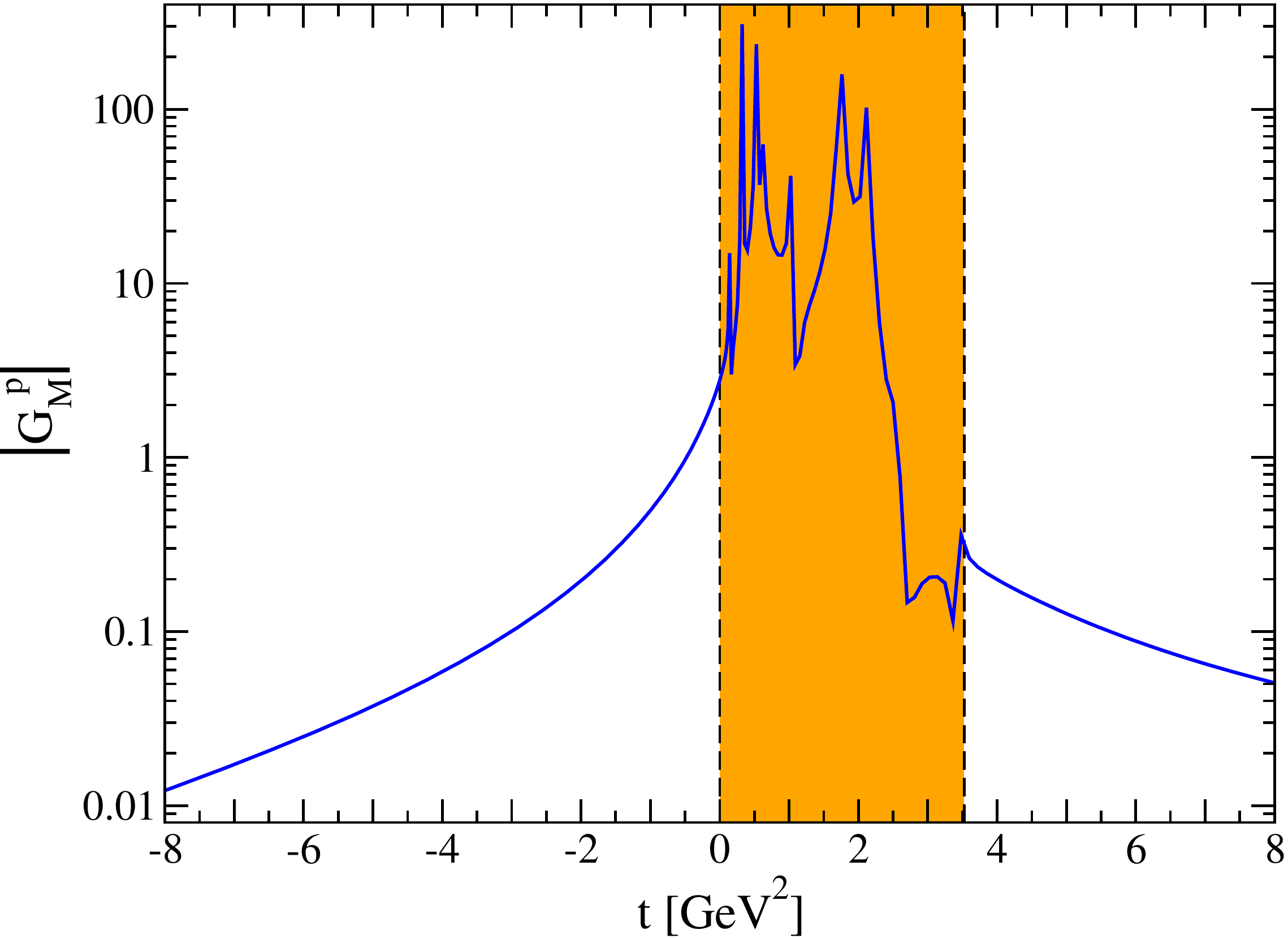}}
\caption{The form factor modulus $|G_M^p(t)|$ for all possible momentum
  transfers based on a fit of space-lie and time-like data around 1996~\cite{Hammer:1996kx}.
  The colored area between the two dashed lines at $t=0$ and $t=4m^2$
  is the unphysical region where the form factor cannot be observed.}
\label{fig:FFgen}
\vspace{-3mm}
\end{figure} 
In Fig.~\ref{fig:FFgen}, we sketch an exemplary form factor (here: $G_M^p(t)$) for all
values of $t$. More precisely, the modulus of the form factor is depicted. For the
space-like region, the threshold is located at $t=0$, whereas the corresponding
threshold in the time-like region is $t=4m^2$. In between these two thresholds,
the various vector meson poles (plus continua) build up the spectral function
to be discussed in detail below. This region cannot be observed. 
We note that for the form factors in the time-like
region, an additional complication arises due to the strong near-threshold nucleon-antinucleon
interactions, which will be considered in Sec.~\ref{sec:FSI}.

\subsection{Elementary cross section and polarization transfer}
\label{sec:XS}

The form factors (FFs) can not be measured directly but are encoded in observables related to
electron scattering. Consider for definiteness electron-proton ($ep$) scattering,
\beq
e\,(p_1) + p\,(p_2) \to e\,(p_3) + p\,(p_4)~,
\eeq
where the four-momenta $p_i$ are subject to the constraint $p_1+p_2 = p_3+p_4$. 
At first order in the electromagnetic fine-structure constant $\alpha$, the Born-approximation,
the differential cross section can be expressed through the Sachs FFs as
\begin{equation}\label{eq:xs_ros}
\frac{d\sigma}{d\Omega} = \left( \frac{d\sigma}
{d\Omega}\right)_{\rm Mott} \frac{1}{\epsilon (1+\tau)}
\underbrace{\left[\tau G_{M}^{2}(Q^{2}) + {\epsilon} G_{E}^{2}(Q^{2})\right]}_{= \sigma_R}\, ,
\end{equation}
where
\beq
\epsilon = [1+2(1+\tau)\tan^{2} (\theta/2)]^{-1}~, ~~ 0\leq \epsilon \leq 1~,
\eeq
is the virtual photon polarization,
$\theta$ is the electron scattering angle in the laboratory frame, and
$({d\sigma}/{d\Omega})_{\rm Mott}$ is the Mott cross section, which corresponds to scattering off
a point-like particle,
\begin{equation}\label{eq:xs_mott}
\left( \frac{d\sigma}{d\Omega}\right)_{\rm Mott} =
\frac{\alpha^2 \cos^2(\theta/2)}{4E_1^2\sin^4(\theta/2)} \frac{E_3}{E_1}~,
\end{equation}
where $E_1\ (E_3)$ is the energy of the incoming (outgoing) electron.
Two quantities out of the energies, momenta and angles suffice to determine this cross
section and are related for such an elastic process. Specifically, in the laboratory frame
with the initial nucleon at rest and neglecting the electron mass, we can write
\begin{align}
Q^2 \approx 4E_1E_3\sin^2\left({\theta}/{2}\right)~.
\end{align}
In experiment, the differential cross section is usually given for a fixed total energy
as a function of the scattering angle, so that a small scattering angle corresponds to
a small momentum transfer. This is exactly the reason why a precise determination of the
em radii is so difficult.  At large momentum transfer, the contribution from the magnetic FF
dominates the cross section.
The contribution from the electric and the magnetic form factor
can be read off form the reduced cross section $\sigma_R$ defined in Eq.~\eqref{eq:xs_ros}.
The reduced cross section $\sigma_R$ depends linearly on $\epsilon$ for a given $Q^2$, with slope $G_E(Q^2)$ and
intercept $\tau G_M^2(Q^2)$. This is called the Rosenbluth separation~\cite{Rosenbluth:1950yq}. 
Two-photon corrections
to this cross section will be discussed in Sect.~\ref{sec:2gamma}.  Also, to investigate the
neutron FFs,
one measures electron scattering of a light nucleus like deuterium or $^3$He. This requires,
however, some accurate few-body technique to disentangle the neutron contribution from the
scattering cross section, as  discussed briefly in App.~\ref{app:nuclei}.

In early $ep$ scattering experiments,  it was found that the form factors could be
well approximated by the dipole form, $G_{\rm dip}(Q^2)$,
\begin{eqnarray}
G_E^p(Q^2) &\simeq& \frac{G_M^p(Q^2)}{\mu_p} \simeq  \frac{G_M^n(Q^2)}{\mu_n} \simeq G_{\rm dip}(Q^2)~,
\nonumber\\
G_{\rm dip}(Q^2) &=& \left(1+ \frac{Q^2}{0.71~{\rm GeV}^2}\right)^{-2}~,
\end{eqnarray}
with $G_E^n(Q^2)=0$ in this approximation.
Employing these dipole FFs in the integrated cross section Eq.~\eqref{eq:xs_ros} defines the
so-called dipole cross section, $\sigma_{\rm dip}$. Often, the form factors or the measured
cross sections are given relative to $G_{\rm dip}(Q^2)$ and $\sigma_{\rm dip}$, respectively.

A method to directly measure the form factor ratio $G_E/G_M$ in polarized electron scattering
off the proton, $\overrightarrow{e}p\to\overrightarrow{e}p$ (or similarly for scattering off
the deuteron or $^3$He), has
been proposed in Refs.~\cite{Akhiezer:1968ek,Arnold:1980zj}. A simultaneous measurement
of the two recoil polarizations (longitudinal, $P_l$, and transverse, $P_t$) allows one to measure
directly the ratio
\begin{equation}
R_p \equiv \mu_p \frac{G_E^p}{G_M^p} = -\mu_p \sqrt{\frac{\tau(1+\epsilon)}{2\epsilon}}\frac{P_t}{P_l}~.
\end{equation}  
While this only determines the form factor ratio (and not the individual FFs), many
systematic uncertainties cancel out and make this observable an important benchmark
for any theoretical form factor calculation.

Let us briefly discuss the determination of the form factors in the time-like region.
They can be extracted from the cross section data $e^+e^- \leftrightarrow \bar{p}p$
and  $e^+e^-\to \bar{n}n$ for the proton and the neutron, respectively.
As only very few differential cross section data exist
in the time-like region, a separation of $G_E$ and $G_M$ is often not possible and
one either makes an assumption like e.g. $G_E = G_M$ in the analysis of the data
or one extracts the effective form factor $|G_{\rm eff}|$, discussed below.
For a review on  the nucleon em form factors in the time-like region,
see Ref.~\cite{Denig:2012by}.

We now consider the process,
\beq
e^+\,(p_1)+e^-\,(p_2)\to
p\, (p_3)+\bar{p}\,(p_4)~,
\eeq
in more detail. It is convenient
to choose the center-of-mass (CM) frame, i.e., $p_{1,2} = (E,\pm k_e)$ and $p_{3,4} = (E,\pm k_p)$.
The photon momentum $q$ then determines the center-of-mass energy by $q^2 = (p_1+p_2)^2=t
= E_{\rm CM}^2 = (2E)^2$. In the metric used here, time-like $q$ implies positive $q^2$.
The three-momenta $k_e,k_p$ appear in the phase-space factor $\beta = k_p/k_e$,  which in the limit 
of neglecting the electron mass yields
\beq
\beta \approx k_p/E = \sqrt{1 - 4m_p^2/q^2}~,
\eeq
the velocity of the proton, and $m_p$ is the proton mass. We denote the emission
angle of the proton by $\theta$. The differential cross section in the one-photon-exchange
approximation then is
\bea
\frac{d\sigma}{d\Omega}
&=& \frac{\alpha^2\beta}{4 q^2}C(q^2)\biggl[(1+\cos^2\theta)|G_M(q^2)|^2 \nonumber\\
&& \qquad\quad + \frac{4m_p^2}{q^2}\sin^2\theta|G_E(q^2)|^2\biggr]~,
\eea
in terms of the electric and magnetic Sachs form factors and $C(q^2)$ is the Sommerfeld-Gamow
factor that accounts for the Coulomb interaction between the final-state particles
\begin{align}
 C(q^2)=\frac{y}{1-e^{-y}},\hspace{8pt} y=\frac{\pi\alpha m_p}{k_p}.
\end{align}
Integrating over the full angular distribution gives the total cross section
\begin{align}
\sigma_{e^+e^- \rightarrow p\bar{p}}(q^2) &= \frac{4\pi\alpha^2\beta}{3q^2}C(q^2)
\left[|G_M(q^2)|^2+\frac{2m_p^2}{q^2}|G_E(q^2)|^2\right]\notag\\
&\equiv \frac{4\pi\alpha^2\beta}{3q^2}C(q^2)\left(1+\frac{2m_p^2}{q^2}\right)|G_{\rm eff}^p(q^2)|^2.
\end{align}
This defines the effective form factor
$G_{\rm eff}$
\begin{align}
\left|G_{\rm eff}\right| \equiv \sqrt{\frac{|G_E|^2+\frac{q^2}{2m_p^2}|G_M|^2}{1+\frac{q^2}{2m_p^2}}}~.
\label{geff}
\end{align}
For neutrons, the formulas are equivalent except for the Sommerfeld-Gamow factor which is
not present in that case. Beyond the Coulomb final-state interactions, higher order QED
corrections are usually neglected.  For the time-reversed process,
the phase space factor is inverted, yielding
\beq
\sigma(e^+e^-\to p\bar{p}) = \beta^2\, \sigma( p\bar{p}\to e^+e^-)~.
\eeq
Taking into account the angular dependence of $p\bar{p}$ production, one can express the
differential cross section via the angular asymmetry $\mathcal{A}$, see
Ref.~\cite{TomasiGustafsson:2001za},
\beq
\frac{d\sigma}{d\Omega} = \frac{d\sigma}{d\Omega}\biggr|_{\theta=90^\circ}
 [1+\mathcal{A}\cos^2\theta],
\eeq
with
\begin{align}
 \mathcal{A} = \frac{{q^2}/{(4m_p^2)} - R^2}{{q^2}/{(4m_p^2)} + R^2}~.
\end{align}
This can be  determined from the FF ratio $R = |G_E/G_M|$.

\subsection{Dispersion relations and spectral decomposition}
\label{sec:specdeco}

\begin{figure}[t] 
\centerline{\includegraphics*[width=0.8\linewidth,angle=0]{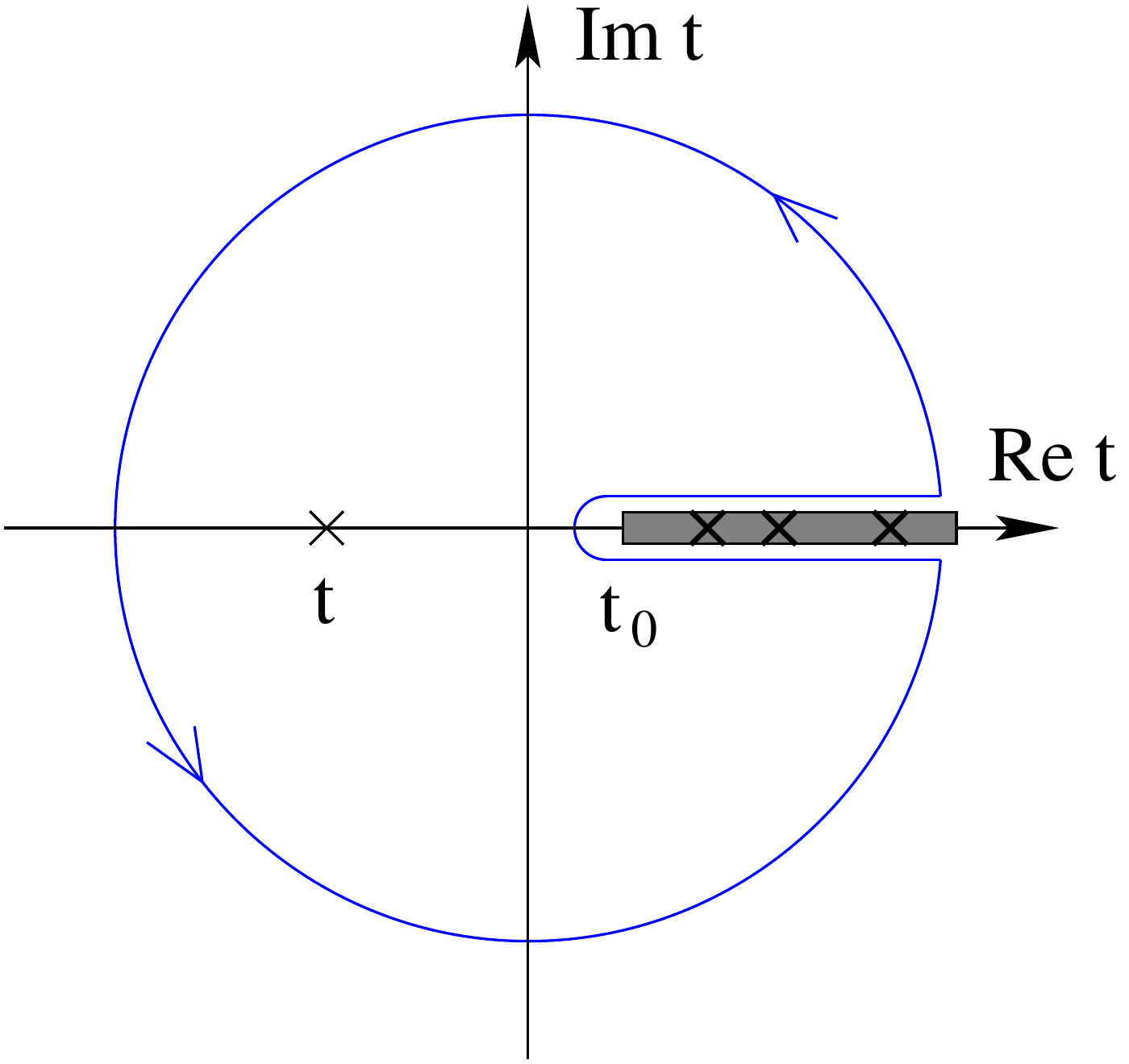}}
\caption{Analytic structure of a generic form factor in the complex-$t$ plane.
Shown are the lowest continuum cut at $t_0$ and a number of vector meson poles
at positive $t$ (crosses). The dispersion integral is calculated for the space-like
value of $t$ also shown by a cross.
}
\label{fig:cauchy}
\vspace{-3mm}
\end{figure}

Dispersion relations (DRs) are based  on unitarity and analyticity. Here, DRs
relate the real and imaginary parts of the electromagnetic nucleon form factors. 
Let $F(t)$ be a generic symbol for any one of the four independent 
nucleon form factors. We write down an unsubtracted
dispersion relation of the form
\begin{equation}
F(t) = \frac{1}{\pi} \, \int_{t_0}^\infty \frac{{\rm Im}\, 
F(t')}{t'-t-i\epsilon}\, dt'\, ,
\label{emff:disp} 
\end{equation}
where $t_0$ is the threshold of the lowest cut of $F(t)$ (see below)
and the $i\epsilon$ defines the integral for values of $t$ on the 
cut. The convergence of an unsubtracted dispersion relation
for the form factors has been assumed. For proofs of such a representation
in perturbation theory, see Ref.~\cite{DZbook} (and references therein).
One could also use 
a once subtracted dispersion relation, since the normalization of the
form factors at $t=0$ is known. However, in what follows, we will only
employ the unsubtracted form give in Eq.~\eqref{emff:disp}.
Most importantly, by  Eq.~(\ref{emff:disp}) the electromagnetic structure
of the nucleon can be related to its absorptive behavior. In Fig.~\ref{fig:cauchy}
we display the analytic structure underlying the dispersion integral
in Eq.~\eqref{emff:disp}. The various ingredients (continuum cuts, vector
meson poles) will be discussed in detail below.

The imaginary part ${\rm Im}\, F$ entering Eq.~(\ref{emff:disp}) 
can be obtained from a spectral decomposition \cite{Chew:1958zjr,Federbush:1958zz}. 
For this purpose, consider the  electromagnetic 
current matrix element in the time-like region ($t>0$), which is 
related to the space-like region ($t<0$) via crossing symmetry.
This matrix element is given by
\begin{eqnarray}
\label{eqJ}
J_\mu &=& \langle N(p_3) \overline{N}(p_4) | j_\mu^{\rm em}(0) | 0 \rangle \\
&=& \bar{u}(p_3) \left[ F_1 (t) \gamma_\mu +i\frac{F_2 (t)}{2 m} \sigma_{\mu\nu}
(p_3+p_4)^\nu \right] v(\bar{p_4})\,,\nonumber
\end{eqnarray}
where $p$ and $\bar{p}$ are the momenta of the nucleon and an\-ti\-nuc\-le\-on
created by the current $j_\mu^{\rm em}$, respectively. 
The four-momentum transfer squared
in the time-like region is $t=(p_3+p_4)^2$. 

\begin{figure}[t] 
\centerline{\includegraphics*[width=0.35\textwidth,angle=0]{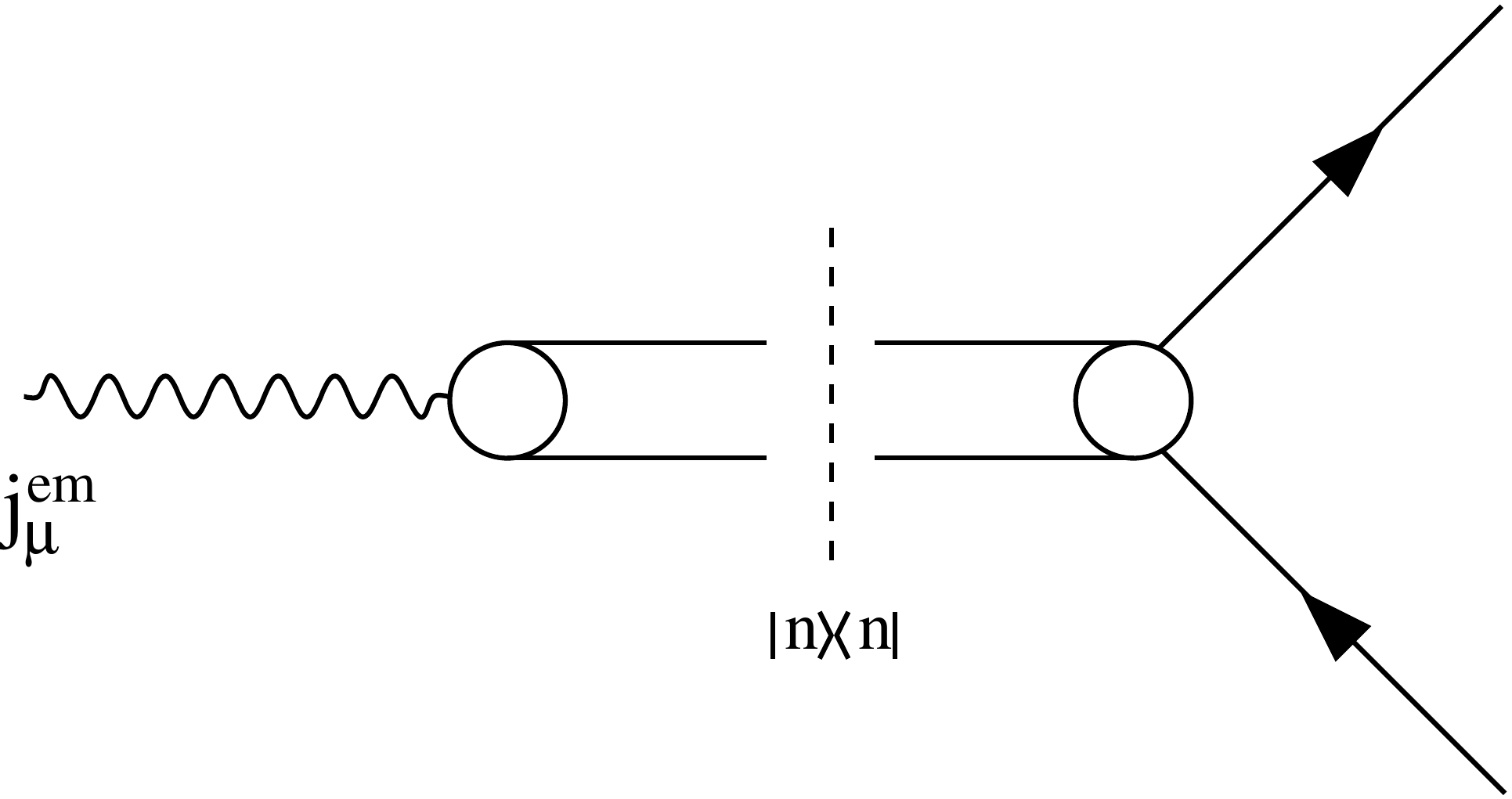}}
\caption{
The spectral decomposition of the 
nucleon matrix element of the electromagnetic current $j_\mu^{\rm em}$.
$| \lambda \rangle $ denotes an hadronic intermediate state.
}
\label{fig:spec}
\vspace{-3mm}
\end{figure}

Using the LSZ reduction formalism, the imaginary part
of the form factors is obtained by inserting a complete set of
intermediate states as \cite{Chew:1958zjr,Federbush:1958zz}
\begin{eqnarray}
\label{spectro}
{\rm Im}\,J_\mu &=& \frac{\pi}{Z}(2\pi)^{3/2}{\cal N}\,\sum_\lambda
\langle p_3 | \bar{J}_N (0) | \lambda \rangle  \nonumber\\
&\times&
\langle \lambda | j_\mu^{\rm em} (0) | 0 \rangle \,v(p_4)
\,\delta^4(p_3+p_4-p_\lambda)\,,
\end{eqnarray}
where ${\cal N}$ is a nucleon spinor normalization factor, $Z$ is
the nucleon wave function renormalization, and $\bar{J}_N (x) =
J^\dagger(x) \gamma_0$ with $J_N(x)$ a nucleon source.
This decomposition is illustrated in Fig.~\ref{fig:spec}.
It relates the spectral function to on-shell matrix elements of other
processes, as detailed below.

The states $|\lambda\rangle$ are asymptotic (observable) states of
momentum $p_\lambda$. They carry the {\em same} quantum numbers as
the current $j^{\rm em}_\mu$:
\begin{eqnarray}
I^G(J^{PC}) &=& 0^-(1^{--})~ \mbox{for the isoscalar component}~,\nonumber\\
I^G(J^{PC}) &=& 1^+(1^{--})~ \mbox{for the isovector component}
\end{eqnarray}
of the current $j^{\rm em}_\mu$. Here, $I$ and $J$ denote the isospin $I=0,1$
and the angular momentum $J=1$ of the photon, whereas $G$, $P$ and $C$ give
the $G$-parity, parity and charge conjugation quantum number, respectively.
Furthermore, these currents  have zero net baryon number. Because of $G$-parity, states
with an odd number of pions only contribute to the iso\-scalar
part, while states with an even number contribute to the 
isovector part.
For the isoscalar part  the lowest mass states are:
\begin{equation}
3\pi, 5\pi, \ldots, 
 K\bar{K}, K\bar{K}\pi, \ldots\ ,
\end{equation}
and for the isovector part they are:
\begin{equation}
2\pi, 4\pi, \ldots\ . 
\end{equation}
Associated with each intermediate state is a
cut starting at the corresponding threshold in $t$ and running to
infinity. As a consequence,
the spectral function ${\rm Im}\, F(t)$ is different from zero along the
cut from $t_0$ to $\infty$, with $t_0 = 4 \, (9) \, M_\pi^2$ for the
isovector (isoscalar) case.

The spectral functions are the central quantities in the 
dispersion-theoretical approach. Using Eqs.~(\ref{eqJ},\ref{spectro}), they
can in principle be constructed from experimental data. 
In practice, this program can only be carried out for 
the lightest two-particle intermediate states.

The longest-range, and therefore at low momentum transfer most 
important continuum contribution comes from the $2\pi$ intermediate state
which contributes to the isovector form factors~\cite{Hohler:1974eq}.
A novel and very precise  calculation of this contribution 
has recently been performed in Ref.~\cite{Hoferichter:2016duk} including
the state-of-the-art pion-nucleon scattering amplitudes from dispersion theory,
as detailed below.
In the isoscalar channel, the  inclusion of the $K\bar{K}$ \cite{Hammer:1998rz,Hammer:1999uf}
and  $\rho\pi$ continua \cite{Meissner:1997qt}
was first introduced in Ref.~\cite{Belushkin:2006qa} in the dispersive analysis of the
em form factors. These important ingredients are discussed in more detail below.
Apart from the continua, there are also single vector-meson pole contributions.
As will become clear in the following, the contributions from the  continua and
the poles are sometimes strongly intertwined, e.g. the $\rho$-meson pole is indeed
generated as part of the $2\pi$-continuum, as known since
long~\cite{Frazer:1959gy,Frazer:1960zza,Frazer:1960zzb}.

\subsection{Two-pion continuum}
\label{sec:2pi}
\begin{figure}[t] 
\centerline{\includegraphics*[width=0.35\textwidth,angle=0]{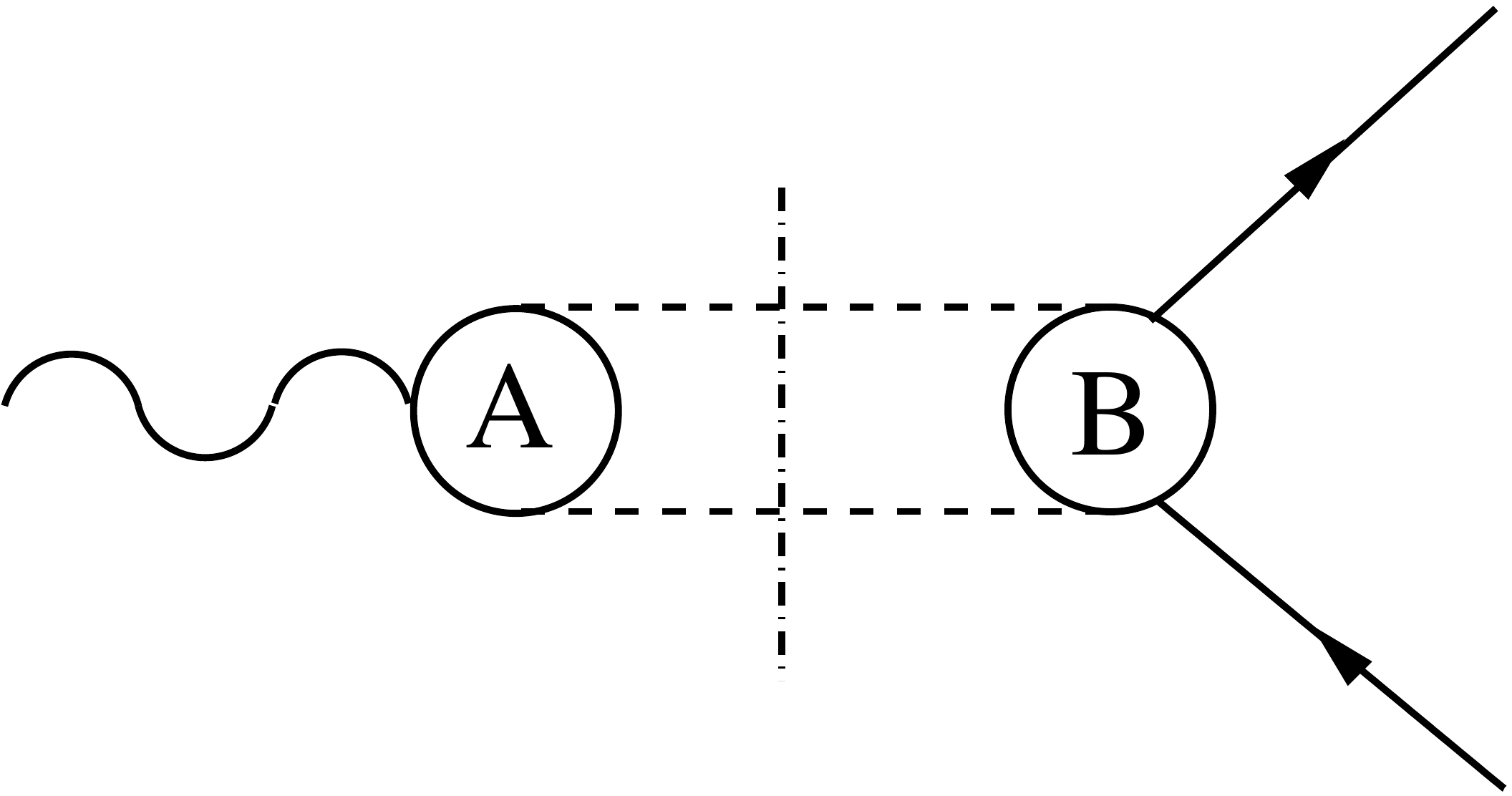}}
\caption{
Two-pion cut contribution to the isovector form factors, given in terms
of the pion vector form factor $F_\pi^V$ (represented by $A$) and  the $\pi\pi\to\bar N N$
P-waves $f^1_\pm$ (represented by $B$). Solid, dashed, and wiggly lines denote nucleons,
pions, and the external photon, respectively,
while the dash-dotted line indicates the cutting of particle propagators.
}
\label{fig:2picont}
\vspace{-3mm}
\end{figure} 
In the isovector channel, the lowest continuum contribution is given by the
two-pion exchange as depicted in Fig.~\ref{fig:2picont}. Therefore,
the unitarity relations for the nucleon form factors reads~\cite{Frazer:1960zzb}
\begin{align}
\label{unitarity_relation}
{\rm Im}~G_E^v(t)&=\frac{q_t^3}{m\sqrt{t}} \,F_\pi^V(t)^*\, f_+^1(t)\, \theta\big(t-4M_\pi^2\big)~,
\notag\\
{\rm Im}~G_M^v(t)&=\frac{q_t^3}{\sqrt{2t}}\, F_\pi^V(t)^*\, f_-^1(t)\, \theta\big(t-4M_\pi^2\big)~,
\end{align}
with $q_t = \sqrt{t/4-M_\pi^2}$, $F_\pi^V(t)$ is the vector form factor of the pion and the
$f_\pm^1(t)$ are the P-wave $\pi N$ partial waves in the t-channel. Watson's theorem  ensures
that the left-hand side of the equations stays real, as long as the same $\pi\pi$ phase
shift is used in the calculation of the pion form factor
and the $\pi\pi\to\bar N N$ partial waves. Therefore, in the most recent determination of the
two-pion continuum, the same three variants of the phase shift $\delta^1_1$ in the data fits
for $F_\pi^V(t)$ were used  as in the Roy-Steiner analysis of pion-nucleon
scattering~\cite{Hoferichter:2015hva}. 
The full consistency among all ingredients entering the unitarity relation that was achieved
in Ref.~\cite{Hoferichter:2016duk} was a key improvement over earlier calculations, and thus
the representation of the two-pion continuum given there will be discussed in what follows.

It is important to discuss the range of validity of the $2\pi$ approximation to the unitarity
relation. Strictly speaking, the $4\pi$ threshold opens at $\sqrt{t}=4M_\pi=0.56$~GeV, but
it is well known from phenomenology that the $4\pi$ contribution is completely negligible below
the $\omega\pi$ threshold at $\sqrt{t}=0.92$~GeV, see e.g.~\cite{Eidelman:2003uh}, and only
becomes sizable once the $\rho'$, $\rho''$ resonances are excited. This can also be
understood from chiral perturbation theory, where the $4\pi$ contribution appears first
at three loop order~\cite{Gasser:1990bv}. For this reason, the two-pion cut contribution
to the isovector form factors is considered up to $\sqrt{t} \simeq \sqrt{50} M_\pi \simeq 1\,$GeV.

Let us now discuss in more detail the various ingredients entering the Eqs.~\eqref{unitarity_relation}.
We start with the vector (em) form factor of the pion. It is given by
\beq
\langle \pi^+(p')|j^\mu_\text{em}|\pi^+(p)\rangle =(p+p')^\mu F_\pi^V(t).
\eeq
The $\pi\pi$ intermediate states produce the unitarity relation
\beq
\label{unitarity_FpiV}
{\rm Im}~F_\pi^V(t)=\sin\delta^1_1(t) e^{-i\delta^1_1(t)} F_\pi^V(t) \theta\big(t-4M_\pi^2\big),
\eeq
with the $\pi\pi$ P-wave phase shift $\delta^1_1$. Eq.~\eqref{unitarity_FpiV} reflects Watson's
final-state theorem~\cite{Watson:1954uc},
which states that the phase of $F_\pi^V$ has to coincide with the $\pi\pi$ scattering phase shift
(up to multiple integers of $\pi$). Neglecting higher intermediate states unitarity determines
$F_\pi^V(t)$ up to a polynomial $P(t)$ in terms of the Omn\`es factor $\Omega^1_1(t)$~\cite{Omnes:1958hv}
\beq
\label{FpiOmnes}
 F_\pi^V(t)=P(t)\Omega^1_1(t)
 =P(t)\exp\Bigg\{\frac{t}{\pi}\int\limits_{4M_\pi^2}^{\infty} dt'\frac{\delta^1_1(t')}{t'(t'-t)}\Bigg\}.
\eeq
In fact, the representation~\eqref{FpiOmnes} provides a very efficient and accurate parameterization of
the experimental data, up to the distortions due to $\rho$--$\omega$ mixing.
This isospin-violating effect can be included via a modification of $F_\pi^V(t)$,
\beq
\label{FpiV}
F_\pi^V(t) = \bigg(1+\alpha t+ \frac{\eps\,t}{M_\omega^2-iM_\omega\Gamma_\omega-t} \bigg)\Omega_1^1(t), 
\eeq
with the $\omega$ mass $M_\omega$ and width $\Gamma_\omega$. The parameters $\alpha$ and $\eps$, where
$\eps$ parameterizes the strength of the $\omega$-$\rho$ mixing, are
fit to recent form factor data, see Refs.~\cite{Aubert:2009ad,Babusci:2012rp,Ablikim:2015orh}
below $\sqrt{t}=1\,$GeV, using the same $\pi\pi$ phase shifts as in the RS
analysis~\cite{Hoferichter:2015hva}. The latter has been determined from Roy and Roy-like equations
by the Bern~\cite{Caprini:2011ky}   and the Madrid-Cracow
group~\cite{GarciaMartin:2011cn}. To get a better handle on the uncertainty estimate for the
final spectral functions from the pion vector FF, in Ref.~\cite{Hoferichter:2016duk}
a variant of the Bernese phase shift was also considered. It includes effects from the $\rho'$
and the $\rho''$ in an elastic approximation~\cite{Schneider:2012ez}.

Next, we discuss the the $t$-channel partial waves $f_\pm^1(t)$, given by~\cite{Frazer:1960zza}
\begin{eqnarray}
\label{tprojform}
f^J_+(t)&=&-\frac{1}{4\pi}\int\limits^1_0 dz_t\;P_J(z_t)\bigg\{\frac{p_t^2}{(p_tq_t)^J}A^I
-\frac{m \, z_t}{(p_tq_t)^{J-1}} B^I\bigg\},\nonumber\\
f^J_-(t)&=&\frac{1}{4\pi}\frac{\sqrt{J(J+1)}}{2J+1}\frac{1}{(p_tq_t)^{J-1}}\nonumber\\
&&\quad\times\int\limits^1_0 dz_t\Big[P_{J-1}(z_t)-P_{J+1}(z_t)\Big]B^I,
\end{eqnarray}
with the $t$-channel scattering angle $z_t=(s-u)/(4p_tq_t)$, the $P_J$ are the Legendre
polynomials, and the momenta are $q_t$ and $p_t = \sqrt{t/4-m^2}$. Further, the standard decomposition
of the $\pi N$ scattering amplitude $T(\pi^a(q)+N(p)\to\pi^b(q')+N(p'))$ in the isospin limit
has been used,
\begin{align}
T^{ba}(s,t)&=\delta^{ba}T^+(s,t)+\frac{1}{2}[\tau^b,\tau^a]T^-(s,t),\\
T^I(s,t)&=\bar{u}(p')\bigg\{A^I(s,t)+\frac{1}{2}(\slashed q + \slashed{q'})B^I(s,t)\bigg\}u(p),\notag
\end{align}
where $a,b$ are isospin indices, the $\tau^a$ are isospin Pauli matrices, $I=\pm$ refers to isoscalar/isovector amplitudes and
$s=(p+q)^2$, $t=(p'-p)^2$, $u=(p-q')^2$ are the Mandelstam variables subject to the
constraint $s+t+u = 2(m^2 + M_\pi^2)$. The best way to determine the pion-nucleon scattering
amplitudes are undoubtedly dispersion relations, as they allow for a systematic continuation
from the physical region into the unphysical ones and further make best use of the
existing scattering data. The most modern and accurate investigations are based on the
Roy-Steiner (RS) equation analysis  of the Bonn group, developed and performed in
Refs.~\cite{Ditsche:2012fv,Hoferichter:2012wf,Hoferichter:2015dsa,Hoferichter:2015hva,RuizdeElvira:2017stg} (for earlier work by the Karlsruhe-Helsinki group, see e.g. Refs.~\cite{Koch:1980ay,Hoehler:1983}).
The RS equations, originally developed in~\cite{Steiner:1968,Hite:1973pm} (and references therein),
are hyperbolic DR of the form
\beq
(s-a)(u-a) = b~, ~~ b =  b(s,t,a)~, ~~ a,b \in \mathbb{R}~,
\eeq
which have a number of advantages compared to other formulations (like e.g. fixed-t DR).
They combine all {\em physical regions}, display an explicit $s \leftrightarrow u$ crossing,
require the absorptive parts only in regions where the corresponding partial wave expansions converge,
and, further, a judicious choice  of the parameter $a$ allows to increase the range of convergence.
The RS equations have a limited range of validity,  $\sqrt{s}\leq\sqrt{s_m} = 1.38\,$GeV and
$\sqrt{t}\leq\sqrt{t_m} = 2.00\,$GeV, where $\sqrt{s_m},\sqrt{t_m}$ denotes the so-called
matching point for the $s$- and $t$-channel partial waves, respectively. The required inputs to
solve the RS equations are
the S- and P-waves above the matching point, the higher partial waves (D-, F-, $\ldots$)
and the inelasticities. An important constraint are the pion-nucleon scattering lengths
deduced from pionic hydrogen and pionic deuterium~\cite{Baru:2011bw} (for a recent update,
see~\cite{Hirtl:2021zqf}). The output of the RS equations are the so-called subthreshold
parameters, which allow one to reconstruct the scattering amplitude in the unphysical region,
such as the $f_\pm^1(t)$ in the pseudophysical region required for the isovector spectral
functions. Some basic definitions of the $\pi N$ scattering amplitude in the unphysical region
are given in App.~\ref{app:piN}.
It should also be mentioned that the results of the RS analysis were given
with theoretical uncertainties, which to our knowledge has been the first time that
a dispersive analysis of pion-nucleon scattering provided these, for details
see~\cite{Hoferichter:2015hva}.

In Ref.~\cite{Hoferichter:2016duk}, the isospin-violating effects beyond the $\rho$-$\omega$
mixing in the pion form factor where also worked out, leading to an improved representation
of the unitarity relations, Eq.~\eqref{unitarity_relation}, namely
\begin{align}
\label{unitarity_final}
{\rm Im}~G_E^v(t)&=\frac{q_t^3}{m\sqrt{t}}|\Omega_1^1(t)||f_+^1(t)|\theta\big(t-4M_\pi^2\big)\notag\\
 &\qquad\times\bigg( 1+\alpha t + \frac{\eps\,t}{M_\omega^2+iM_\omega\Gamma_\omega-t} \bigg)\notag\\
&+\eps \,{\rm Im}~\bigg(\frac{t}{M_\omega^2-i M_\omega  \Gamma_\omega-t}\bigg)\notag\\
&\qquad\times\frac{1}{\pi}\int\limits_{4M_\pi^2}^\infty d t'\frac{\frac{q_t'^3}{m\sqrt{t'}}|\Omega_1^1(t')||f_+^1(t')|}{t'-t-i\eps},\notag\\
{\rm Im}~G_M^v(t)&=\frac{q_t^3}{\sqrt{2t}}|\Omega_1^1(t)||f_-^1(t)| \theta\big(t-4M_\pi^2\big)\notag\\
 &\qquad\times \bigg( 1 +\alpha t+ \frac{\eps\,t}{M_\omega^2+iM_\omega\Gamma_\omega-t} \bigg)\notag\\
 &+\eps \,{\rm Im}~\bigg(\frac{t}{M_\omega^2-i M_\omega  \Gamma_\omega-t}\bigg)\notag\\
 &\qquad\times\frac{1}{\pi}\int\limits_{4M_\pi^2}^\infty d t'\frac{\frac{q_t'^3}{\sqrt{2t'}}|\Omega_1^1(t')||f_-^1(t')|}{t'-t-i\eps}.
\end{align}
For a more detailed discussion of this representation, the reader is referred
to Ref.~\cite{Hoferichter:2016duk}.

\begin{figure}[t]
\centering
\includegraphics[width=0.95\linewidth,clip]{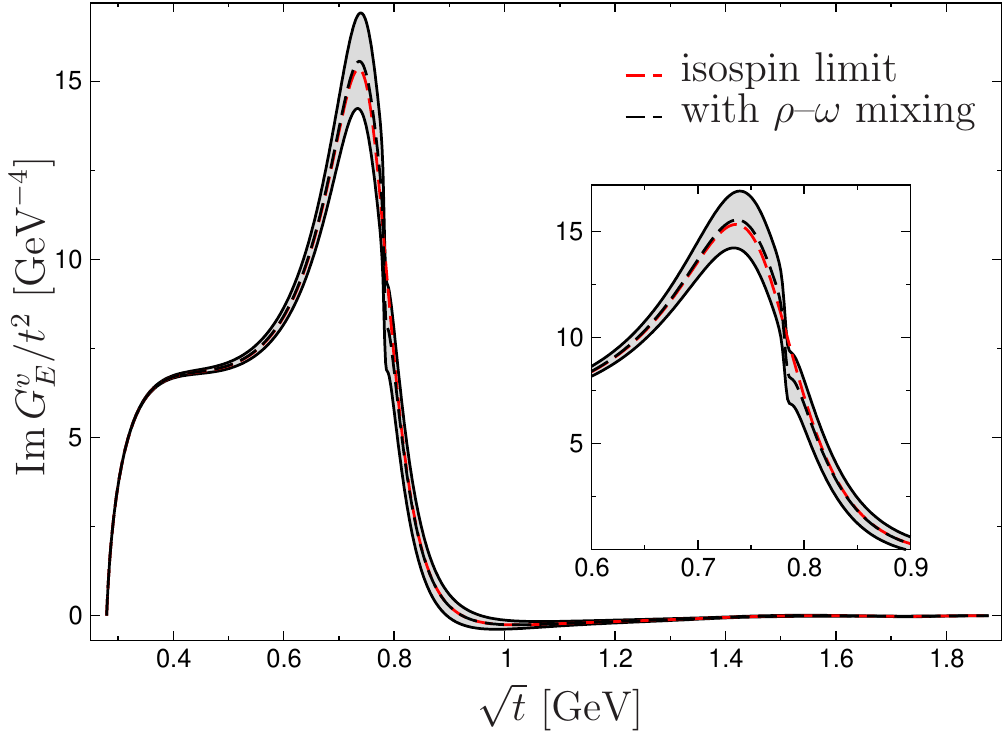}\\[0.2cm]
\includegraphics[width=0.95\linewidth,clip]{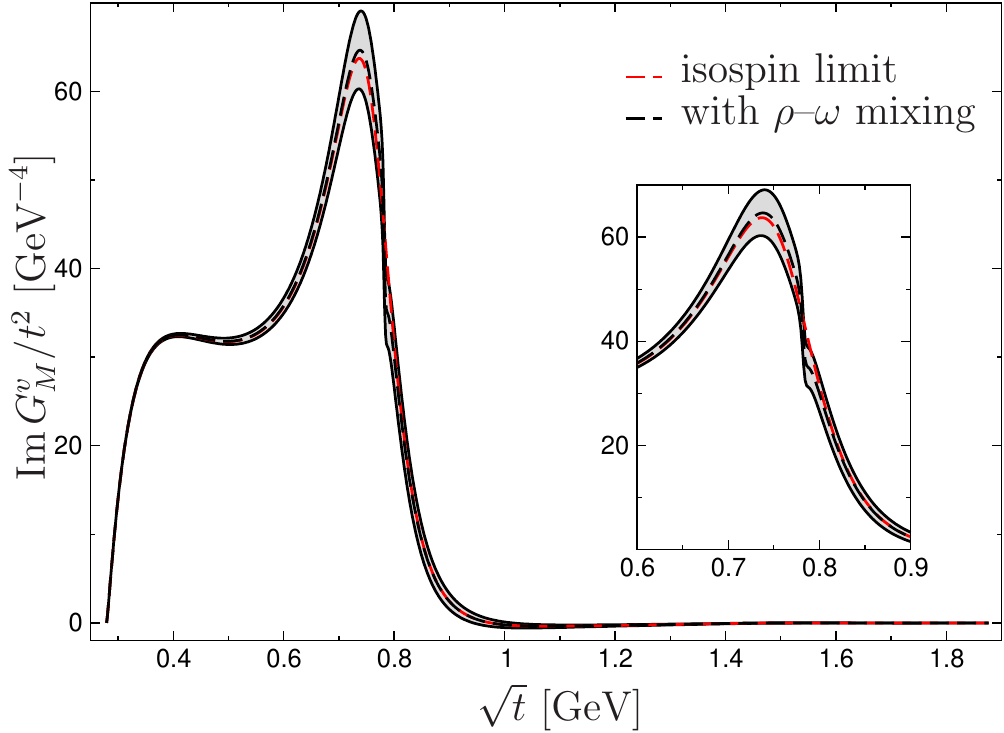}
\caption{Weighted isovector spectral functions for $G_E^v(t)/t^2$ and $G_M^v(t)/t^2$. The black
dashed line gives our central solution, the gray band the uncertainty estimate, and the
red dashed line the result if $\rho$--$\omega$ mixing is turned off. The insets magnify
the region around the $\rho$ peak. }
\label{fig:ImG}
\vspace{-3mm}
\end{figure}

Putting pieces together, the isovector  spectral functions divided by $t^2$
based on Eq.~\eqref{unitarity_final} are shown in Fig.~\ref{fig:ImG}. These nicely
exhibit the $\rho$-resonance at $\sqrt{t}=0.77\,$GeV as well as a remarkable
enhancement on the left shoulder of the resonance. This shows that the $\rho$ is indeed
generated by unitarity~\cite{Frazer:1959gy} and thus no explicit $\rho$-meson is
required in the isovector spectral function.  The visible enhancement on the left shoulder
of the $\rho$ can be traced back
to the fact that the partial wave amplitudes $f_\pm^1(t)$ have a singularity on the
second Riemann sheet \cite{Hoehler:1983} (origi\-nating from the projection of the nucleon
pole terms in the invariant pion-nucleon scattering amplitudes) located at
\beq
t_c = 4M_\pi^2 - \frac{M_\pi^4}{m^2} = 3.98\,M_\pi^2~,
\eeq
very close to the physical threshold at $t_0 = 4M_\pi^2$. The isovector form factors
inherit this singularity (on the second sheet) and the closeness to the physical
threshold leads to the pronounced enhancement  between $\sqrt{t} = 0.3 - 0.6\,$GeV
shown in Fig.~\ref{fig:ImG}. This issue will be taken up below.
The uncertainties displayed in  Fig.~\ref{fig:ImG} originate from three different
sources: 1) the subthreshold parameters  $b_{00}^-$, $b_{01}^-$, $a_{00}^-$ and $a_{01}^-$
(as defined in App.~\ref{app:piN}), 2) the pion-pion phase shift $\delta_1^1(t)$
and 3)  the data for the pion form  factor $F_\pi^V(t)$. In fact, the uncertainty
of the subthreshold parameters from the RS analysis is in fact the dominating effect
below 1~GeV.
We note that the effect of the $\rho$-$\omega$ mixing is small, as the comparison
of the black and red dashed lines in  Fig.~\ref{fig:ImG} shows. Note also that this
consistent inclusion of isospin-breaking effects in the pion em form factor and the
$\pi N$ partial waves constitutes a major achievement compared to earlier analyses.
The two-pion continuum contribution to the isovector form factors is displayed below
in Fig.~\ref{specff} .

Based on the DR, Eq.~\eqref{emff:disp}, it is is straightforward to derive
sum rules  for the normalizations and radii of the isovector form factors.
These were first considered in Ref.~\cite{Hohler:1974eq} for the various
nucleon radii, see also~\cite{Hoferichter:2016duk},
\begin{align}\label{eq:SR1}
\frac{1}{2}(r^v_E)^2&=\frac{6}{\pi}\int\limits_{4M_\pi^2}^\infty  dt \frac{{\rm Im}~G_E^v(t)}{t^2}
=\frac{1}{2}\Big[(r^p_E)^2- (r^n_E)^2\Big]~,\notag\\
\mu^v(r^v_M)^2&=\frac{6}{\pi}\int\limits_{4M_\pi^2}^\infty dt \frac{{\rm Im}~G_M^v(t)}{t^2}\notag\\
 &=\frac{1}{2}\Big[(1+\kappa_p)(r^p_M)^2-\kappa_n(r^n_M)^2\Big]~,
\end{align}
where $\mu^v=(1+\kappa_p-\kappa_n)/2\approx 2.353$ is the isovector magnetic
moment of the nucleon.
Note that the sum rules for the radii remain unchanged if a once-subtracted dispersion relation
is used instead of the unsubtracted one. Cutting the integrals at $\Lambda = 2m$, one
finds
\bea
\label{eq:SR}
\frac{1}{2}(r^v_E)^2 &=& 0.405(36)~{\rm fm}^2~, \nonumber\\
\mu^v (r^v_M)^2 &=& 1.81(11)~{\rm fm}^2~.
\eea
It is remarkable that just using a simple $\rho$-exchange using e.g. a Breit-Wigner or a
Gounaris-Sakurai form \cite{Gounaris:1968mw}, the corresponding isovector radii would be
sizeably  underestimated (by about 40\%),
as inspection of Fig.~\ref{fig:ImG} reveals. Thus, any dispersive analysis that does not
include the full two-pion continuum but only the $\rho$-resonance in the isovector spectral
function below 1~GeV will simply miss important physics.
We will come back later to these sum rules.

\subsection{Three-pion continuum}
\label{sec:3pi}

\begin{figure}[t] 
\centerline{\includegraphics*[width=0.35\textwidth,angle=0]{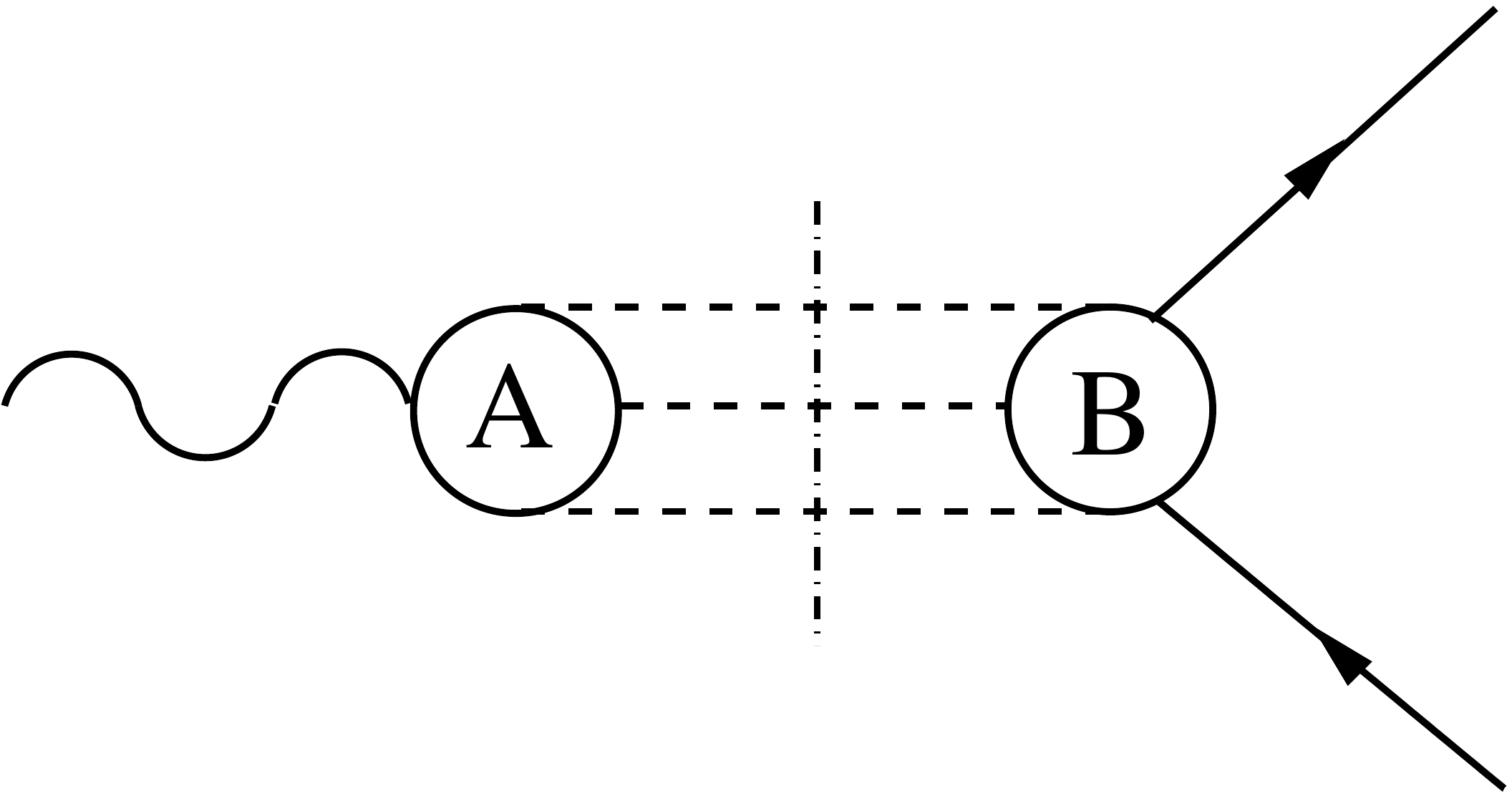}}
\caption{
Three-pion cut contribution to the isoscalar form factors, given in terms
of the $\gamma \to 3\pi$ (represented by $A$) and the $3\pi \to\bar N N$
(represented by $B$) transition amplitudes.
Solid, dashed, and wiggly lines denote nucleons, pions, and the external photon, respectively,
while the dash-dotted line indicates the cutting of particle propagators.
}
\label{fig:3picont}
\end{figure}

The lowest isoscalar continuum is given by 3-pion exchange as depicted in 
Fig.~\ref{fig:3picont}. There, $A$ refers to the $\gamma \to 3\pi$ transition amplitude,
that is given at low energies by the anomalous Wess-Zumino-Witten  Lagrangian~\cite{Wess:1971yu,Witten:1983tw}
and $B$ corresponds to the $3\pi \to \bar{N}N$ amplitude. An analysis based on unitarity alone
of this contribution does not exist, but it has been shown in chiral perturbation theory
at leading \cite{Bernard:1996cc}  and subleading \cite{Kaiser:2019irl} orders, that there is 
no enhancement on the left wing of the $\omega$ resonance. This argument is outlined in
App.~\ref{app:3pi}.
Thus, the usual inclusion of the $\omega$ as a vector meson pole is justified. In case of the $\phi$, the
situation is, however, more complicated as discussed next.

\subsection{{\boldmath$K\bar{K}$} continuum}
\label{sec:KKbar}

The first important continuum contribution to the isoscalar
 spectral function is the one from $K\bar{K}$ states, as
evaluated in Refs.~\cite{Hammer:1998rz,Hammer:1999uf}
from an analytic continuation of $KN$ scattering data.
The $K\bar{K}$ contribution to the
imaginary part of the isocalar form factors is given by 
\cite{Hammer:1998rz,Hammer:1999uf}
\begin{eqnarray}
\label{imf1}
{\rm Im}\, 
F_1^{(s,K\bar{K})}(t)&=&\theta(t-4M_K^2){\rm Re}\,\biggl\{\biggr({m q_t\over 4 p_t^2}\biggr)\nonumber\\
\times && \!\!\!\!\!\!\!\biggl[{\sqrt{t}\over 2\sqrt{2}m}\bpm(t)-\bpp(t)\biggr]
F_K^V (t)^{\ast}\biggr\}\,,\nonumber\\ &&\\\label{imf2}
{\rm Im}\, 
F_2^{(s,K\bar{K})}(t) &=&{\theta(t-4M_K^2)\rm Re}\,
\biggl\{\biggl({m q_t\over 4 p_t^2}\biggr)\nonumber\\
\times&& \!\!\!\!\!\!\!\biggl[\bpp(t)-{\sqrt{2}m\over\sqrt{t}}\bpm(t)\biggr]F_K^V (t)^{\ast}\biggr\}\,,
\nonumber\\ &&
\end{eqnarray}
where $p_t=\sqrt{t/4-m^2}$ and  $q_t=\sqrt{t/4-M_K^2}$, with $M_K$ the charged kaon mass.
Further, $F_K^V (t)$ is the kaon form factor, defined via
\begin{equation}
\langle K^+(p')|j^\mu_\text{em}|K^+(p)\rangle =(p+p')^\mu F_K^V(t)~,
\end{equation}  
whereas the $\bppm (t)$ are the $J=1$ partial wave amplitudes for 
$K\bar{K}\to N\bar{N}$ \cite{Hammer:1998rz,Hammer:1999uf}. Having determined these imaginary
parts, the contribution of the $K\bar{K}$-continuum 
to the form factors is obtained from the dispersion relation
Eq.~(\ref{emff:disp}). 

The $\bpp(t)$ and $\bpm(t)$ in the above equations
are the kaon-nucleon partial wave amplitudes with total angular momentum
$J=1$ (for definitions, see App.~\ref{app:KN}).
For $t \geq 4m^2$ the partial waves are bounded by unitarity,
\begin{equation}
\label{ubs}
\sqrt{p_t / q_t}\,  |\bppm(t)|\leq 1 \,.
\end{equation}
In the unphysical region $4M_K^2 \leq t \leq 4m^2$, however,  they are
not constrained by unitarity.
In Ref.~\cite{Hammer:1998rz}, the amplitudes $\bppm (t)$ 
in the unphysical region
have been determined from an analytic continuation of $KN$-scattering 
amplitudes.
The contribution of the physical region $t\geq  4m^2$ in the dispersion
integral (\ref{emff:disp}) is suppressed for small momentum transfers
and bounded because of Eq.~(\ref{ubs}).
Using the analytically continued amplitudes in the unphysical region
and the unitarity bound in the physical region, the contribution of
the $K\bar{K}$ continuum can therefore be calculated.
Strictly speaking  this calculation provides an upper bound on the 
spectral function since one replaces the amplitudes and the 
form factor in Eqs.~(\ref{imf1},\ref{imf2}) by their absolute
values.

The striking feature in the spectral function is a clear 
$\phi$ resonance structure just above the $K\bar{K}$ threshold.
The resonance emerges in the partial wave amplitude
$b_1^{1/2,\,1/2}$ as well as in the kaon form factor $F_K$. 
In contrast to the $2\pi$ continuum, there is no strong enhancement on the 
left wing of the $\phi$ resonance which sits directly at the 
$K\bar{K}$ threshold. This is completely analogous to the
lowest-lying isoscalar resonance, the $\omega(782)$, which
also does not exhibit any enhancement on its left shoulder.

\begin{figure}[t]
\centerline{\includegraphics*[width=0.5\textwidth,angle=0]{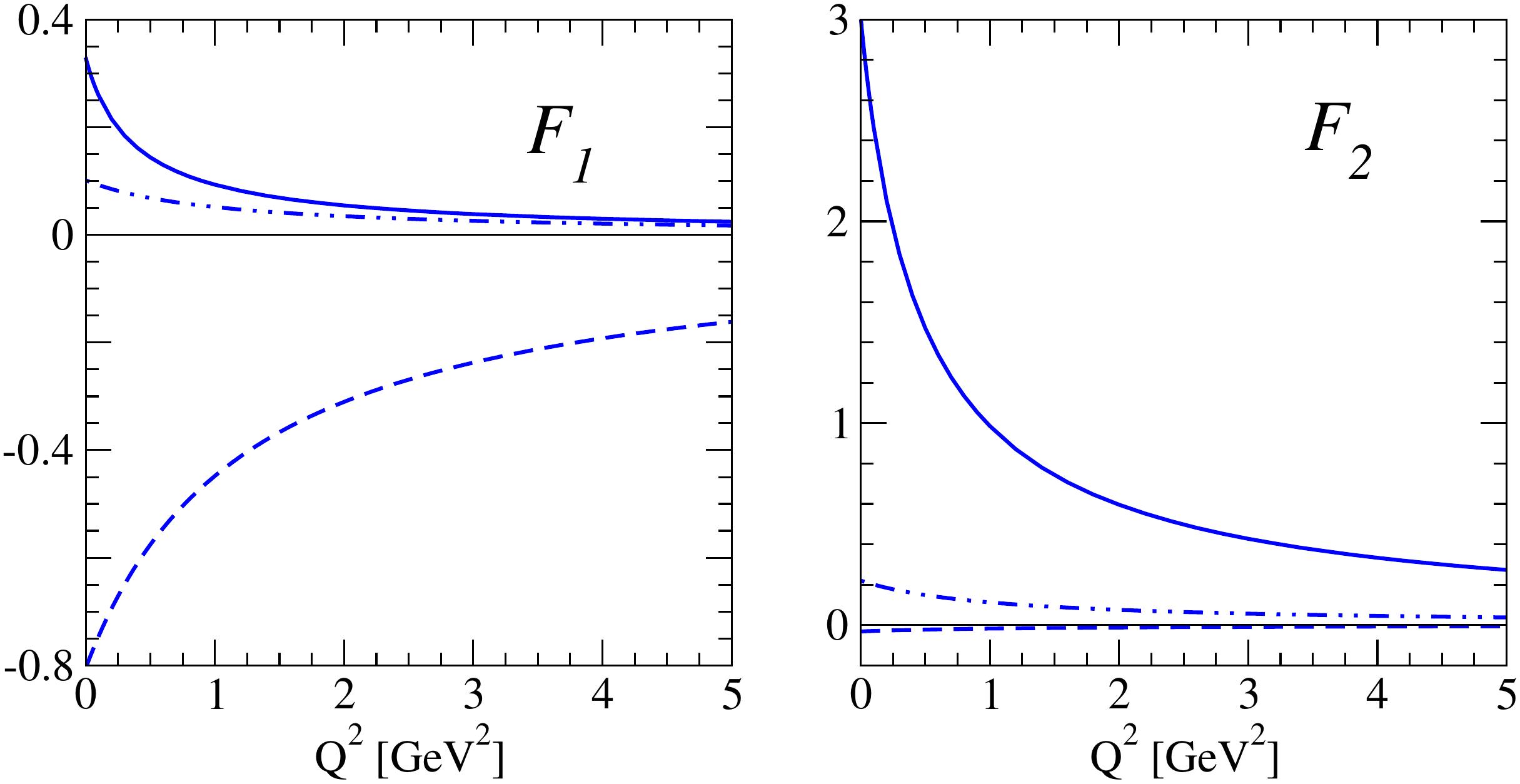}}
\caption{\label{specff} The continuum contributions to the 
nucleon form factors $F_1$ (left panel) and $F_2$ (right panel)
in the space-like region.
The contribution of the $2\pi$ continuum to the isovector form factors 
is given by the solid line, while the contribution of the  
$K\bar{K}$ and $\rho\pi$ continua to the isoscalar form factors are
given by the dash-dotted and dashed lines, respectively.
}
\vspace{-3mm}
\end{figure}

The resulting contribution to the nucleon form factors can be 
parameterized by a pole term at the $\phi$ mass~\cite{Belushkin:2006qa}:
\bea
F_i^{(s,K\bar{K})}(t) &=&
\frac{1}{\pi} \int_{4M_K^2}^\infty
\frac{{\rm Im} \, F_i^{(s,K\bar{K})}(t')}{t'-t}dt'\nonumber\\
&\simeq&
\frac{a_i^{K\bar{K}}}{M_\phi^2-t}\,,
\qquad i=1,2\,,
\label{KKcont}
\eea
with $a_1^{K\bar{K}}=0.1054$ GeV$^2$ and $a_2^{K\bar{K}}=0.2284$
GeV$^2$.
As a consequence, the contribution of the $K\bar{K}$ continuum 
to the electromagnetic nucleon form factors can conveniently 
be included in the analysis via Eq.~(\ref{KKcont}).
The form factor contributions from Eq.~(\ref{KKcont}) are also shown
in Fig.~\ref{specff}.

\subsection{{\boldmath$\rho\pi$} continuum}
\label{sec:rhopi}

Another important contribution to the isoscalar spectral function
is the correlated  $\rho\pi$  exchange, that was investigated in the Bonn-J\"ulich
nucleon-nucleon interaction  mo\-del in Ref.~\cite{Janssen:1996kx}. Since
in that work cancellations between $\phi$-exchange contribution and this correlated 
$\pi\rho$-exchange was found, the $\rho\pi$ contribution to the isoscalar spectral function was
worked out in Ref.~\cite{Meissner:1997qt}. 
This continuum contribution was evaluated
in terms of a dispersion integral which in turn can be represented
by an effective pole term for a fictitious $\omega'$ meson with 
a mass $M_{\omega'} = 1.12$~GeV \cite{Meissner:1997qt}:
\bea
F_i^{(s,\rho\pi)}(t)&=&
\frac{1}{\pi} \int_{(M_\pi +M_\rho)^2}^\infty
\frac{{\rm Im} \, F_i^{(s,\rho\pi)}(t')}{t'-t}dt' \nonumber\\
&\simeq&
\frac{a_i^{\rho\pi}}{M_{\omega'}^2-t}\,, \qquad i=1,2
\eea
with $a_1^{\rho\pi}=-1.01$ GeV$^2$ and $a_2^{\rho\pi}=-0.04$ GeV$^2$.
In the form factor analysis, one uses this effective pole instead of the full 
spectral function.

There is very little sensitivity
in the dispersive fits to $a_{2}^{\rho\pi}$, which can vary between $-0.04$ and
$-0.4$ without affecting the outcome of the fit. 
If the $\omega^{\prime}$ pole is treated as a real resonance,
the latter value is consistent with $f_{\omega^{\prime}}\sim 10$ for
$a_{1}^{\rho\pi}=-1.01$ if the coupling constants $g^i_{\omega^{\prime}NN}$
($i=1,2$) from Ref.~\cite{Meissner:1997qt} are used as input (for a
precise definition of these couplings, see Sec.~\ref{sec:specf}).

In Fig.~\ref{specff}, we show the contribution of the 
$2\pi$, $K\bar{K}$, and $\rho\pi$ continua to the electromagnetic 
nucleon form
factors $F_1$ and $F_2$. The  $2\pi$ contributes to the isovector
form factors while the $K\bar{K}$ and $\rho\pi$ continua contribute
to the isoscalar form factors. The $K\bar{K}$ and $\rho\pi$
contributions have opposite sign and partially cancel each other.
The dominant contribution to $F_1^s$ comes from the $\rho\pi$ continuum
while for $F_2^s$ the $K\bar{K}$ contribution is larger.
While the $K\bar{K}$ and $\rho\pi$ contributions can be represented
by simple pole terms, the expressions for the $2\pi$ continuum
Eq.~(\ref{unitarity_final}) are  more complicated. This is related to the 
strong enhancement close to the $2\pi$ threshold on the left wing of the
$\rho$ resonance discussed above.
Finally, note that these continuum contributions enter as an independent 
input in the dispersive analysis. They are not fitted to cross section or
form factor data.

\subsection{Vector meson poles}
\label{sec:specf}

\begin{figure}[tb] 
\centerline{\includegraphics*[width=0.25\textwidth,angle=0]{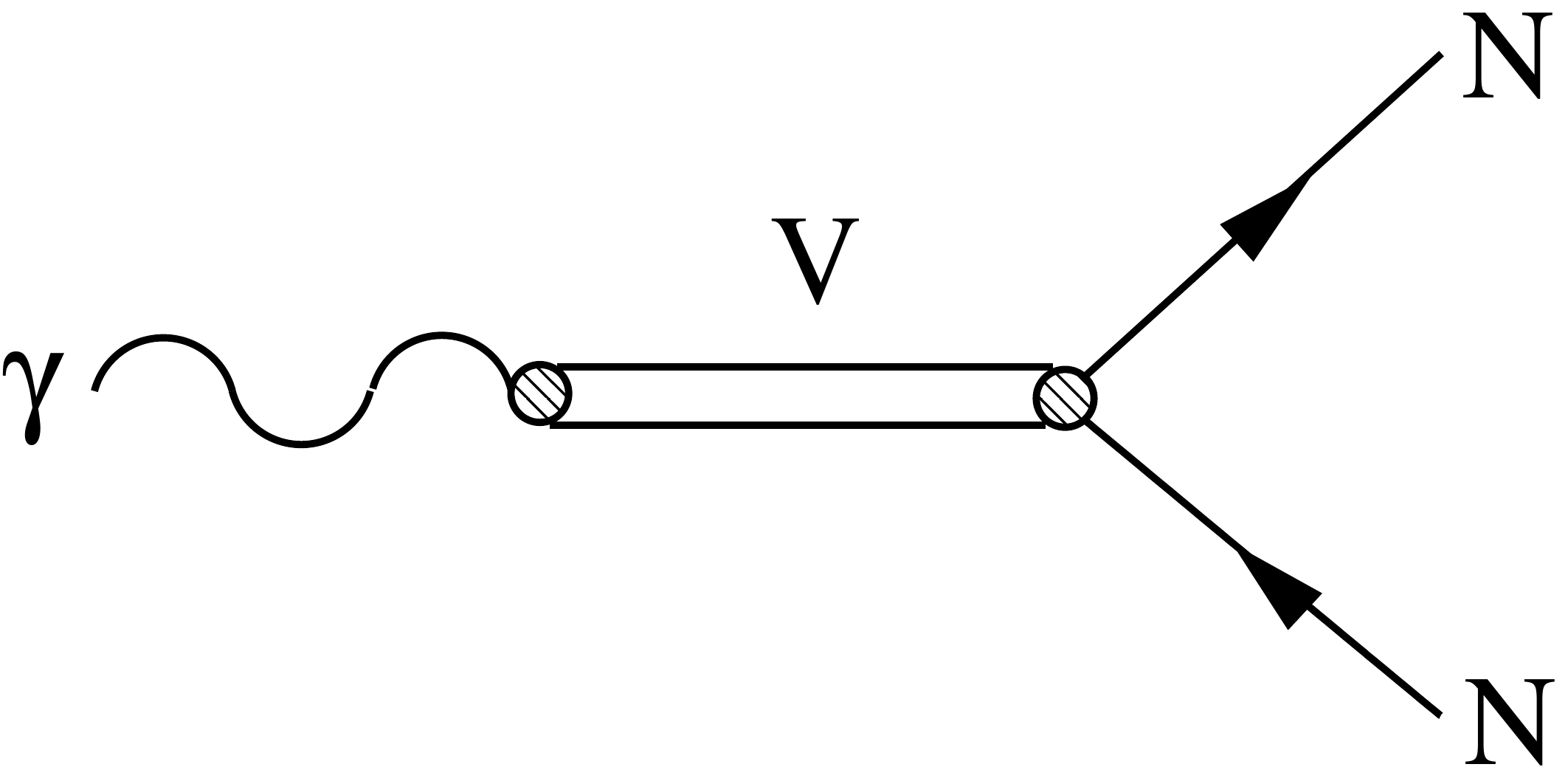}}
\caption{Vector meson dominance: The photon only couples through
  vector mesons, $V=\rho,\omega,\phi,\ldots$, to the nucleon.
}
\label{fig:VMD}
\end{figure}

In the most simple picture, the photon couples to the nucleon through
vector mesons only (i.e. there is no direct photon-nucleon coupling), 
the so-called {\em vector meson dominance} (VMD) picture,
see e.g. \cite{Sakurai:1960ju,Kroll:1967it,Bando:1987br,Meissner:1987ge},
as depicted in Fig.~\ref{fig:VMD}.
In this picture, the form factors take the simple form
\begin{eqnarray}
  F_i^s &=& \sum_{V=s_1,s_2,\ldots} \frac{a_i^V}{M_V^2-t}~, \nonumber\\
  F_i^v &=& \sum_{V=v_1,v_2,\ldots} \frac{a_i^V}{M_V^2-t}~,
\end{eqnarray}  
with 
\beq\label{eq:Vcouplings}
a_i^V = \frac{M_V^2}{f_V} g_i^{VNN}~, ~~ V= \rho, \omega, \phi, \ldots~, 
\eeq
and the couplings $f_V$ can be deduced from the leptonic decay widths $V\to e^+e^-$,
\beq
\frac{f_V^2}{4\pi} = \frac{\alpha^2}{3}\frac{M_V}{\Gamma(V\to e^+e^-)}~.
\eeq
Also, we have identified  $s_1,s_2$  with the $\omega, \phi$ and
$v_1$ with the $\rho$. Each such vector meson comes with two couplings,
the vector coupling $a_1^V$ and the tensor coupling $a_2^V$. 
One also employs the ratio of the tensor to the vector coupling, $\kappa_V$,
defined via
\beq
\kappa_V = \frac{g_2^{VNN}}{g_1^{VNN}}.
\eeq
While for some
resonances these couplings can be deduced from nucleon-nucleon scattering data,
in the dispersive analysis, they are considered as fit parameters (with the exception
of the $\rho$, which is completely determined from the $2\pi$ continuum).
In the pure VMD picture with only $\rho$ and $\omega$ vector mesons
contributing,
one can relate the tensor-to-vector coupling ratio to
the isovector and isoscalar anomalous magnetic moments of the nucleon, such
that
\bea
\kappa_\rho^{\rm VMD} &=& \kappa_p-\kappa_n \simeq 3.71~, \nonumber\\
\kappa_\omega^{\rm VMD} &=& \kappa_p+\kappa_n \simeq -0.12~.
\eea
However, extracting $\kappa_\rho$ from the two-pion continuum leads to
a larger value, $\kappa_\rho \simeq 6$, consistent with extractions from
nucleon-nucleon scattering, see e.g. the discussion in~\cite{Mergell:1995bf}.

The corresponding imaginary part, i.e. the contribution to the
spectral function for any vector meson reads:
\beq
{\rm Im}~F^V(t) = \pi \, a_i^V\, \delta(t-M_V^2)~.
\eeq

As already discussed in Sec.~\ref{sec:2pi}, the $\rho$-meson is entirely
generated by the two-pion continuum, so that an explicit $\rho$ will never
appear in the spectral function. Different to that, the lowest isoscalar
mesons are the $\omega$ and the $\phi$, which are explicitly taken into
account. As noted before, the related $3\pi$ continuum has a very small
nonresonant contribution, that can be safely neglected~\cite{Bernard:1996cc},
see also App.~\ref{app:3pi}.
Also, in the isoscalar region around 1~GeV, we consider the $K\bar{K}$ and $\rho\pi$
continua, which tend to cancel, and an additional residual $\phi$ pole.
Because of the complicated structure of the isoscalar spectral function
around 1~GeV, it is no longer possible to extract useful $\phi NN$ couplings,
as it was done in earlier works, where one just had the $\phi$-pole in this region.
The large $\phi$-couplings found in these earlier studies are clearly an
artifact of the simplified isoscalar spectral function assumed in this region.

\subsection{Structure of the spectral functions}

As discussed above, the spectral function can at present only be obtained
from unitarity arguments and experimental data for 
the lightest two-particle intermediate sta\-tes ($2\pi$ and $K\bar{K}$).
Furthermore, the $\rho\pi$ continuum contribution has been calculated in the Bonn-J\"ulich
$NN$ model.

The remaining contributions to the spectral function
can be parameterized by vector meson poles. On the one hand, 
the lower mass poles can be identified with physical vector 
mesons such as the $\omega$ and the $\phi$.
The higher mass poles on the other hand, are simply an effective way
to parameterize higher mass strength in the spectral function.
These effective poles  at higher momentum transfers appear 
in the isoscalar ($s_1,s_2,...$) and isovector channels ($v_1,v_2,...$)
It should be noted that we are dealing with an ill-posed problem here
\cite{sabba,SabbaStefanescu:1978hvt}, that means increasing the number of poles will
from some point on not improve the description of the data. 
Therefore, the strategy has always been to use as few poles as possible.
We come back to this issue  in Sec.~\ref{sec:error}.

\begin{figure}[tb] 
\centerline{\includegraphics*[width=0.50\textwidth,angle=0]{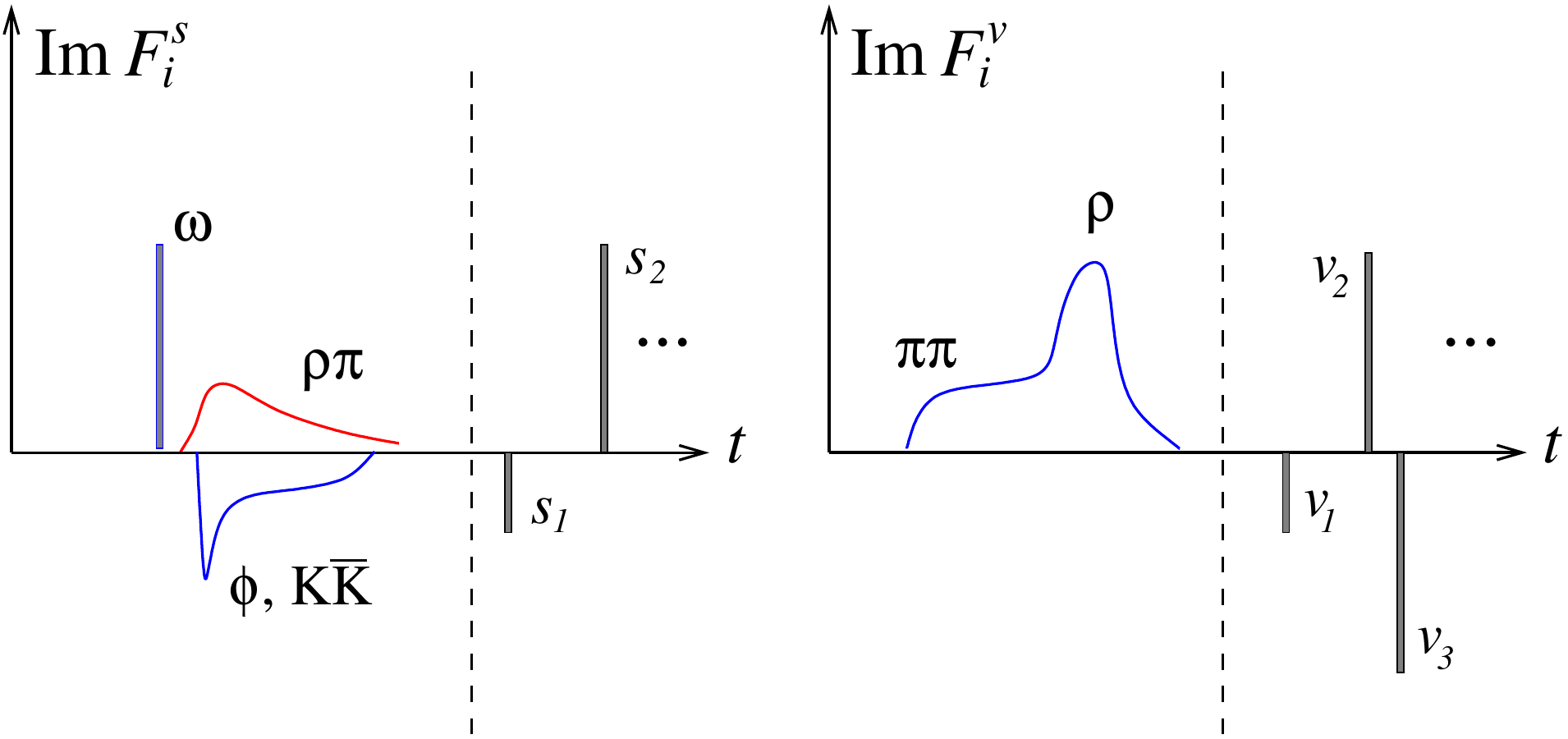}}
\caption{Cartoon of the isoscalar (left) and isovector (right) spectral function
  in terms of continua and (effective) vector meson poles. The vertical dashed
  line separates the well-constrained low-mass region from
  the high-mass region which is parameterized by effective poles.}
\label{fig:cartoon}
\vspace{-3mm}
\end{figure}

Putting all pieces together, the spectral function has the general structure
\bea
{\rm Im }\,F_i^{s} (t) &=& {\rm Im }\,F_i^{(s,K\bar{K})} (t)
+ {\rm Im }\,F_i^{(s,\rho\pi)} (t)  \nonumber \\ 
&+& \sum_{V=\omega,\phi,s_1,...} \pi a_i^{V}
\delta (M^2_{V}-t) \,, \quad i = 1,2 \, ,
\label{emff:s} \\
{\rm Im }\,F_i^{v} (t) &=& {\rm Im }\,F_i^{(v,2\pi)} (t)
\nonumber \\  
&+& \sum_{V= v_1,...} \pi a_i^{V}
\delta (M^2_{V}-t) \,, \quad i = 1,2 \, .
\label{emff:v}
\eea
For the light isoscalar vector mesons, the residua in the pole terms can be
related to their couplings. Only rough estimates exist for these: $0.5\,\mbox{GeV}^2
<a_{1}^\omega < 1\,\mbox{GeV}^2$, $|a_{2}^\omega| < 0.5\,\mbox{GeV}^2$~\cite{Grein:1977mn}
and $|a_{1}^\phi| < 2\,\mbox{GeV}^2$, $|a_{2}^\phi| < 1\,\mbox{GeV}^2$~\cite{Meissner:1997qt}.
Note that the dominant vector $\omega NN$ coupling is taken to be positive, consistent
with one-boson-exchange in the nucleon-nucleon interaction.
These ranges are used as constraints in the fits.
The masses of the effective poles ($s_1, s_2, \- \ldots , v_1,v_2, \ldots$) are fitted to the  data.
We remark that to ensure the stability of the fit \cite{SabbaStefanescu:1978hvt},
we demand that the residua of the vector meson poles are bounded, $|a_i^V| < 5\,$GeV$^2$
(this can also be considered a naturalness argument for the couplings), and
that no effective poles with masses below 1~GeV appear. Furthermore, the masses
of these effective poles should also be smaller than 5~GeV.
We generally do not include widths for the effective poles. However,
if one wants to mimic the imaginary part of the form factors in the 
time-like region, one can e.g.  allow for a large width for the highest
mass effective pole (see, e.g., Ref.~\cite{Belushkin:2006qa}). A cartoon of the
resulting (isoscalar and isovector) spectral functions is shown in Fig.~\ref{fig:cartoon}.  The vertical dashed
  line separates the phenomonologically well-constrained low-mass region from
  the effective vector meson poles at higher masses.

\subsection{Constraints}
\label{sec:con}

The number of parameters in the spectral function (i.e. the various
meson couplings $a_i^V$ ($i=1,2$) and the masses of the effective poles)
is reduced by enforcing various constraints.

The first set of constraints concerns the low-$t$ behavior of the
form factors. We enforce the correct normalization of the form factors
as given in Eq.~(\ref{norm}). The nucleon radii, however, are not 
included as a constraint. The exception to this is the squared neutron
charge radius, which in some dispersive fits has been constrained
to the value from low-energy neutron-atom scattering
experiments \cite{Kopecky:1995zz,Kopecky:1997rw}. In the new fits discussed
later, we implement this constraint using the high-precision determination
of the neutron charge radius squared based on a chiral effective field theory
analysis of electron-deuteron scattering~\cite{Filin:2019eoe,Filin:2020tcs},
\beq\label{eq:r2En}
\langle r_n^2\rangle = 0.105^{+0.005}_{-0.006}~{\rm fm}^2~.
\eeq

Another set of constraints arises at large momentum transfers.
Perturbative QCD (pQCD) constrains the behavior of the nucleon
electromagnetic form factors for large momentum transfer.
Brodsky and Lepage \cite{Lepage:1980fj} worked out the behavior  for $Q^2 \to \infty$,
\begin{equation}
F_i (t) \to \frac{1}{Q^{2(i+1)}} \, 
\left[ \ln\left(\frac{Q^2}{ \Lambda_{\rm QCD}^2}\right)
\right]^{-\gamma} \, , \quad i = 1,2 \, ,
\label{emff:fasy1}
\end{equation}
with
\beq
\gamma = 2 + \frac{4}{3\beta}~, ~~~ \beta = 11 - \frac{2}{3}N_f~,
\eeq
in terms of the leading order QCD $\beta$-function.
The anomalous dimension $\gamma\approx 2$ depends weakly on the number of
flavors, $N_f$ \cite{Lepage:1980fj}.
The power behavior of the form factors at large $Q^2$ can be easily 
understood from perturbative gluon exchange. In order to distribute the 
momentum transfer from the virtual photon
to all three quarks in the nucleon, at least two massless
gluons have to be exchanged. Since each of the gluons has a propagator 
$\sim 1/Q^2$, the form factor has to fall off as $1/Q^4$. In the case
of $F_2$, there is additional suppression by $1/Q^2$ since a quark spin 
has to be flipped. The analytic continuation of the logarithm in Eq.~\eqref{emff:fasy1}
to time-like momentum transfers $-Q^2\equiv t>0$ yields an additional term,
$\ln(-t/\Lambda^2) =
\ln(t/\Lambda^2) - i\pi$ for $t> \Lambda^2$. Employing the Phragmen-Lindeloef
theorem~\cite{Hoehler:1983}, it follows that  the imaginary part has to vanish
in the asymptotic limit. Taking these facts into account, the proton effective
FF can be described for large time-like momentum transfer $t$
by~\cite{Bianconi:2015owa}
\begin{align}
 |G_{\rm eff}^p(t)| = \frac{A}{t^2(\ln^2(t/\Lambda^2)+\pi^2)},\label{eq:pqcd}
\end{align}
with the parameters from a fit to data prior to the 2013 measurement by the
BaBar collaboration~\cite{Lees:2013ebn}, given as $A=72\,$GeV$^{-4}$ and $\Lambda=0.52\,$GeV.

The power behavior of the form factors leads to superconvergence relations of the form
\begin{equation}
\label{eq:scr}
\int_{t_0}^\infty {\rm Im}\, F_i (t) \;t^n dt =0\, , \quad i = 1,2 \, ,
\end{equation}
with $n=0$ for $F_{1}$ and $n=0,1$ for $F_{2}$. These will be employed
in the current analysis. In earlier DR analyses, modifications of the
superconvergence relations
were used including e.g. some higher order corrections. These should be,
however, abandoned as the data are simply not sensitive to such corrections.
We note that these  superconvergence relations have already been used in
Ref.~\cite{Hohler:1976ax}, i.e.  before the pQCD analysis.

Consequently, the number of effective poles in
Eqs.~(\ref{emff:s}, \ref{emff:v}) is determined
by the stability criterion mentioned before, that is,
we take the minimum number of poles necessary to fit the data.
The number of free parameters is then strongly reduced by the 
various constraints (unitarity, normalizations, superconvergence
relations). These constraints can be implemented as what is called
``hard constraints'' or ``soft constraints'', respectively. In the
former case, one solves a system of algebraic equations relating the
various parameters (couplings, masses), thus reducing the number of free
parameters in the fit (for an explicit representation,
see e.g.~\cite{Mergell:1995bf}). In the latter case, the $\chi^2$ is
augmented by a Lagrange multiplier enforcing the corresponding constraints, see
Sec.~\ref{sec:error}. Both options are viable and have been used.

It is straightforward to enumerate the nunber of fit parameters, which is given
by the couplings and masses of the vector meson, $N_V = 4 + 3N_s +3N_v$,
with $N_s (N_v)$ the number of the effective isoscalar (isovector) poles
and the $4$ represents the $\omega$ and $\phi$ couplings, minus the
number of constraints, given by $N_C = 4 + 6 + 1$, referring to the low-$t$,
the high-$t$ constraints and the neutron charge radius squared, respectively.
If the latter in not included, $N_C =10$. Putting pieces together, we
have in total $N_F = N_V - N_C = 3(N_s+N_v) - 7$ or $N_F = 3(N_s+N_v) - 6$ fit
parameters (including or excluding the $(r_E^n)^2$-constraint).

\subsection{Two-photon effects}
\label{sec:2gamma}
\begin{figure}[tb] 
\centerline{\includegraphics*[width=0.550\linewidth,angle=0]{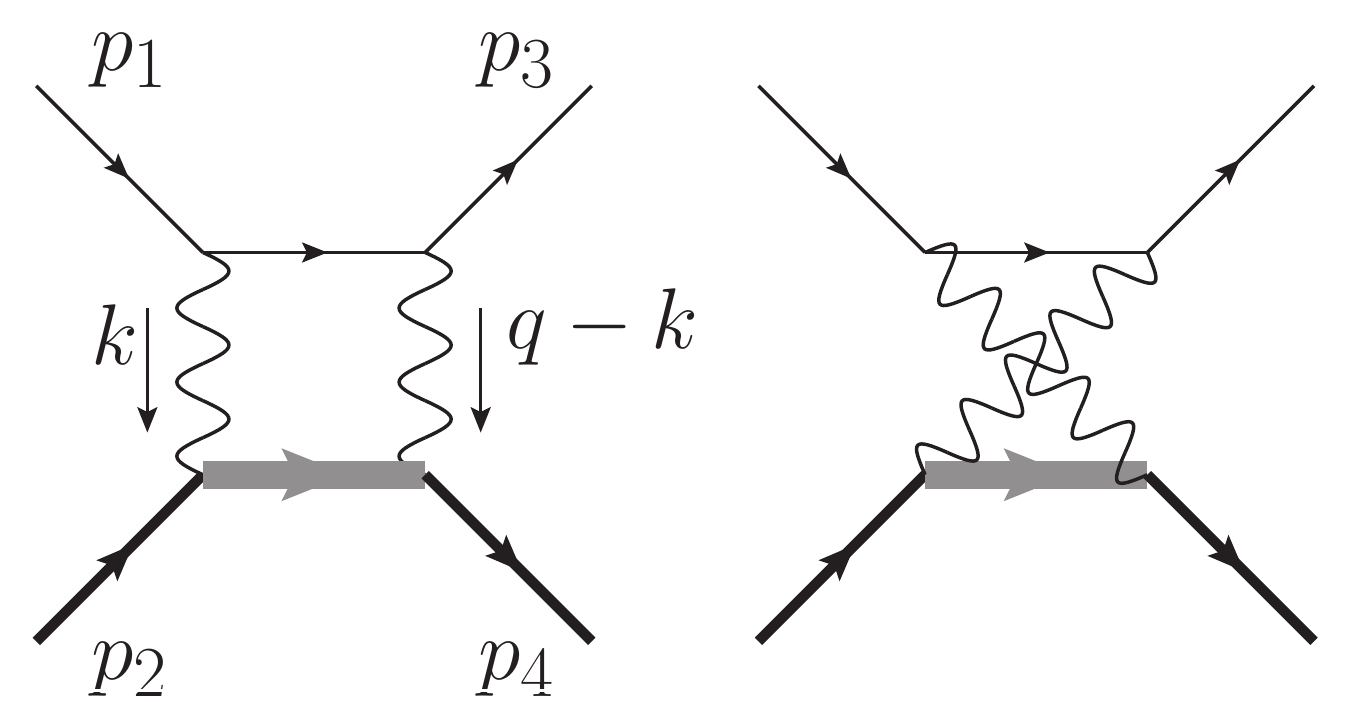}}
\caption{The two-photon exchange box and crossed box graphs. The
thin solid line denotes electron, the wiggly line photons, and the
soild line represents nucleons in the incoming and outgoing states
and the fat grey line denotes the nucleon or the $\Delta$-resonance
in the intermediate state.}
\label{fig:box}
\vspace{-3mm}
\end{figure}

The interest in two-photon corrections was triggered by the high precision measurements
of the form factor ratio  $G_{E}/G_{M}$ using the polarization transfer method reported in Refs.~\cite{Jones:1999rz,Gayou:2001qd}.
These results were found to be in striking disagreement with the world data based on the Rosenbluth separation. 
Even after removing inconsistencies from the Rosenbluth  data base~\cite{Arrington:2003df}, this discrepancy 
remained, triggering a flurry of works on two-photon corrections beyond the work of 
Refs.~\cite{Meister:1963zz,Mo:1968cg,Maximon:2000hm}, which neglected, however, the effects of the 
structure of the nucleon in the calculation of the two-photon box and
crossed-box diagrams, see Fig.~\ref{fig:box}. These diagrams were calculated in various
approaches, like in hadronic models, using generalized parton distributions or
using dispersive methods, see e.g. Refs.~\cite{Arrington:2011dn,Afanasev:2017gsk}
for reviews and very recent work in  Ref.~\cite{Ahmed:2020uso}.
Here, we concentrate on the work presented in~\cite{Lorenz:2014yda},
because the two-photon corrections given there are  applied in the DR analyses since then.

The corrections to the electron-proton cross sections at order $\alpha^3$ are given by
the interference of the one-photon-exchange amplitude $\mathcal{M}_{1\gamma}$ and the
amplitudes from vacuum polarization, vertex corrections, self-energy corrections and the
two-photon-exchange amplitude $\mathcal{M}_{2\gamma}$ and additionally the contribution
from Bremsstrahlung. The main data set that are  considered in the DR analyses already
contains a set of calculations of such corrections by Maximon and Tjon \cite{Maximon:2000hm}.
This calculation contains improvements towards earlier works by Mo and Tsai \cite{Mo:1968cg}
but still uses a soft-photon approximation, particularly relevant for the two-photon exchange (TPE)
contribution. This contribution to the corrected cross section can be expressed through a
factor of $(1 + \delta_{2\gamma})$ as
\bea
\frac{d\sigma_{\rm corr}}{d\Omega} &=& (\mathcal{M}^\dagger_{1\gamma} + \mathcal{M}^\dagger_{2\gamma}
+ \ldots) (\mathcal{M}_{1\gamma}^{} + \mathcal{M}_{2\gamma}^{} + \ldots)\nonumber\\
&=& \frac{d\sigma_{1\gamma}}{d\Omega}(1 + \delta_{2\gamma} + \ldots)~,
\eea
so that
\beq
\delta_{2\gamma} \underbrace{\approx}_\text{$\mathcal{O}(\alpha)$}
\frac{2\text{Re}(\mathcal{M}^{\dagger}_{1\gamma}\mathcal{M}_{2\gamma}^{})}{|\mathcal{M}_{1\gamma}|^2}~.
\label{tpe}
\eeq
We briefly discuss the soft-photon approximation by Maximon and Tjon since only the difference
between any new evaluation of the $2\gamma$ corrections and this approximation is required for
the purification of the $ep$ scattering data. Ref.~\cite{Maximon:2000hm} separates the
IR-divergent part of the TPE-ampli\-tude
by considering the poles in the photon propagators, i.e. one vanishing photon momentum. The
resulting factor is 
\begin{align}
\delta_{2\gamma, \text{IR}}^{\text{MT}}
= -\frac{2\alpha}{\pi}\ln\frac{E_1}{E_3}\ln\frac{Q^2}{\lambda^2}
\end{align}
where $\lambda$ is an infinitesimal photon mass and $E_1~(E_3)$  the incoming~(outgoing)
electron energy. The logarithmic infrared singularity in $\lambda$ is canceled by a term in the
Bremsstrahlung correction, so that the complete cross section is $\lambda$-independent.
The same cancellation takes place, if both $\delta_{2\gamma, \text{IR}}$ and the Bremsstrahlung
correction are calculated in the older approximation scheme by Mo and Tsai.

In Ref.~\cite{Lorenz:2014yda}, the interference  between the $1\gamma$-amplitude 
\begin{align}
\mathcal{M}_{1\gamma} = -\frac{e^2}{q^2}\bar{u}_e(p_3)\gamma_{\mu}u_e(p_1)\bar{u}_N(p_4)
\Gamma^{\nu}u_N(p_2)
\end{align}
and the $2\gamma$-amplitude
\begin{align}
\mathcal{M}_{2\gamma}^{\rm box} = -ie^4 \int \frac{d^4k}{(2\pi)^4}L_{\mu\nu}^{\rm box}
H_{N/\Delta}^{\mu\nu}D(k)D(q-k).
\end{align}
was calculated. In this notation, the metric tensor from the photon propagator has already
been contracted. Then, $\mathcal{M}_{1\gamma}$ is given in terms of the conventional lepton
spinors $u(p)$  and the elastic  nucleon-vertex  $\Gamma^\nu(q)$ from Eq.~\eqref{eq:NME}.
The $2\gamma$-amplitude contains the lepton tensor
\beq\label{eq:lept}
 L_{\mu\nu}^{\rm box} = \bar{u}_e(p_3)\gamma_{\mu}S_F(p_1-k,m_e)\gamma_{\nu}u_e(p_1)~,
\eeq
whereas the hadronic tensor for nucleon or $\Delta$ intermediate states are
\begin{align}\label{eq:hadr}
 H_N^{\mu\nu} &= \bar{u}_N(p_4)\Gamma^{\mu}(q-k)S_F(p_2+k,m_N)\Gamma^{\nu}(k)u_N(p_2)\notag\\
 H_{\Delta}^{\mu\nu} &= \bar{u}_N(p_4)(p_4)\Gamma_{\gamma\Delta\rightarrow N}^{\mu\alpha}(p_2+k,q-k)
 S_{\alpha\beta}\notag\\ &\times (p_2+k)\Gamma_{\gamma N\rightarrow\Delta}^{\beta\nu}(p_2+k,k)u_N(p_2),
\end{align}
respectively. Here,  $\Gamma_{\gamma\Delta\to N}^{\mu\alpha}(p,k)$ is the transition vertex 
\begin{align}
\langle \Delta(p')|J^{\nu}_{em}|N(p) \rangle
= \Psi_{\mu}(p')\Gamma^{\mu\nu}_{\gamma N\to\Delta}(p',q)u(p)
\end{align}
in terms of the Raita-Schwinger spinor field $\Psi_{\mu}^{(a)}(p)$ \cite{Rarita:1941mf}.
Various parameterizations of this transition matrix element exist, see e.g.
Refs.~\cite{Jones:1972ky,Kondratyuk:2001qu}. The corresponding electric $G_E$, magnetic $G_M$
and Coulomb $G_C$ transition form factors can be related to the helicity 
amplitudes measured in pion electroproduction off the nucleon, see
e.g.~Ref.~\cite{Lalakulich:2006sw}.
Accounting for this momentum dependence also in the $\Delta N \gamma$ vertices in the box
and crossed box diagrams was the main improvement in  Ref.~\cite{Lorenz:2014yda} compared
to some earlier calculations. Further, in the denominator of the photon propagator
for the pure nucleon graph, one includes an infinitesimal photon mass $\lambda$
\begin{align}
 D(k) = \frac{1}{k^2-\lambda^2+i\epsilon},
\end{align}
to regulate the infrared divergences. The loop containing the $\Delta$ is not IR divergent
because of the mass of the $\Delta$. The $S_F$ and $S_{\alpha\beta} $ in
Eqs.~(\ref{eq:lept},{\ref{eq:hadr}) are the conventional spin-1/2 and spin-3/2 propagators,
respectively. The calculation of the crossed box graph proceeds accordingly. 

In Fig.~\ref{fig:nuc},  the $\epsilon$-dependence at $Q^2$ = 3\,GeV$^2$ is shown, which
allows for a comparison to previous calculations such as~\cite{Blunden:2005ew}. In this case,
the dependence on the nucleon FF parameterization largely cancels out. The use of the
pole fit parameterization from  Ref.~\cite{Blunden:2005ew}  indeed reproduces their result.
Lowering the $Q^2$-value in the calculation decreases the nucleon-TPE correction.  
\begin{figure}[t]
\centering
\includegraphics[width=0.65\linewidth, angle=270]{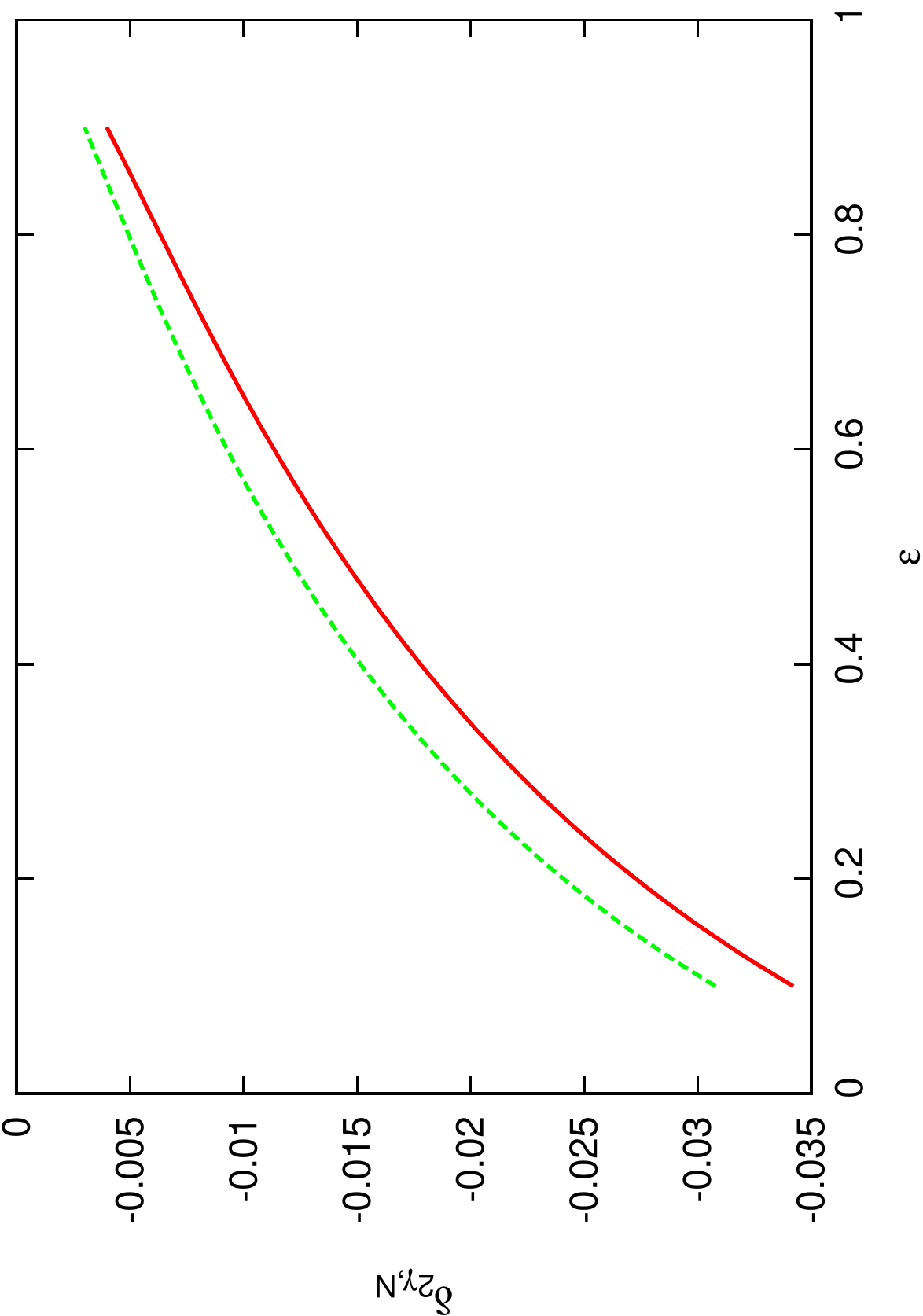}
\caption{Dependence of the TPE with nucleon intermediate state on the nucleon form factors at
$Q^2$ = 3~GeV$^2$. The correction factor $\delta_{2\gamma,N}$ is calculated once with dipole
Sachs FFs (green dashed line) and once with the simplified pole fit from
Ref.~\cite{Blunden:2005ew} (red solid line). Displayed
is the difference of the calculation in~\cite{Lorenz:2014yda}  
to the soft-photon approximation by Maximon and Tjon~\cite{Maximon:2000hm}. 
\label{fig:nuc}}
\end{figure}
\begin{figure}[t]
\centering
\includegraphics[width=0.65\linewidth, angle=270]{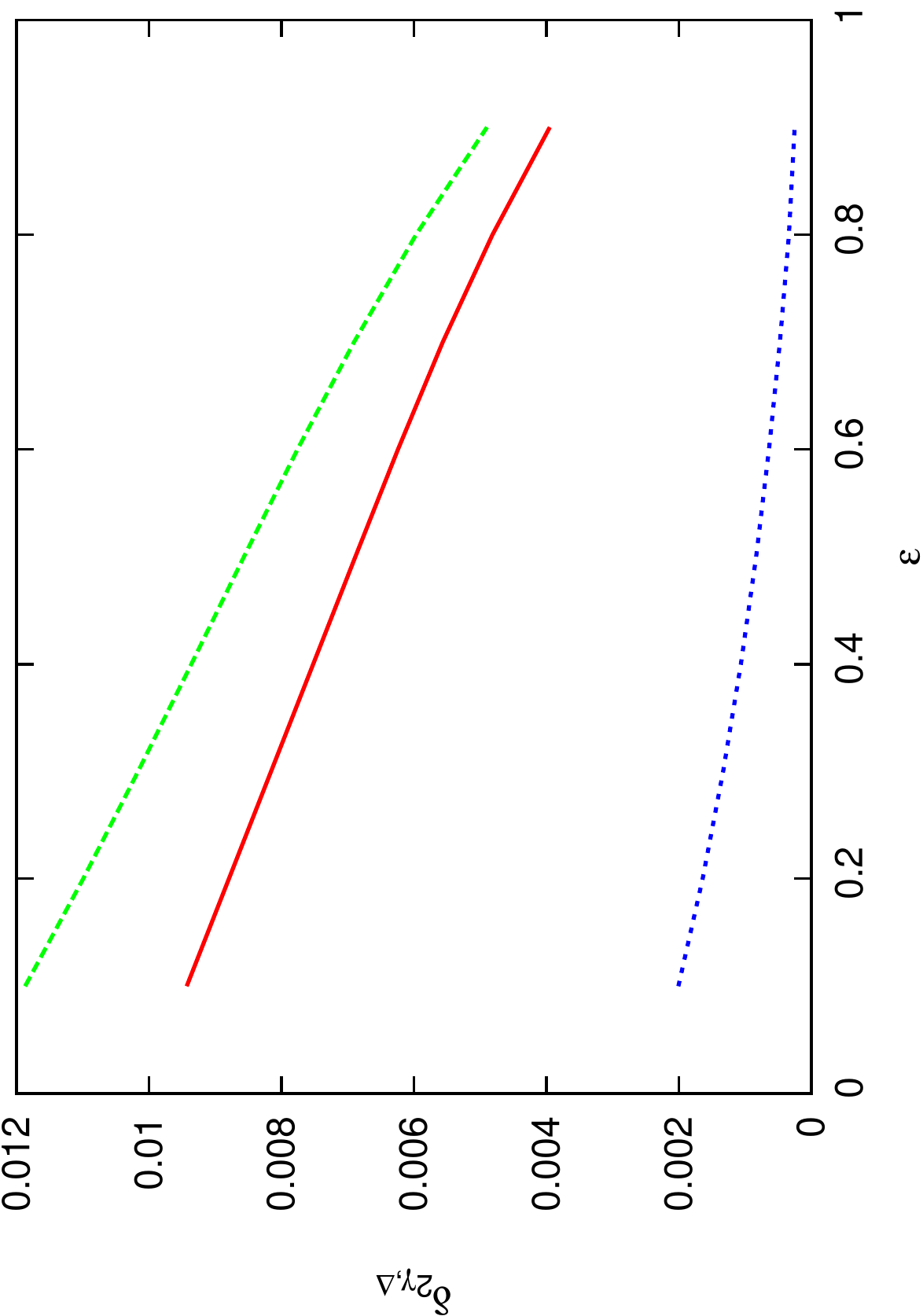}
\caption{Dependence of the TPE with $\Delta$ intermediate state on the nucleon form factors
at $Q^2$ = 3~GeV$^2$ with the $N\Delta\gamma$-vertex directly matched to helicity amplitudes
from electroproduction of nucleon resonances. The red solid, green dashed and blue dotted
lines refer to the calculation with the FFs from the dispersive approach, Sachs dipole FFs
and Sachs monopole FFs, respectively. Note that the result based on  monopole FFs is only shown
because they have been used in some earlier analyses. For more details, see
Ref.~\cite{Lorenz:2014yda}. 
\label{fig:del}}
\end{figure}
%
For the intermediate $\Delta$, the situation is different, one finds a stronger dependence
on the FF parameterizations.
In Fig.~\ref{fig:del}  the results of the calculation that employs
the helicity amplitudes  obtained from data on electroproduction of nucleon
resonances~\cite{Tiator:2003uu} are displayed. These corrections can be parameterized
conveniently by a set of FFs, determined in Ref.~\cite{Lalakulich:2006sw} and used in
Ref.~\cite{Graczyk:2013pca} for a similar calculation albeit without realistic NFFs. This
form of the $\gamma N\Delta$-vertex does not deviate significantly from recent data and
is numerically well-treatable.  These results are similar to the ones of
Ref.~\cite{Zhou:2014xka},
where different transition form factors are employed.

As stated before, the sum of $\delta_{2\gamma, N}$ and $\delta_{2\gamma, \Delta}$ from
Ref.~\cite{Lorenz:2014yda}   constitute the two-photon corrections employed in the
DR analysis of the Bonn-Darmstadt group. Their effect on the high-precision data
from Mainz~\cite{Bernauer:2010wm} is displayed in Fig.~\ref{fig:app}. Note that
the original data contain an approximation of the two-photon correction
given by~\cite{McKinley:1948zz}.
\begin{equation}
\delta_F = Z\alpha\pi\frac{\sin ({\theta}/{2})
-\sin^2({\theta}/{2})}{\cos^2({\theta}/{2})}~,
\end{equation}
where $Z$ is the nuclear charge (here, $Z=1$).
As pointed out in Ref.~\cite{Arrington:2011kv}, this approximation is only valid
as $Q^2\to 0$ and has the wrong sign for some kinematical regions. Thus, this
contribution is subtracted from the data and the two-photon corrections from
 Ref.~\cite{Lorenz:2014yda} are added.   The differences are quite visible.
 The corrections from  Ref.~\cite{Lorenz:2014yda}  will be
 also employed in the calculations presented in the next section.
Nevertheless, an  updated calculation of these corrections  would be welcome.
\begin{figure}[t]
\centering
\includegraphics[width=0.45\textwidth]{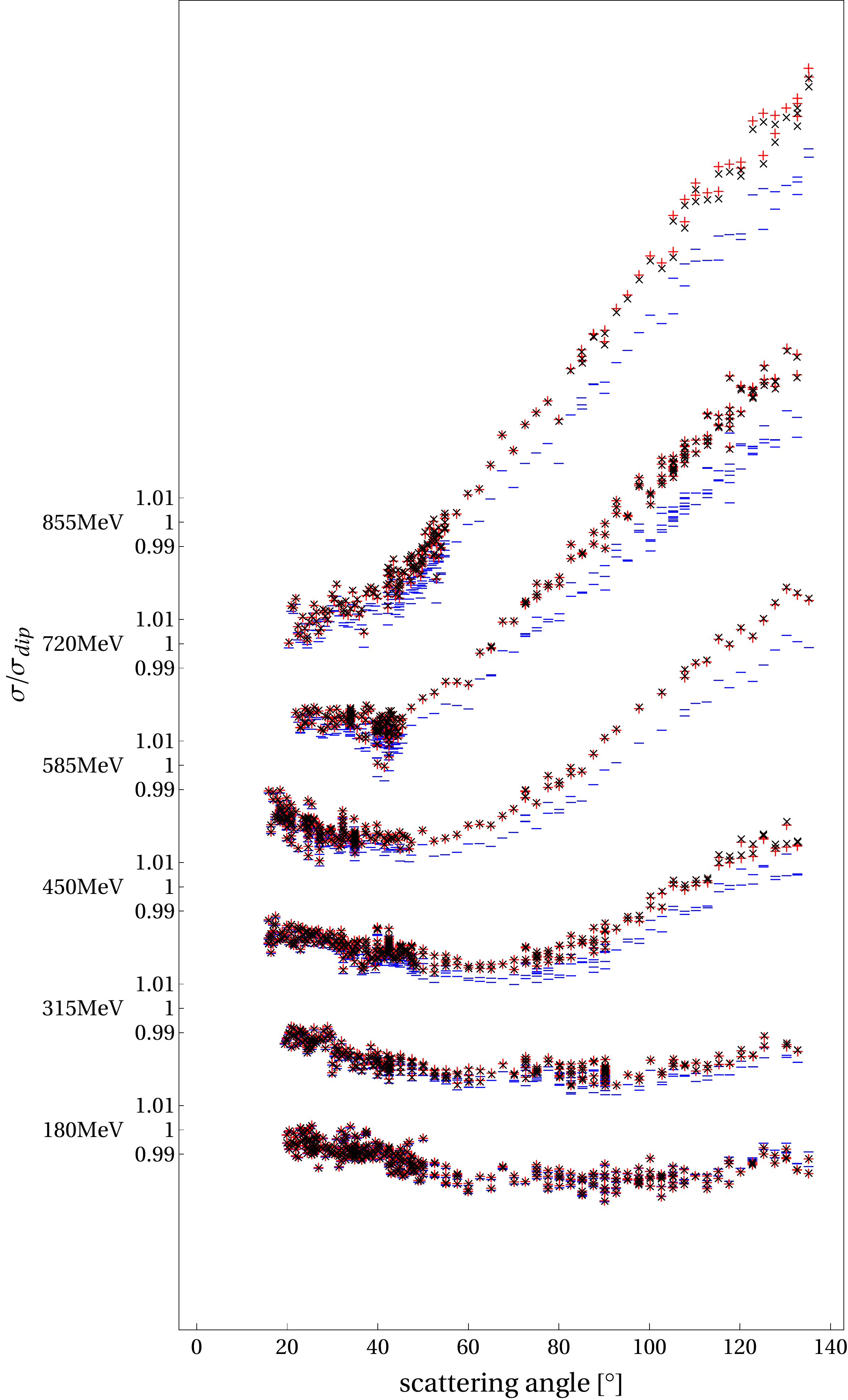}
\caption{The impact of TPE-corrections on the electron-proton scattering cross sections
  from Ref.~\cite{Bernauer:2010wm}. From the original data, the Feshbach approximation
  is subtracted  and the corrections from Ref.~\cite{Lorenz:2014yda} are added.
  Shown are the cross section data divided by that one calculated by dipole Sachs FFs
  to make the deviations clearer. Blue lines: McKinley-Feshbach--approximation, red crosses:
  TPE with intermediate nucleon states only, black crosses: TPE with intermediate nucleons
  and $\Delta$ resonances.
  \label{fig:app}}
\end{figure}

\subsection{Fit strategies and error analysis}
\label{sec:error}

In this section, we briefly describe how the fits of the spectral functions
to data are performed and how the statistical and systematic errors can be
determined. 

First, we discuss the quality of the fits, which  is measured in terms of the
total (traditional) $\chi^2$,
\begin{equation}
\chi^2_1 = \sum_i\sum_k\frac{(n_k C_i - C(Q^2_i,\theta_i,\vec{p}\,))^2}{(\sigma_i+\nu_i)^2}~,
\label{eq:chi1}
\eeq
where $C_i$ are the cross section data at the points $Q^2_i,\theta_i$ and
$C(Q^2_i,\theta_i,\vec{p}\,)$ are the cross sections for a given FF parameterization
for the parameter values contained in $\vec{p}$. Moreover, the $n_k$ are normalization
coefficients for the various data sets (labeled by the integer $k$),
while  $\sigma_i$ and $\nu_i$ are their statistical and systematical errors, respectively.
A more refined definition of the $\chi^2$ is given by~\cite{Lorenz:2014yda}
\bea
\chi^2_2 &=& \sum_{i,j}\sum_k(n_k C_i - C(Q^2_i,\theta_i,\vec{p}\,))[V^{-1}]_{ij}\nonumber\\
&& \qquad\qquad \times (n_k C_j - C(Q^2_j,\theta_j,\vec{p}\,))~,
\label{eq:chi2}
\eea
in terms of the covariance matrix $V_{ij} = \sigma_i\sigma_j\delta_{ij} + \nu_i\nu_j$.
This latter definition accounts for the correlation between the various fit parameters.
A fit to form factor data uses the same definitions, except for the absence of the
normalization factors.

One also considers the {\em reduced} $\chi^2$, which is given by:
\beq
\chi^2_{\rm red} = \frac{\chi^2_i}{N_{D} - N_{F}}~, ~~i =1,2~,
\eeq
with $N_D$ the number of fitted data points and $N_F$ the
number of independent fit parameters, see Sec.~\ref{sec:con}.

As noted in Sec.~\ref{sec:con} the various constraints on the form factors can be implemented
algebraically (hard constraints) or by modifying the $\chi^2$ (soft constraints).
The latter type of constraints are implemented as additive terms to the
total $\chi^2$ in the following form
\beq
\chi^2_{\rm add.} = p\, [x-\langle x\rangle]^2 \,\exp\left(p\, [x-\langle x\rangle]^2\right)~,
\eeq
where $\langle x\rangle$ is the desired value and $p$ is a strength parameter,
which regulates the steepness of the exponential well and helps to stabilize the
fits~\cite{Hammer:2006mw,Belushkin:2006qa}.

One method to estimate the fit (statistical) errors is the  bootstrap procedure, see e.g.
Ref.~\cite{Efron:1993qfh}. One simulates a large number of data sets compared to the number of
data points by randomly varying the points in the original set within the given
errors assuming their normal distribution. Let us consider the radius extraction. In that
case, one fits to each of these data sets separately, extracts the radius from each fit
and consider the distribution of these radius values, which is sometimes denoted as
bootstrap distribution. The artificial data sets represent many real samples.
Therefore, this radius distribution represents the probability distribution that one
would get from fits to data from a high number of measurements. The precondition for
using this method are independent and identically distributed data points which is
fulfilled when  the $\chi^2$ sum does not depend on the sequential order of the
contributing points. For $n$ simulated data sets, the errors thus scale with $1/\sqrt{n}$.
However, to get a more realistic uncertainty, we exclude one percent of the data
points from the sample and can so determine the lowest and highest value of the
extracted radius. The same procedure can, of course, also be applied to the full
form factors. In Fig.~\ref{fig:bootPRad},
we again use the 71 PRad data points to show the bootstrap extraction of the
proton charge radius and its statistical uncertainty based on 1000~samples.
The extracted error thus reads (a similar plot is obtained for the magnetic
radius) \cite{Lin:2021umk}
\beq
\delta (r_E^p)_{\rm stat.} = \pm 0.012~{\rm fm}~, ~~
\delta (r_M^p)_{\rm stat.} = \pm 0.005~{\rm fm}~.
\eeq
We note that the bootstrap  error for $r_M^p$ for the PRad data given
in~\cite{Lin:2021umk} is corrected here.
\begin{figure}[t]
\centering
\includegraphics[width=0.45\textwidth]{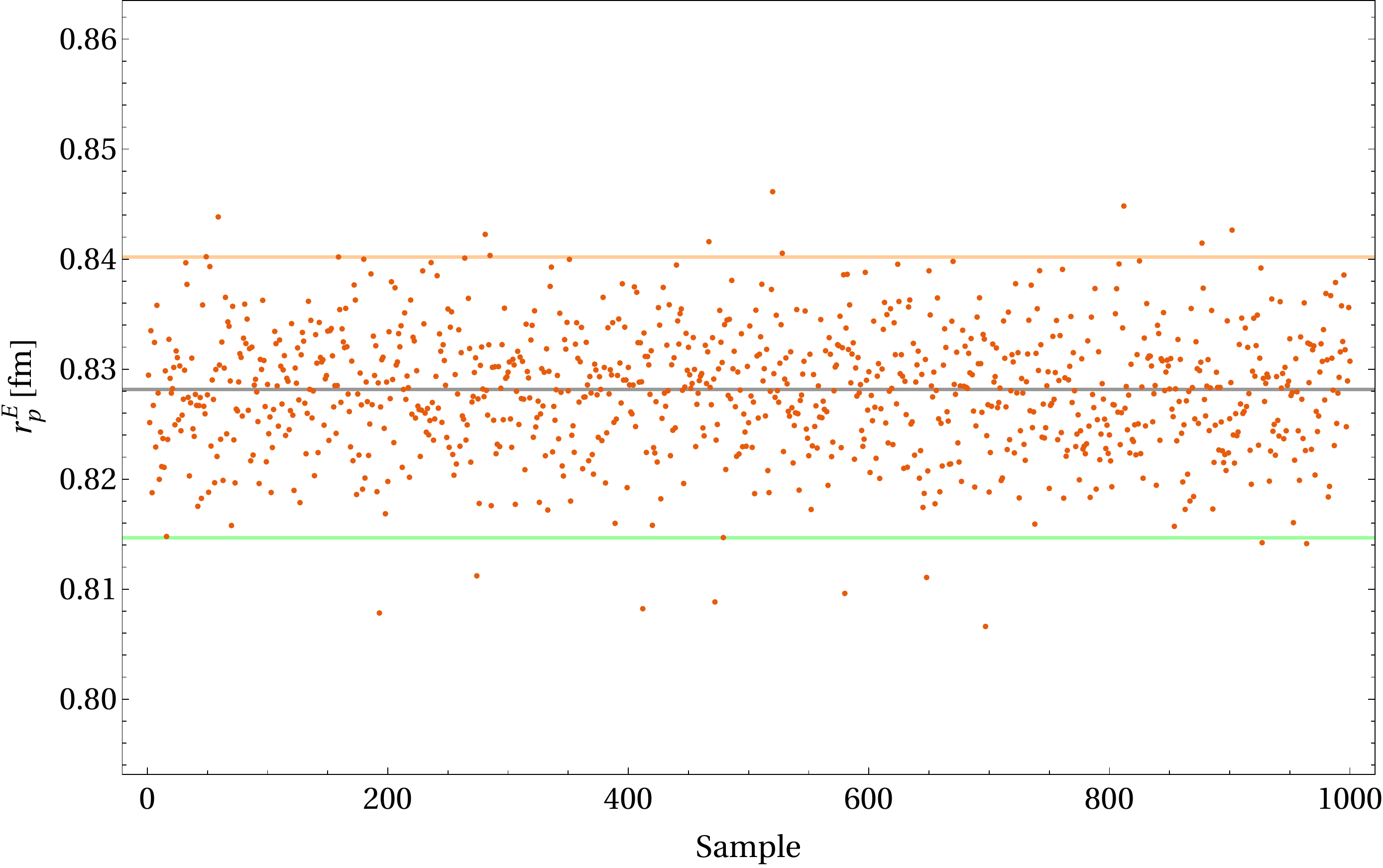}
\caption{The bootstrap procedure exemplified on the example of the
  proton charge radius extraction from the PRad data. See text for more
  details.
  \label{fig:bootPRad}}
\end{figure}


Another statistical tool to estimate the error intervals of our model parameters
is the Bayesian approach, see e.g. Ref.~\cite{Schindler:2008fh}
(and references therein).
In contrast to the interpretation of probabilities in the
classical (also called frequentist)  approach,
where the  probability is the frequency of an event to occur over a large
number of repeated trials, the Bayesian method
uses probabilities to express the current state of knowledge about the unknown parameters,
which allows one to estimate the uncertainty as a statement about the parameters.
The key ingredients to a Bayesian analysis are
the prior distribution, which quantifies what is known about the model parameters prior to
data being observed, and the likelihood function, which describes information about the
parameters contained in the data. The prior distribution and likelihood can be combined
to derive the posterior distribution by means of Bayes' theorem:
\begin{equation}
P(\mathrm{paras}|\mathrm{data})=\frac{P(\mathrm{paras})
	P(\mathrm{data}|\mathrm{paras})}{P(\mathrm{data})}~,
\end{equation}
where ``paras'' denotes the parameters.

\begin{figure*}[t!] 
\centerline{\includegraphics*[width=0.895\textwidth,angle=0]{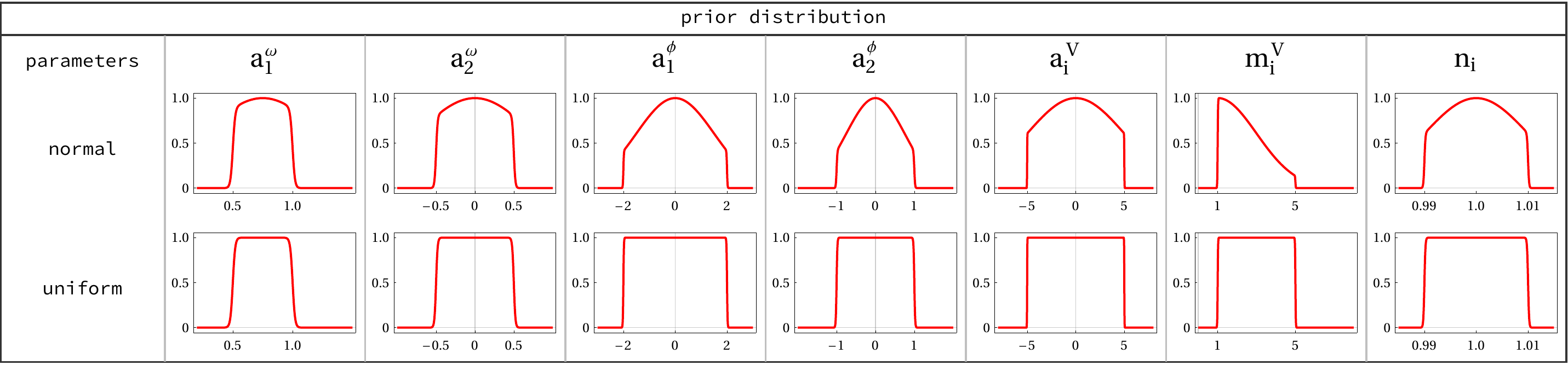}}
\caption{
Prior distributions used for the various couplings and masses in the analysis of
the PRad data. Upper/lower panel: Normal/uniform distribution. 
}
\label{fig:PRAdPriors}
\end{figure*}

It is the main goal of a  Bayesian statistical analysis to obtain the posterior
distribution of the model parameters. The posterior distribution contains the total
knowledge about the model parameters after the data have been observed. From a
Bayesian perspective, any statistical inference of interest can be obtained through
an appropriate analysis of the posterior distribution. For example, point estimates of
parameters are commonly computed as the mean of the posterior distribution and interval
estimates can be calculated by producing the end points of an interval that correspond
with specified percentiles of the posterior distribution. A powerful and easy-to-implement
method to access posterior distribution is  Markov Chain Monte Carlo (MCMC) algorithm. A
systematic illustration of Bayesian analysis application
can be found in Ref.~\cite{Wesolowski:2015fqa}.

\begin{figure}[t!] 
	\begin{center}
	\includegraphics*[width=0.35\textwidth,angle=0]{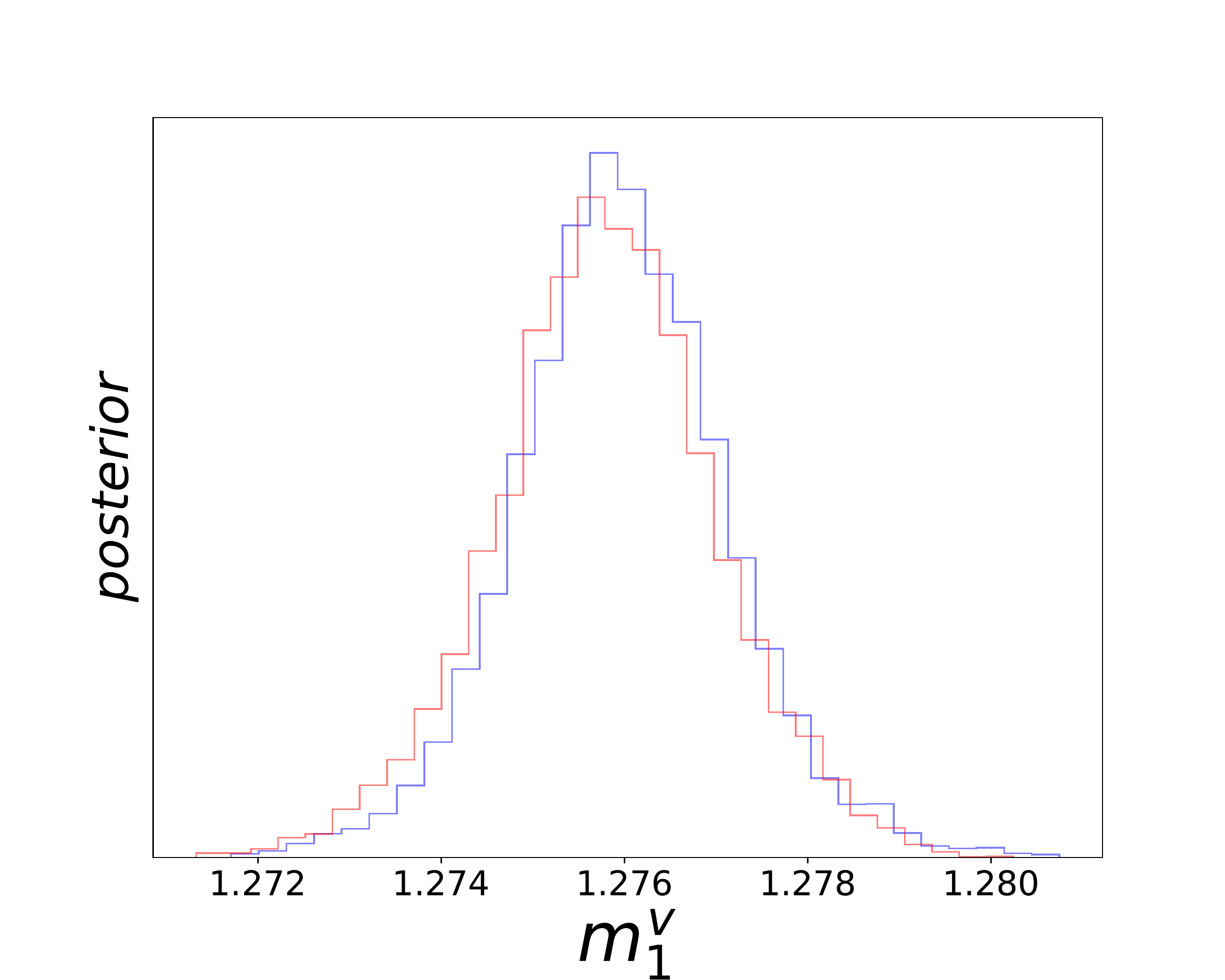}
	\includegraphics*[width=0.35\textwidth,angle=0]{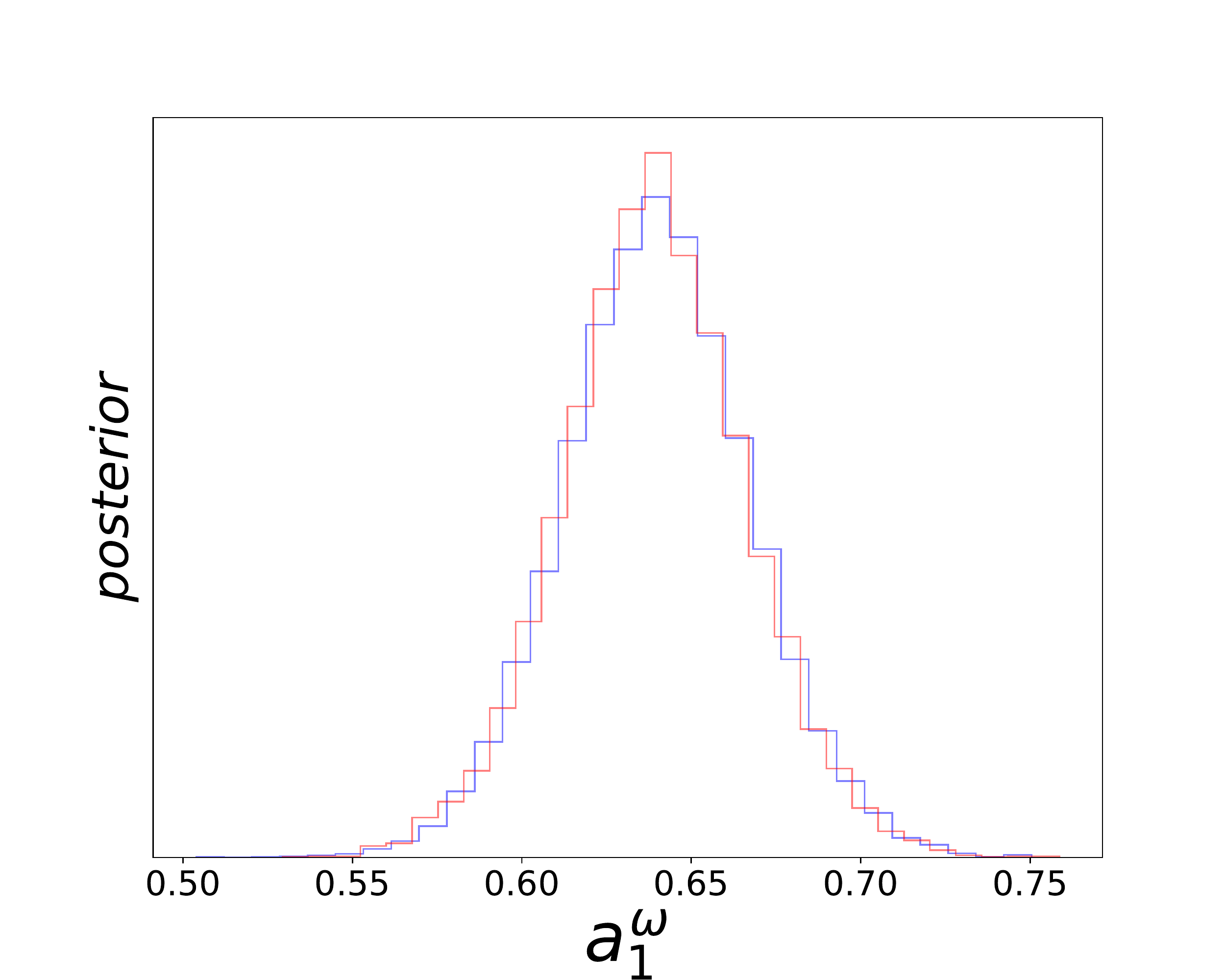}
	\end{center}
	\caption{Posterior distributions of $m_1^V$(upper panel) and $a_1^\omega$ (lower panel) 
	  from the Bayesian analysis of the fits to the PRad data.
          Blue/Red lines: Normal/uniform prior.
	}
	\label{fig:poster}
	\vspace{-0.3mm}
\end{figure}
As an example, we implement a Bayesian analysis for the fit to PRad data where the $2s+2v$
configuration of the spectral function is used. The likelihood function is given by
\begin{equation}
	L(D|\vec{p})=\frac{1}{N}e^{-\chi^2/2},
\end{equation}
with the $\chi^2$ objective function defined in Eq.~\eqref{eq:chi2}.
Here, $\vec{p}$ contains the model parameters and $D=\{d_i\}$ denotes the PRad data points.
$N$ is the normalization constant. Two different prior distributions shown in
Fig.~\ref{fig:PRAdPriors} are considered to test the stability of the obtained statistical
outputs from our Bayesian analysis.
We apply a particular MCMC sampling algorithm called ParaMonte~\cite{ParaMonte}
to acquire a Monte Carlo sample from the posterior distribution. The obtained
posteriors of the parameters $m^V_1$  and $a_1^\omega$ are taken as an example to show the
equivalence of normal and uniform priors we used 
as shown in Fig.~\ref{fig:poster}. The statistical estimates of
form factor and radius errors from our Bayesian analysis are discussed in next section.

Next, we discuss the extraction of the {\em systematic} uncertainties, which is
always the most difficult task. Our strategy is similar to what was already done
in Ref.~\cite{Hohler:1976ax}, namely to vary the number of isoscalar and isovector
poles around the values  corresponding to the best solution, where the total $\chi^2$
does not change by more than 1\%. An example of this is given in Tab.~\ref{tab:syst}
taken from Ref.~\cite{Lin:2021umk}. Here, only the PRad data~\cite{Xiong:2019umf}
are considered. The best fit corresponds to 2 isoscalar and 2
isovector poles, so we can read off the systematic errors in this case as
\cite{Lin:2021umk}
\beq
\delta (r_E^p)_{\rm syst.} = \pm 0.001~{\rm fm}~, ~~ \delta (r_M^p) = {}^{+0.018}_{-0.012}~{\rm fm}~.
\eeq
We note that while the absolute $\chi^2$ does not change, the reduced one
worsens as the number of fit parameter increases. As expected, the systematic
error is larger for the magnetic radius as at low $Q^2$, the electric FF dominates.
More detailed results will be given below. 
%
\begin{table}[t]
\centering  
\begin{tabular}{|ccc|cc|}
\hline
eff. poles & tot. $\chi^2$ & red. $\chi^2$  &   $r_E^p$~[fm]    &  $r_M^n$~[fm] \\
\hline
$2s+2v$*     &  88.5   & 1.321   &   0.829    &   0.843 \\
$3s+2v$      &  88.5   & 1.383   &   0.829    &   0.860 \\
$3s+3v$      &  88.5   & 1.451   &   0.828    &   0.848 \\
$4s+3v$      &  88.5   & 1.526   &   0.829    &   0.843 \\
$4s+4v$      &  88.5   & 1.609   &   0.829    &   0.845 \\
$5s+4v$      &  88.5   & 1.702   &   0.829    &   0.837 \\
$5s+5v$      &  88.5   & 1.806   &   0.828    &   0.861 \\
\hline
\end{tabular}
\caption{Fit to the PRad data with varying numbers of isoscalar ($s$)
  and isovector ($v$) effective poles. Given are the total and the
  reduced $\chi^2$ and the resulting values for the proton radii. The *
  marks the best solution which defines the central values for the radii.
}
\label{tab:syst}
\vspace{-3mm}
\end{table}

\section{Physics results}
\label{sec:phys}

In this section, we display a number of physics results, in particular
we discuss fits including proton polarization transfer and neutron form factor data,
and present new uncertainty analyses, thus extending and deepening
the work of Ref.~\cite{Lin:2021umk}. We also discuss the inclusion of
data for the time-like form factors and the related physics. First, however,
we want to sharpen and validate our toolbox to pin down the errors on the
example of the PRad data.

\subsection{Detailed analysis of the PRad data}

The PRad data~\cite{Xiong:2019umf} are given at two beam energies, $E=1.1, 2.2\,$ GeV,
covering squared momentum transfers in the range $Q^2 = 2\cdot 10^{-4}-6\cdot 10^{-2}\,$GeV$^2$,
in total 71 differential cross section data points.  Using this data set, we will make a detailed
comparison of the bootstrap and the Bayesian methods to extract the statistical
uncertainty. The extraction of the systematic uncertainty for these data was already discussed
in Sec.~\ref{sec:error}.

Before continuing, it is worth noting that from the proton data alone, the isospin
of a given pole is not determined. One can, however, simply assign a given number of
isoscalar and isovector poles  besides the continuum contributions, which have a given
isospin, as well as the $\omega$  and $\phi$ mesons. This ambiguity will be
resolved once neutron data are also fitted, see Sec.~\ref{sec:newfits}.

We consider first the Bayesian analysis described in Sec.~\ref{sec:error}.
We assume two sets of priors, the normal and the uniform distributions depicted
in Fig.~\ref{fig:PRAdPriors}. In both cases, the constraints on the various
couplings and masses discussed in Sec.~\ref{sec:con} are already included. Note
further that in case of the uniform distribution, the prior for the unknown mass
$m_1^v$ is biased towards smaller values. The resulting proton em radii are
equal within 3 significant digits for these two different prior distributions,
see Tab.~\ref{tab:protonradiiPRad}. Note that the systematic errors of these data
have already discussed in Sec.~\ref{sec:error}.
\begin{table}[h]
\centering  
\begin{tabular}{|c|cc|}
\hline
Method                      &   $r_E^p$~[fm]         &  $r_M^p$~[fm] \\
\hline
Bayesian normal           & $0.828\pm 0.004$       &  $0.843\pm 0.002$ \\
Bayesian uniform          & $0.828\pm 0.004$       &  $0.843\pm 0.002$ \\
\hline
Bootstrap                 & $0.828\pm 0.012$       &  $0.843\pm 0.005$ \\
\hline
\end{tabular}
\caption{Statistical uncertainty in the proton electromagnetic radii from
  the PRad data  using two different Bayesian distributions and the bootstrap
  approach.}
\label{tab:protonradiiPRad}
\vspace{1mm}
\end{table}

Next, we compare the radius extraction from the Bayes\-ian and the bootstrap
method, which are shown in Fig.~\ref{fig:PRAdBB} for the proton charge radius.
\begin{figure}[h!] 
\centerline{\includegraphics*[width=0.4\textwidth,angle=0]{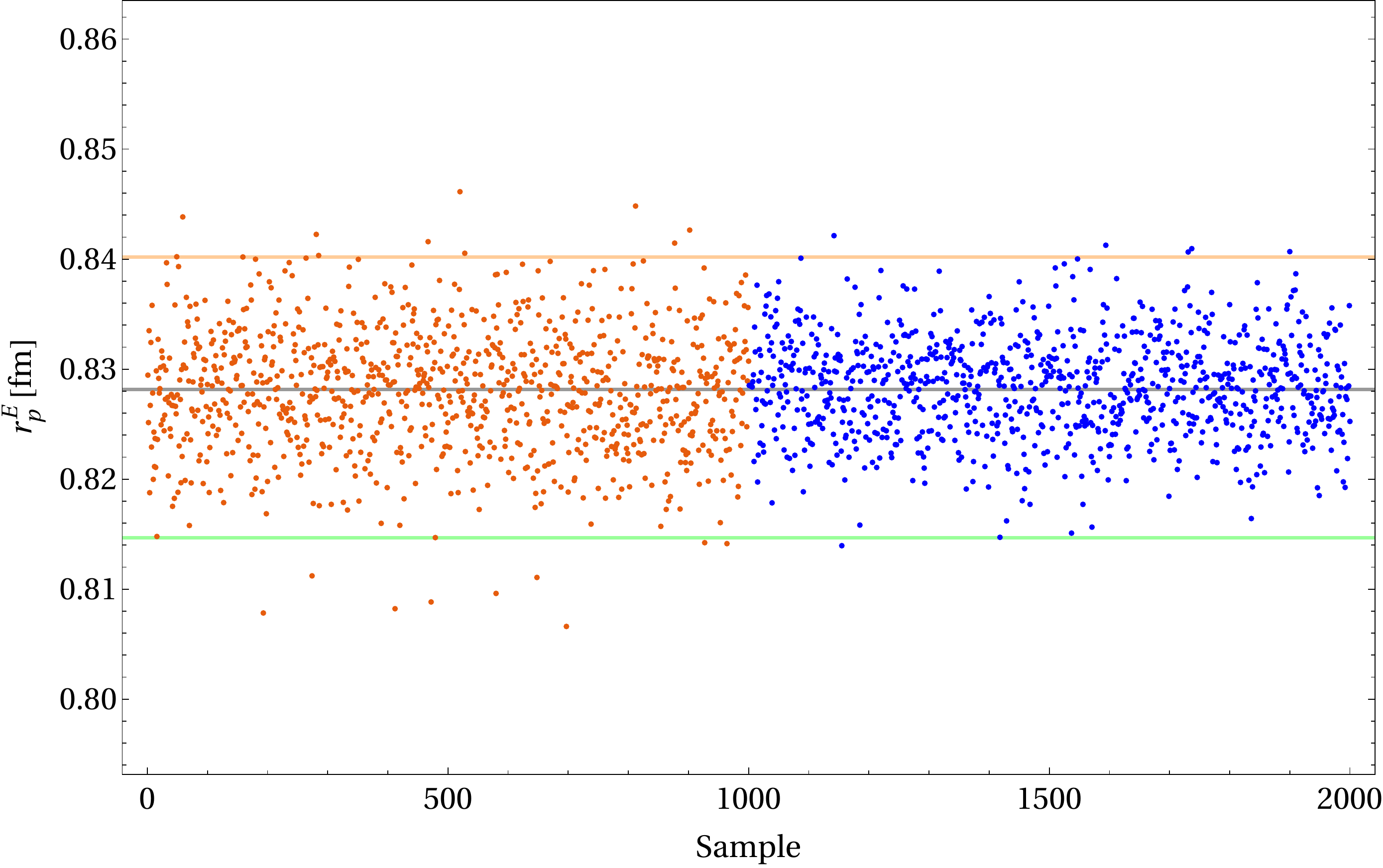}}
\caption{
Distributions of $r_E^p$ from the PRad data using the bootstrap approach
(left, orange points) and
from the Bayesian method (right, blue points).}
\vspace{1mm}
\label{fig:PRAdBB}
\end{figure}
\begin{figure}[h!] 
\centerline{\includegraphics*[width=0.4\textwidth,angle=0]{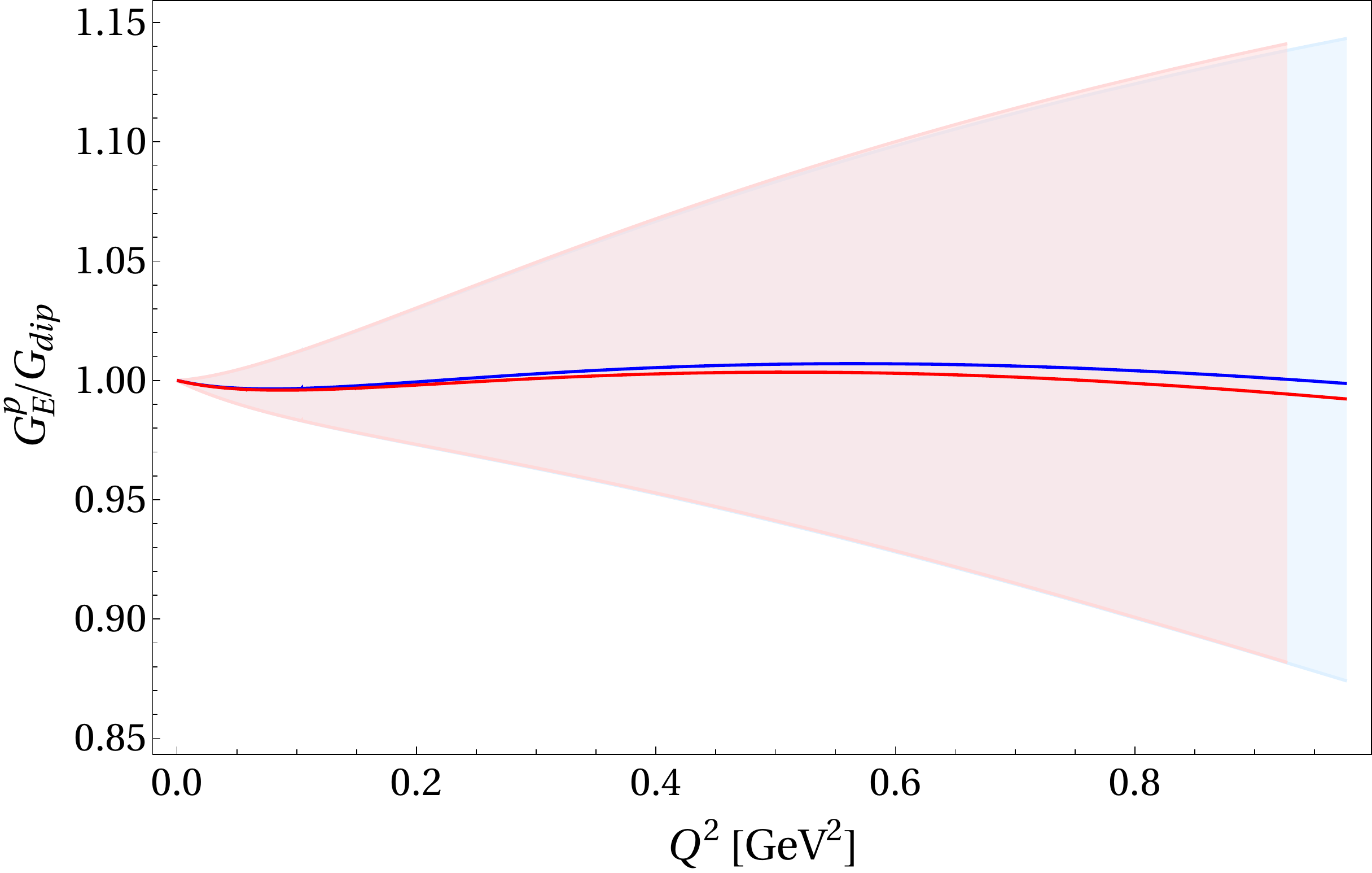}}
\centerline{\includegraphics*[width=0.4\textwidth,angle=0]{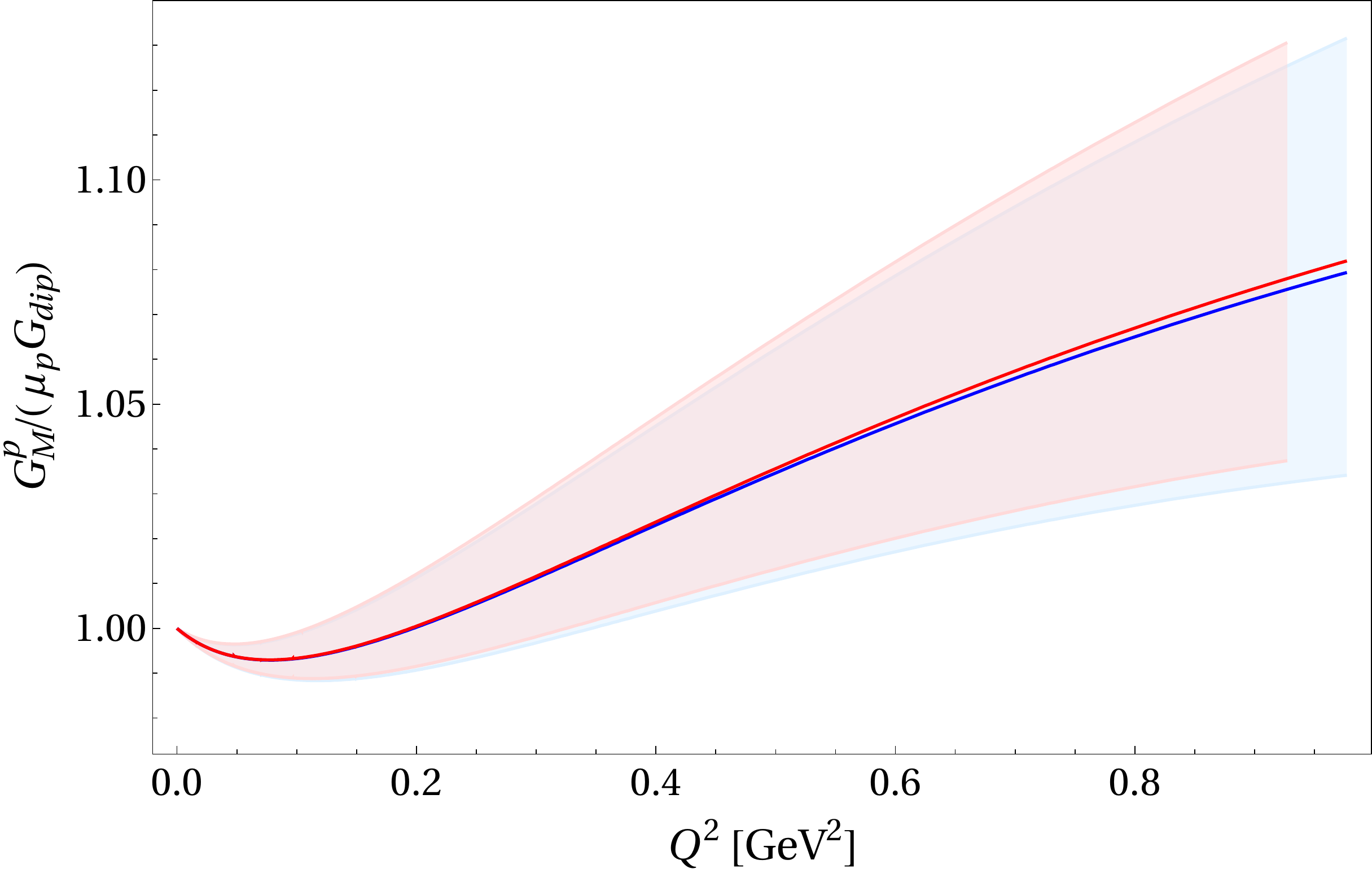}}
\caption{Electric (open panel) and magnetic (lower panel) form factor extracted
from the PRad data with the uncertainties determined from the Bayesian method
(normal prior distribution) (red area) and form the bootstrap approach (blue area).
Both form factors are normalized to the dipole FF and systematic uncertainties are
not shown.
}
\label{fig:PRAdFFs}
\vspace{-3mm}
\end{figure}
As can be read off from this figure and also seen in Tab.~\ref{tab:protonradiiPRad},
the results are very similar, with the bootstrap having slightly larger errors,
which is due to our conservative choice considering the 99\% quantile.
The resulting normalized form factors $G_E(Q^2)/G_{\rm dip}(Q^2)$ \\
$G_M(Q^2)/(\mu_p G_{\rm dip}(Q^2))$
as well as their uncertainties for the two methods are shown in Fig.~\ref{fig:PRAdFFs}.
The differences are negligible. The form factor ratio $\mu_p G_E^p/G_M^p$ measured
below $Q^2 =1\,$GeV~\cite{Ron:2011rd,Zhan:2011ji} is also well described, as
already displayed in Fig.~2 of Ref.~\cite{Lin:2021umk} 
Note also that the magnetic form factor does not display any bump-dip structure
below $Q^2<1$~GeV$^2$ as found in the MAMI analysis~\cite{Bernauer:2013tpr}.

Having shown the equivalence of both methods here, in what follows we will
stick to the bootstrap procedure, which is easier to implement in case of
large data sets with a larger number of fit parameters. 

\subsection{Fits to proton and neutron data}
\label{sec:newfits}

We are now in the position to analyze the the full data set. To be more
precise, for the proton we fit to the cross section data from PRad~\cite{Xiong:2019umf} 
and from MAMI-C~\cite{Bernauer:2013tpr} as well as to the polarization transfer data from
Jefferson Lab~\cite{Punjabi:2005wq,Puckett:2010ac,Meziane:2010xc,Puckett:2011xg}
above $Q^2 = 1\,$GeV$^2$ (note that the data from Refs.~\cite{Jones:1999rz,Gayou:2001qd} are
updated in Refs.~\cite{Punjabi:2005wq,Puckett:2011xg}, respectively, and thus do not appear
in the data base)
together with the neutron form factor world data base already used in~\cite{Lorenz:2012tm}.
The size of the data base and the $Q^2$-ranges we are fitting is provided in Tab.~\ref{tab:dbase}.
\begin{table}[t]
\vspace{2mm}
\centering  
\begin{tabular}{|c|c|c|}
\hline
Data type                 &  range of $Q^2$ [GeV$^2$] & \# of data   \\
\hline
$\sigma(E,\theta)$, PRad  &  $0.000215 - 0.058$  & 71        \\
$\sigma(E,\theta)$, MAMI  &  $0.00384  - 0.977$     & 1422      \\
$\mu_P G_E^p/G_M^p$, JLab  &  $1.18 - 8.49$     & 16        \\
$G_E^n$, world            &  $0.14 - 1.47$     & 25        \\
$G_M^n$, world            &  $0.071- 10.0 $     & 23        \\
\hline
\end{tabular}
\caption{Data base used in the fits.}
\label{tab:dbase}
\vspace{-3mm}
\end{table}
We also include the constraint on the neutron charge radius squared, updated to the
latest value given in Eq.~\eqref{eq:r2En} from Ref.~\cite{Filin:2019eoe}.
Ultimately, we need to reassess the neutron data base by performing chiral EFT analyses
of electron scattering of the deuteron and (polarized) $^3$He. This, however, goes beyond
the scope of the present work.

Before showing the results of the best fit and the corresponding statistical and systematic
uncertainties, it is worth pointing out we made extensive searches for solutions with altogther
36 combinations of isoscalar (is) and isovector (iv) poles, ranging from $3+3$ to $8+8$ is+iv
poles, with the reduced $\chi^2$ varying by less than 5\%, in most cases even by less than 1\%.
We noticed that the fits with a larger number of is than iv poles turned out to be slightly
better.

The best solution has $6+4$ is+iv poles, and the fits to the $ep$ cross section data with $Q^2<1\,$GeV$^2$,
the proton form factor ratio with $Q^2>1\,$GeV$^2$, the electric form factor
of the neutron and the magnetic form factor of the neutron are shown in
Figs.~\ref{fig:XSbestfit},~\ref{fig:FFratbestfit},~\ref{fig:GEnbestfit},~\ref{fig:GMnbestfit}, respectively.
\begin{figure}[t]
\centering
\includegraphics[width=0.45\textwidth]{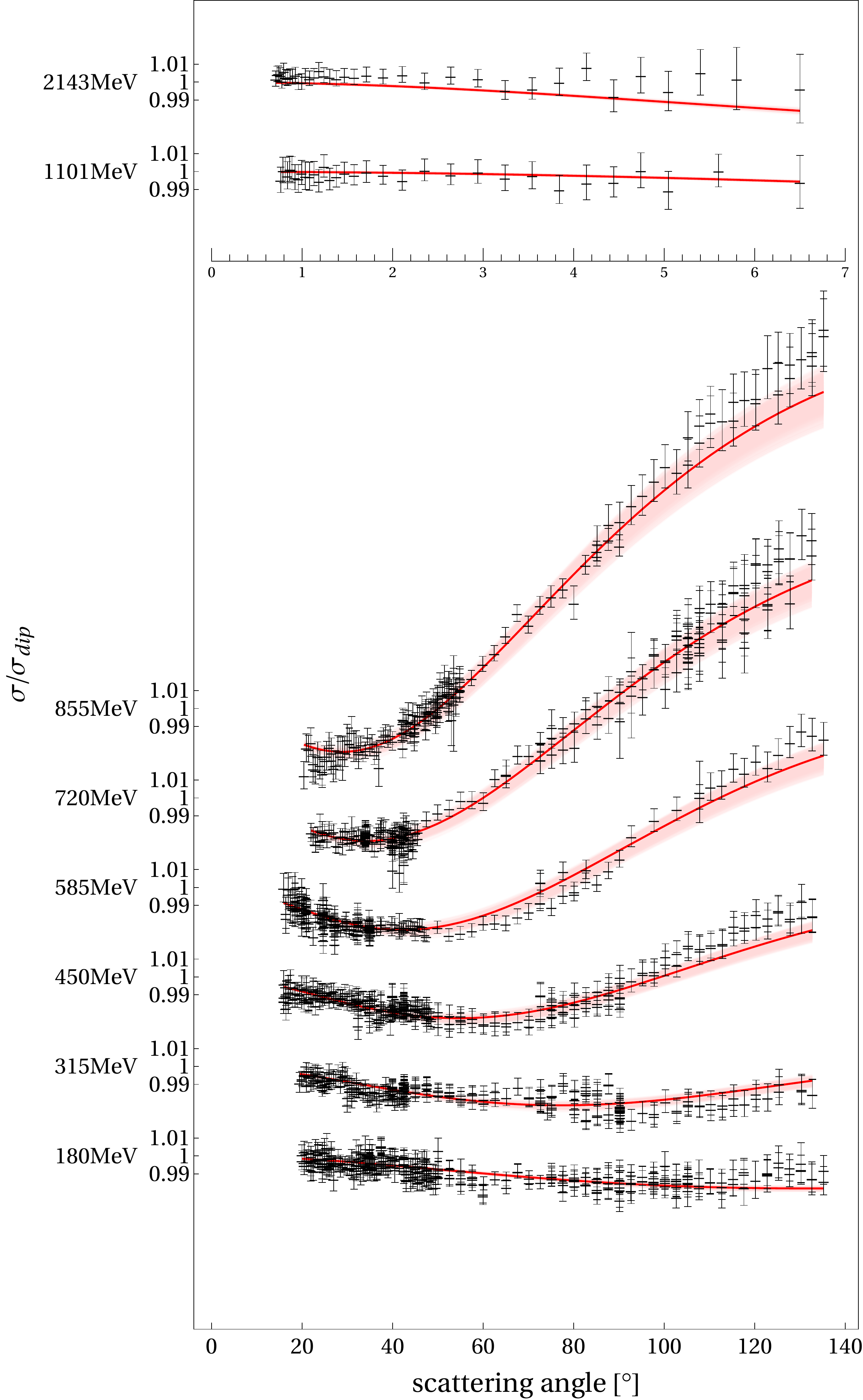}
\caption{Best fit (solid red line ) to the $ep$ cross section data from PRad (upper panel)
  and MAMI (lower panel) including the two-photon corrections discussed in Sec.~\ref{sec:2gamma}. 
  The red bands give the uncertainty due to the bootstrap procedure. Systematical uncertainties
  are not shown.
  \label{fig:XSbestfit}}
\vspace{-3mm}
\end{figure}
\begin{figure}[t]
\centering
\includegraphics[width=0.45\textwidth]{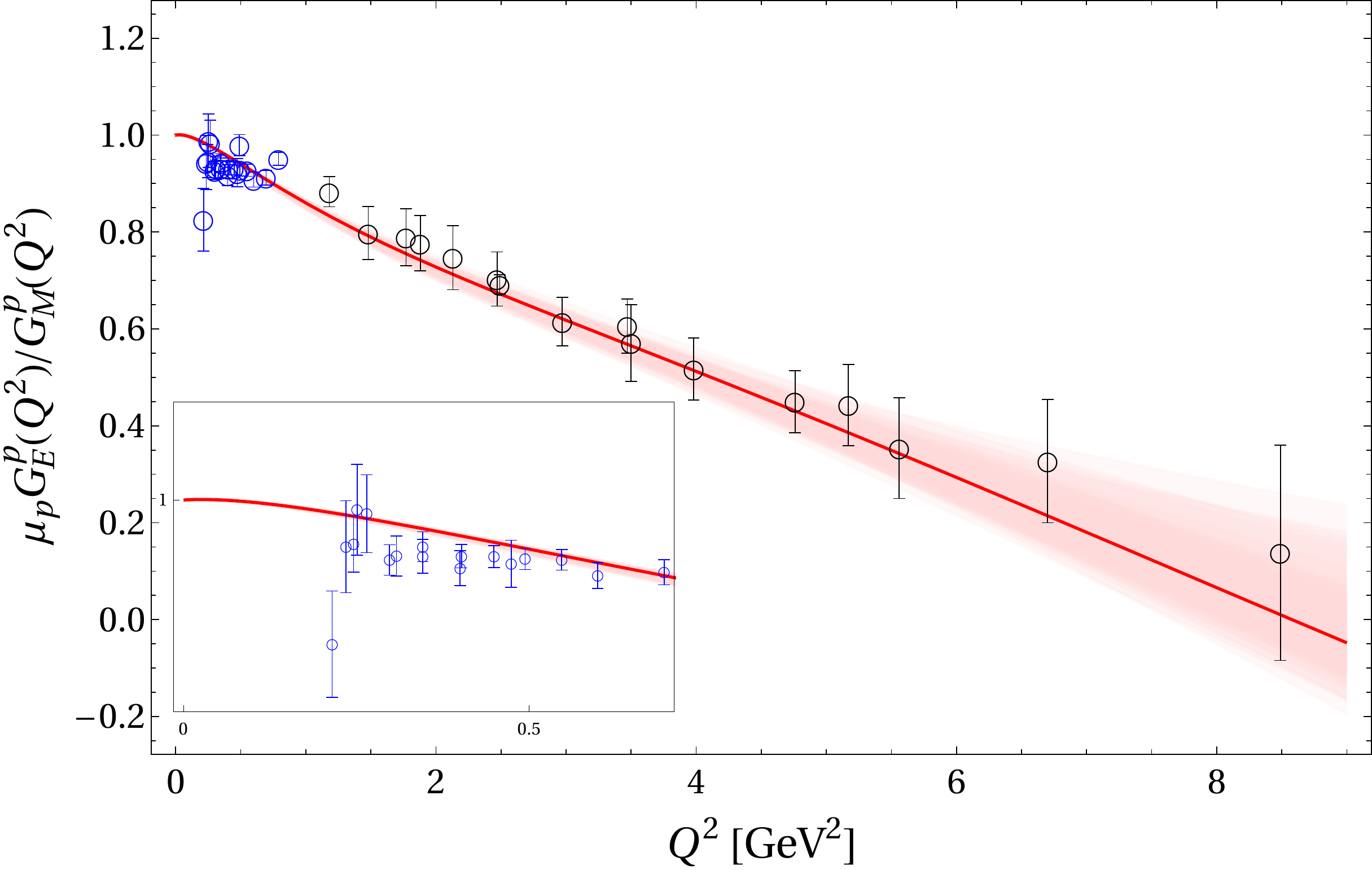}
\caption{Best fit to the proton form factor ratio data from JLab.
  Note that the blue data (also shown for $Q^2<0.7\,$GeV$^2$ in the inset)
  are not fitted. For notations, see Fig.~\ref{fig:XSbestfit}.
  \label{fig:FFratbestfit}}
\vspace{4mm}
\includegraphics[width=0.45\textwidth]{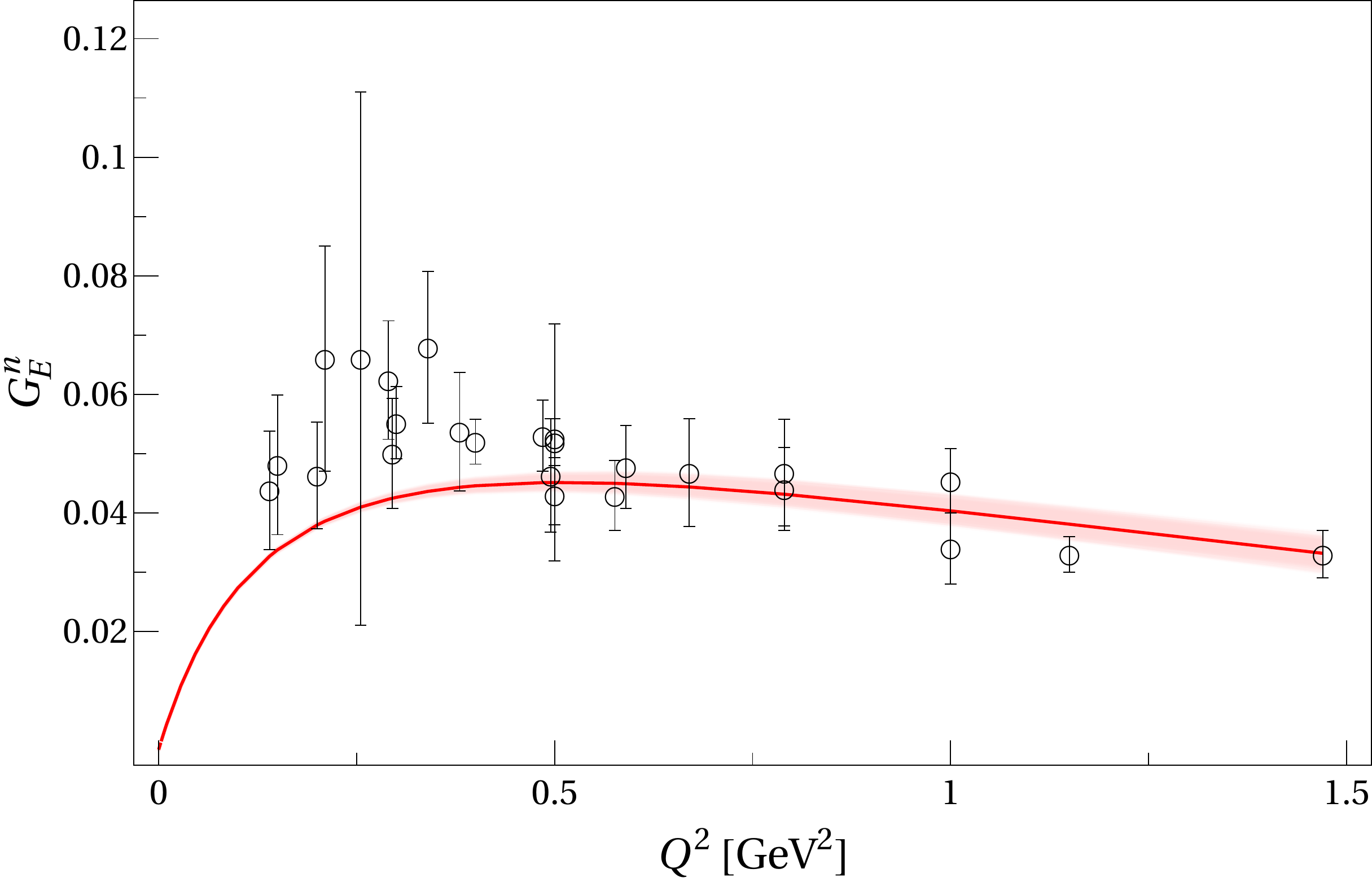}
\caption{Best fit to the neutron electric form factor data.
  For notations, see Fig.~\ref{fig:XSbestfit}.
  \label{fig:GEnbestfit}}
\vspace{4mm}
\includegraphics[width=0.45\textwidth]{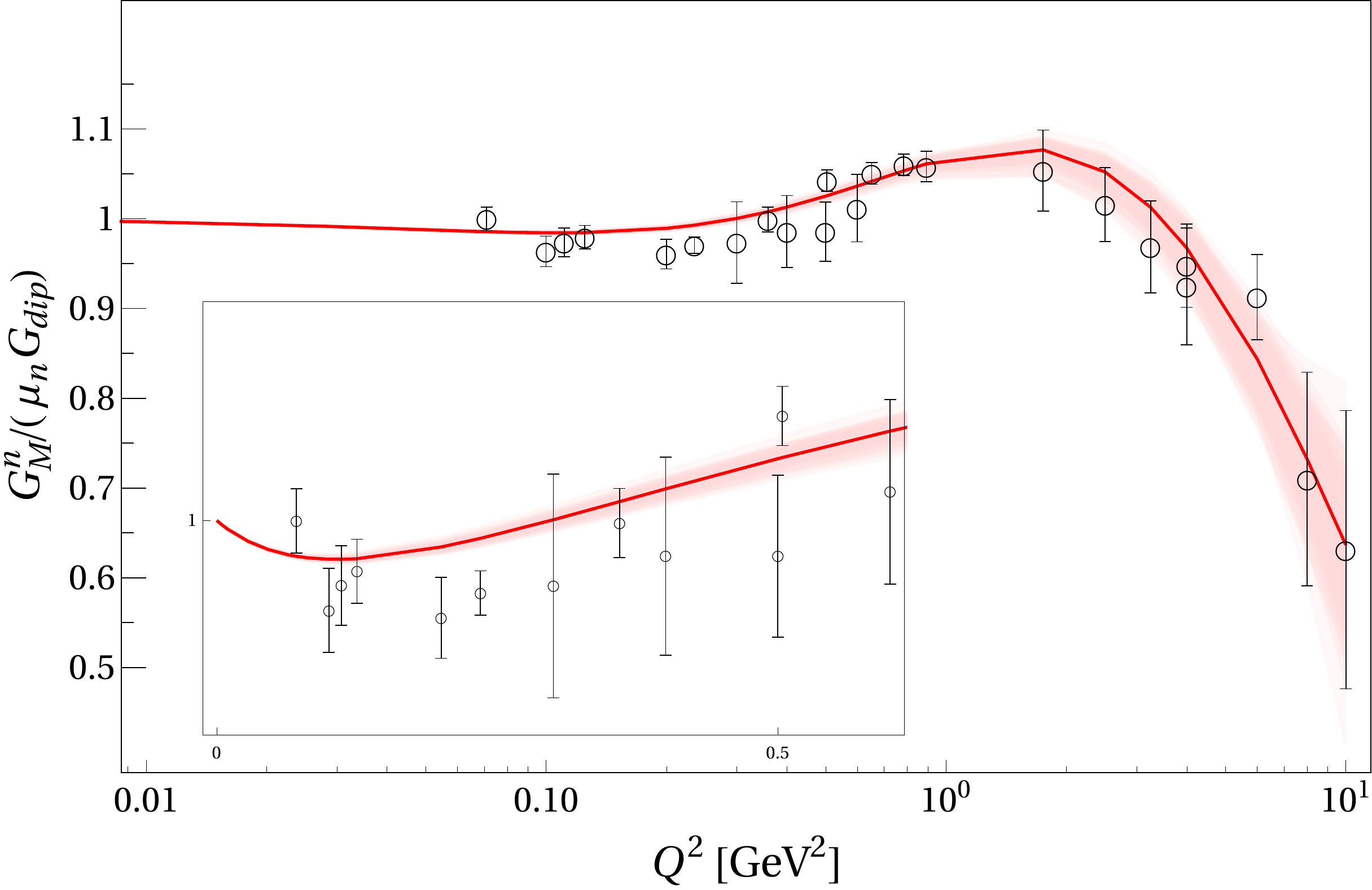}
\caption{Best fit to the neutron magnetic form factor data.
  For notations, see Fig.~\ref{fig:XSbestfit}.
  \label{fig:GMnbestfit}}
\vspace{-3mm}
\end{figure}
The corresponding central values for the various vector mesons masses, vector meson
couplings and the normalization constants of the MAMI and PRad data are collected
in Tab.~\ref{tab:values} in App.~\ref{app:newresults}. It is remarkable that while
the isoscalar spectral functions requires a number of high mass effective poles,
the effective isovector poles all have masses below 2.3~GeV. 
We note that the vector
coupling of the residual $\phi$ comes out small, consistent with expectations from the
OZI rule. The tensor coupling is, however, quite large, but the next effective
isoscalar pole has a comparable tensor coupling of opposite sign, see also the
discussion in Sec.~\ref{sec:vecco}. We also note that the various normalization
factors are deviating from one by less than 1\%.

Let us now discuss the predictions of and physics related to these fits.
First, we extract the various radii from these fits,
\bea\label{eq:radii_fin}
r_E^p &=& 0.839\pm 0.002{}^{+0.002}_{-0.003}{\rm fm}, \nonumber\\
r_M^p &=& 0.846\pm 0.001{}^{+0.001}_{-0.005}~{\rm fm}, \nonumber\\
r_M^n &=& 0.866\pm 0.002{}^{+0.010}_{-0.005}~{\rm fm},
\eea
where the first error is statistical (based on the bootstrap procedure
explained in Sec.~\ref{sec:error}) and the second one is systematic,
based on the variations in the spectral functions discussed before.
The values for the radii are completely consistent with earlier
determination, cf. Tables~\ref{tab:protonradii},\ref{tab:neutronradii},
but with a much improved uncertainty estimation.
Clearly, given the data set we fitted, the systematic uncertainty is largest
for the neutron magnetic radius. The statistical uncertainty is small for
all three radii.

It is also interesting to give the radii of the Dirac and Pauli from factors in 
the isospin basis, in particular for the comparison with lattice QCD results.
The reason for this is that the contribution of the so-called disconnected diagrams
to the isoscalar FFs, which are notoriously difficult to calculate.  These radii are
given by:
\bea\label{eq:Fsv_radii_fin}
r_1^s &=& 0.778^{+0.002}_{-0.001}{}^{+0.002}_{-0.003}{\rm fm}, \nonumber\\
r_2^s &=& 0.585^{+0.071}_{-0.087}{}^{+0.306}_{-0.123}~{\rm fm}, \nonumber\\
r_1^v &=& 0.751^{+0.002}_{-0.001}{}^{+0.002}_{-0.003}~{\rm fm}, \nonumber\\
r_2^v &=& 0.880\pm 0.001 \pm 0.003~{\rm fm}~.
\eea
Similarly, the electric and magnetic radii in the isospin basis are
(using the conventions given in Eqs.~(\ref{def:r1},\ref{def:r2}))
\bea\label{eq:Gsv_radii_fin}
r_E^s &=& 0.773\pm 0.002{}^{+0.002}_{-0.003}{\rm fm}, \nonumber\\
r_E^v &=& 0.900\pm 0.002 \pm 0.002~{\rm fm}, \nonumber\\
r_M^s &=& 0.801 \pm 0.008{}^{+0.010}_{-0.038}~{\rm fm}, \nonumber\\
r_M^v &=& 0.854 \pm 0.001{}^{+0.003}_{-0.002}~{\rm fm}~.
\eea
We note that the central values in Eq.~(\ref{eq:Gsv_radii_fin})
lead to the squared isovector radii,
of $(1/2)(r_E^v)^2= 0.405\,$fm$^2$
and $\mu^v(r_M^v)^2 = 1.72\,$fm$^2$, which are perfectly consistent with the
sum rule estimates in Eq.~\eqref{eq:SR} but have, of course, much smaller
uncertainties. 
It is also interesting to compare the isovector radii 
We also note that the value for the squared isovector with a recent state-of-the-art lattice QCD calculation
at physical pion masses \cite{Djukanovic:2021cgp},
\bea
(r_E^v)_{\rm lat} &=& 0.894(14)_{\rm stat}(12)_{\rm sys}~{\rm fm}~, \nonumber\\
(r_M^v)_{\rm lat} &=& 0.813(18)_{\rm stat}(7)_{\rm sys}~{\rm fm}~.
\eea
While the value of the isovector charge radius
is consistent with ours, the latttice value  for the isovector magnetic radius 
is smaller than ours,  that is there
is some tension. It remains to be seen what future lattice calculations
will give.

As in earlier fits~\cite{Lorenz:2014yda,Lin:2021umk}, the data for the
proton form factor ratio $\mu_PG_E^p/G_M^p$ for $Q^2<1\,$GeV$^2$, which
do not participate in the fit, are well described, see the inset in
Fig.~\ref{fig:FFratbestfit}. This points towards consistency between the
two-photon corrected cross section data and the ratio data, that are
not affected by such corrections. The situation is, however, different
for larger momentum transfers. In Figs.~\ref{fig:GEpbestfit},\ref{fig:GMpbestfit}
we display $G_E^p(Q^2)$ and $G_M^p(Q^2)$, that did not participate in the fits.
Because the proton form factor ratio tends to zero at $Q^2 \simeq 8\,$GeV$^2$,
marked deviations from the dipole form are observed. Only at very large momentum transfer,
the fall-off required by pQCD is observed. More precisely, we find that $Q^4 F^{p,n}_1(Q^2)$
starts to level off beyond $30\,$GeV$^2$, whereas that is not the case yet for $Q^6 F^{p,n}_2(Q^2)$.
Clearly, in this region of momentum transfer, more data are needed to pin down the form
factors more precisely and to eventually see the onset of perturbative QCD.
This is entirely consistent with earlier findings, see e.g.
Refs.~\cite{Mergell:1995bf,Belushkin:2006qa}.

\begin{figure}[t!]
\centering
\includegraphics[width=0.45\textwidth]{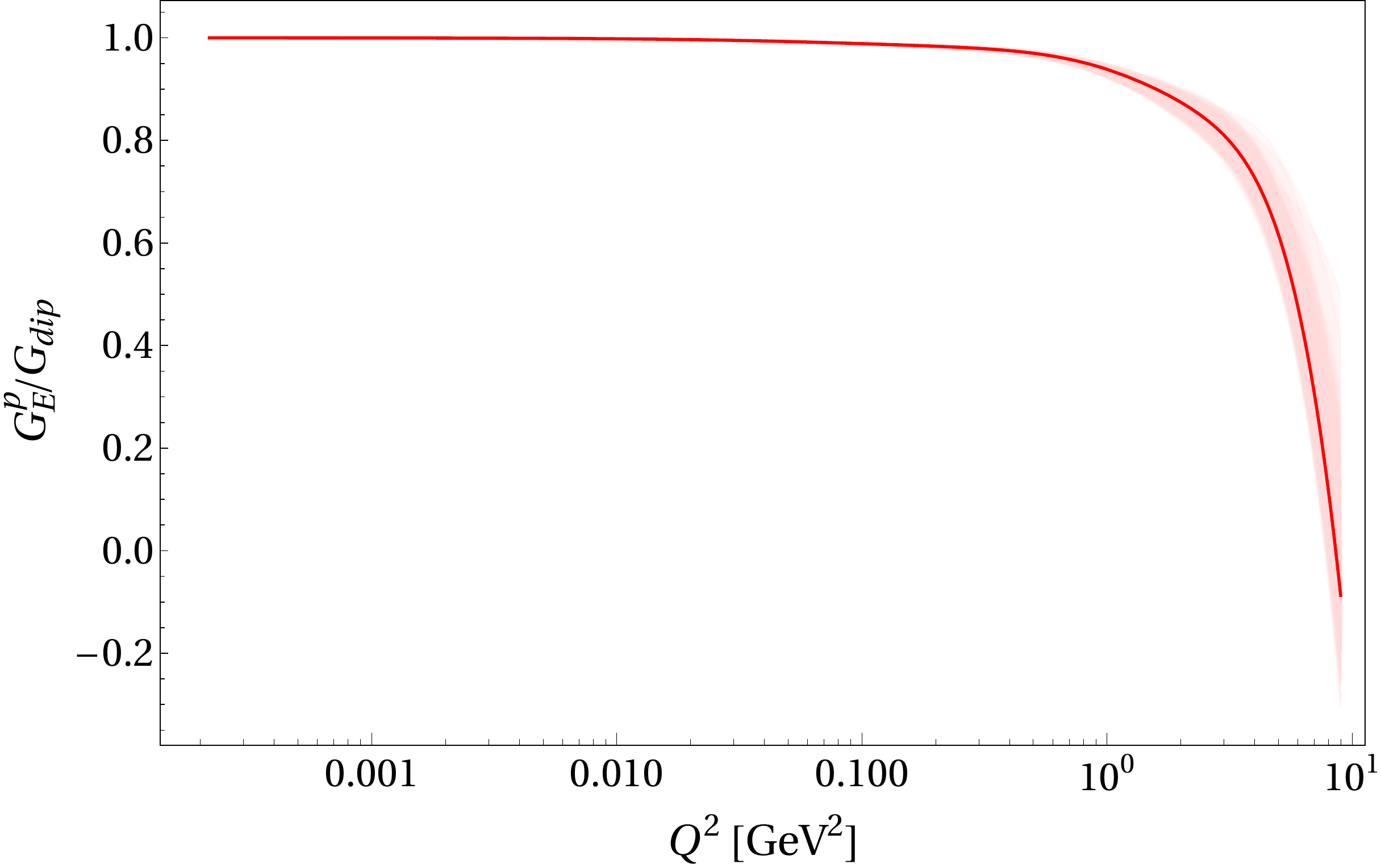}
\caption{Proton electric form factor divided by the  dipole FF
  from the best fit (red solid line)
  with the bootstrap uncertainties displayed by the light red band.
  \label{fig:GEpbestfit}}
\vspace{2mm}
\includegraphics[width=0.45\textwidth]{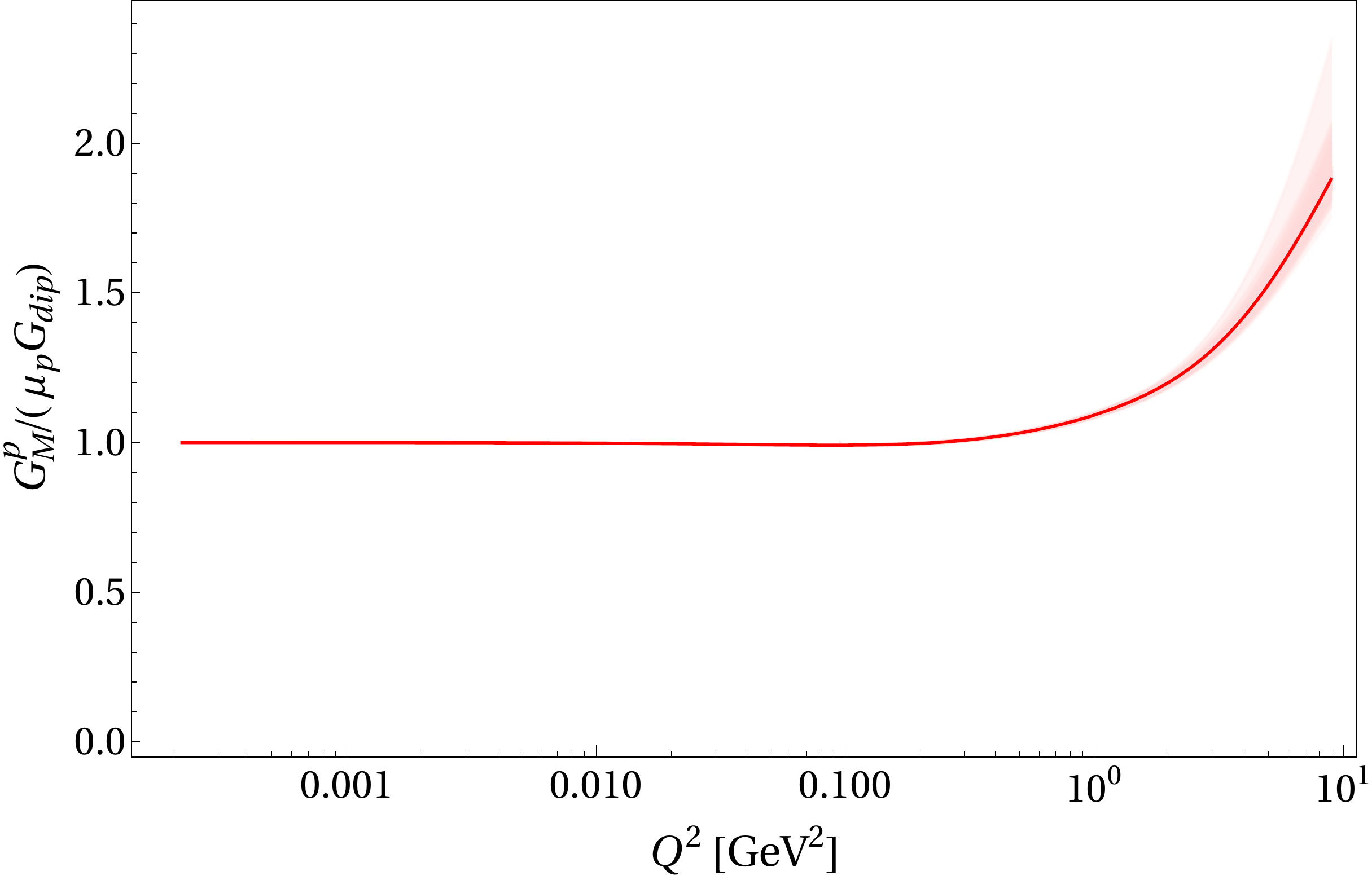}
\caption{Proton magnetic form factor divided by the  dipole FF
  from the best fit (red solid line)
  with the bootstrap uncertainties displayed by the light red band.
  \label{fig:GMpbestfit}}
\vspace{-3mm}
\end{figure}

Note that the long-range part of the Breit-frame
charge and magnetization distributions that follows from the
Sachs form factors 
can be interpreted in terms of a ``pion cloud'' and some additional
short-range contributions from the $\rho$ and other short-ranged
physics. However, we emphasize that this separation is scale-dependent
and thus not unique. A general discussion of the pion cloud can be
found in Refs.~\cite{Hammer:2003qv,Meissner:2007tp}.

\subsection{Vector meson couplings}
\label{sec:vecco}

As noted before, in the earlier DR analyses, in the isoscalar spectral function
below 1~GeV, only the $\omega$ and the $\phi$ mesons were retained and thus
using Eq.~\eqref{eq:Vcouplings}, one was able to extract the vector and the
tensor couplings of these mesons. However, we have shown that in the region
of the $\phi$, the isoscalar spectral function is much more complicated,
preventing one from extracting $\phi$-meson couplings. In what follows,
we will thus only consider the $\omega$-meson. In this case, we have
\beq
g_i^{\omega NN} = \frac{f_\omega}{M_\omega^2} \, a_i^\omega~,~~i=1,2~,
\eeq
with
\beq
f_\omega = 16.7~.
\eeq
The earlier fits, which had no restrictions on the residua 
led to a large vector and a small tensor coupling,
\bea
g_1^{\omega NN} &=& 19.4\pm 1.0~,~~ \kappa_\omega = -0.17~,
\nonumber\\
g_1^{\omega NN} &=& 20.9\pm 0.3~,~~ \kappa_\omega = -0.16\pm 0.01~,
\eea
from Ref.~\cite{Hohler:1976ax} and Ref.~\cite{Mergell:1995bf}, respectively.
This vector coupling is sizeably larger than from the determination using
forward dispersion relations in nucleon-nucleon scattering, $g_1^{\omega NN} = 10.1\pm 0.9$ 
\cite{Grein:1977mn}. This smaller value is, however, inconsistent with the
approximate dipole behavior of $F^s_1(Q^2)$~\cite{Hohler:1994rt}. Note, however,
that in one-boson-exchange models of the NN interaction, one
typically finds values of $g_1^{\omega NN}(M_\omega^2) \simeq 20$ which for typical
strong form factors translates to  $g_1^{\omega NN}(0) \simeq 10$~\cite{Ericson:1988gk}.

Starting with the work of BHM~\cite{Belushkin:2006qa}, the isoscalar spectral
function was considerably improved. In that work, the vector coupling was still
large, but the tensor coupling could not be pinned down so precisely,
\beq
g_1^{\omega NN} = 16.7\ldots 23.1~,~~
g_2^{\omega NN} = -3.6\ldots 10.3~.
\eeq
Values within this range where also found in the analysis of the
MAMI-C data combined with the proton form factor data for $Q^2>1\,$GeV$^2$
and the neutron FF data base~\cite{Lorenz:2012tm}
\beq
g_1^{\omega NN} = 20.4~,~~
g_2^{\omega NN} = 0.1~,
\eeq
where only central values were given. Finally, the analyses that concentrated
mostly on the high-precision $ep$ data from MAMI-C and PRad, the $\omega$ couplings
took the values
\bea
g_1^{\omega NN} &=& 13.6~,~~ g_2^{\omega NN} = -5.2~,
\nonumber\\
g_1^{\omega NN} &=& 23.4~,~~ g_2^{\omega NN} = 0.3~,
\eea
from Ref.~\cite{Lorenz:2014yda} and Ref.~\cite{Lin:2021umk}, respectively.
Finally, we present the results of our new fits, including the statistical
and the systematical uncertainty:
\bea
g_1^{\omega NN} &=& 18.81^{+0.44}_{-0.48}{}^{+1.35}_{-3.16}~,\nonumber\\
g_2^{\omega NN} &=& 1.18^{+1.41}_{-0.92}{}^{+0.79}_{-5.74}~.
\eea
For the central values, the tensor-to-vector  coupling ratio is small, $\kappa_\omega = 0.06$.
We note that the uncertainties on the vector coupling are modest, they are much
larger for the  suppressed tensor coupling. Similar to the findings in BHM, the sign
of the tensor coupling is not determined and the range of allowed values is sizeable.

\subsection{Time-like form factors and final-state interactions}
\label{sec:FSI}

\begin{figure}[t!] 
\centerline{\includegraphics*[width=0.4\textwidth,angle=0]{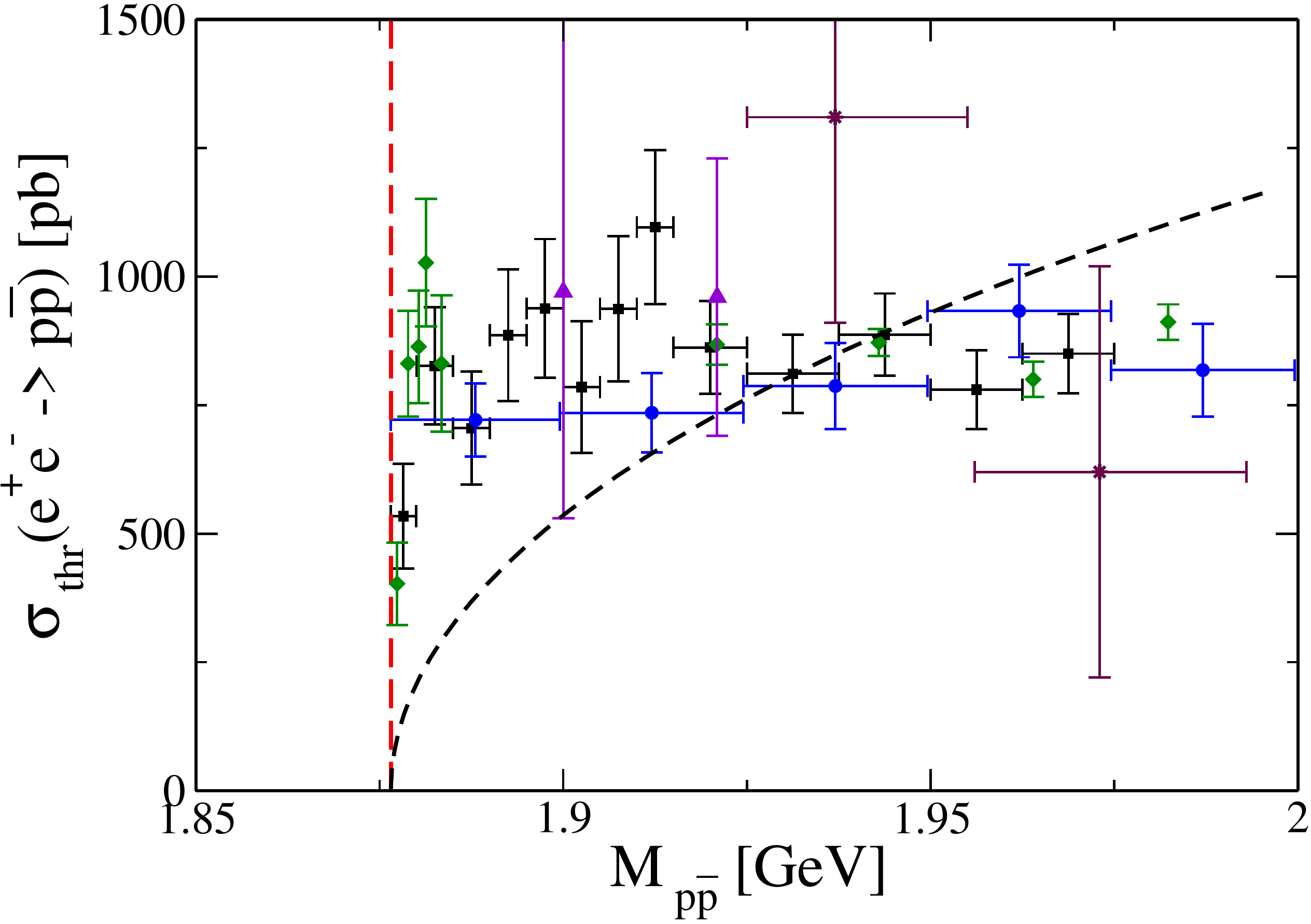}}
\centerline{\includegraphics*[width=0.4\textwidth,angle=0]{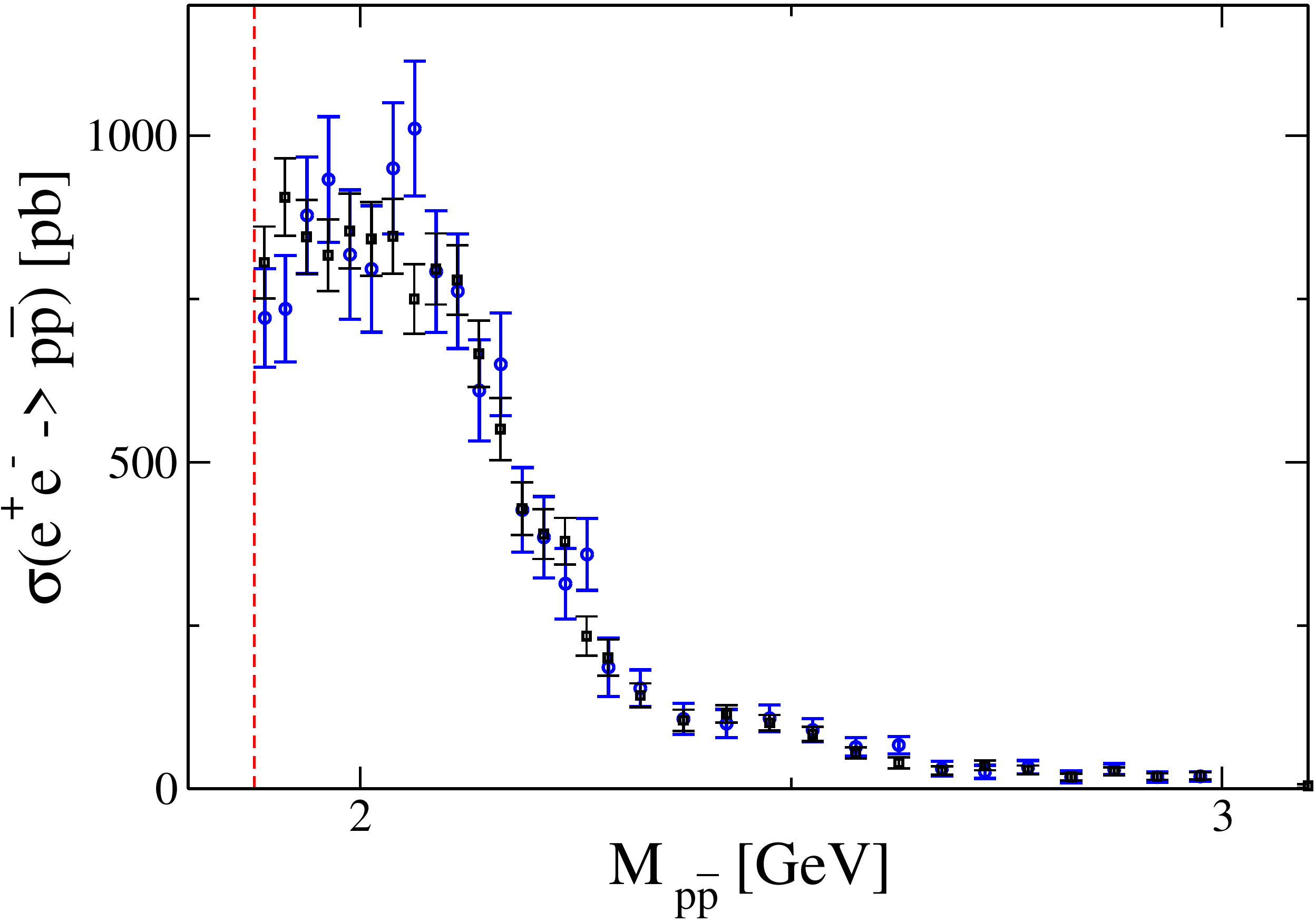}}
\caption{Upper panel: Cross section for $e^+e^-\to p\bar{p}$ in the threshold
region. The data are from BESIII~\cite{Ablikim:2021kjh} (blue circles),
BaBar~\cite{Lees:2013ebn} (black square), CMD-3~\cite{CMD-3:2018kql} (green diamonds),
DM1~\cite{Delcourt:1979ed}  (maroon crosses) and FENICE~\cite{Antonelli:1998fv} (violet triangles).
The dashed black line denote the phase space, normalized to the data at 50~MeV
excess energy.
Lower panel: Cross section for $e^+e^-\to p\bar{p}$ for invariant masses
below 3~GeV. The blue circles/black squares represent the data from
BESIII~\cite{Ablikim:2021kjh}/BaBar~\cite{Lees:2013ebn}. In both panels, the
vertical red dashed line marks the $p\bar{p}$ threshold.
}
\label{fig:datatlpFF}
\vspace{-3mm}
\end{figure}
Before discussing the DR fits including the data from the time-like region,
it is worth noting two very intriguing experimental findings related to the
cross section $\sigma(e^+e^-\to p\bar{p})$ (and the reversed reaction) 
and the corresponding effective
form factor $|G_{\rm eff}^p|$. First, as shown in the upper panel of
Fig.~\ref{fig:datatlpFF}, there is a strong enhancement in the close-to-threshold
region, as comparison with the phase space behavior (normalized to the data at about 50~MeV
excess energy) clearly reveals. Note also that due to the
Coulomb interaction between the proton and the antiproton, the cross section
does not start at zero. No such effects are seen in  $\sigma(e^+e^-\to n\bar{n})$.
We note that such threshold enhancements are also observed in other processes like
e.g. $J/\Psi\to xp\bar{p}$, $\Psi'(3686)\to xp\bar{p}$ with
$(x = \gamma, \omega, \rho, \pi, \eta)$ and $B^+\to K^+p\bar{p}$, see e.g.
Refs.~\cite{Bai:2003sw,Aubert:2005gw,BESIII:2011aa}. 
Second, extending further out in momentum transfer, the BaBar data~\cite{Lees:2013ebn}
and also the BESIII data~\cite{Ablikim:2021kjh} exhibit some oscillating structures,
most pronounced for invariant masses $M_{p\bar{p}}$ below 2.5~GeV, see the lower panel
in Fig.~\ref{fig:datatlpFF}. The corresponding neutron
data  from FENICE~\cite{Antonelli:1998fv} and SND~\cite{Achasov:2014ncd} are less
precise than the proton data, but show a similar behavior for $q^2 \lesssim 4\,$GeV$^2$.
For recent fits to the time-like proton effective FF accounting for these structures,
see~\cite{Tomasi-Gustafsson:2020vae}.

DR fits including space- and time-like data were performed in
Refs.~\cite{Hammer:1996kx,Belushkin:2006qa,Lorenz:2015pba}.
Here, we focus on the work done in the latest paper. Though that
work investigated some issues related to FSI in an exploratory way, it provides the most detailed information
on the physics contained in the time-like FFs. In this work,
the spectral functions was enlarged to account for the coupling
to the newly established  $\phi(2170)$ vector meson, as
well as baryonic triangle graphs with virtual $NN\pi$,
$N\Delta\pi$ and $\Delta\Delta\pi$ particles, the first of these giving
a simple representation of the strong final-state interactions (FSI),
see the discussion below.

\begin{figure}[t!]
\centering
\includegraphics[width=0.4\textwidth]{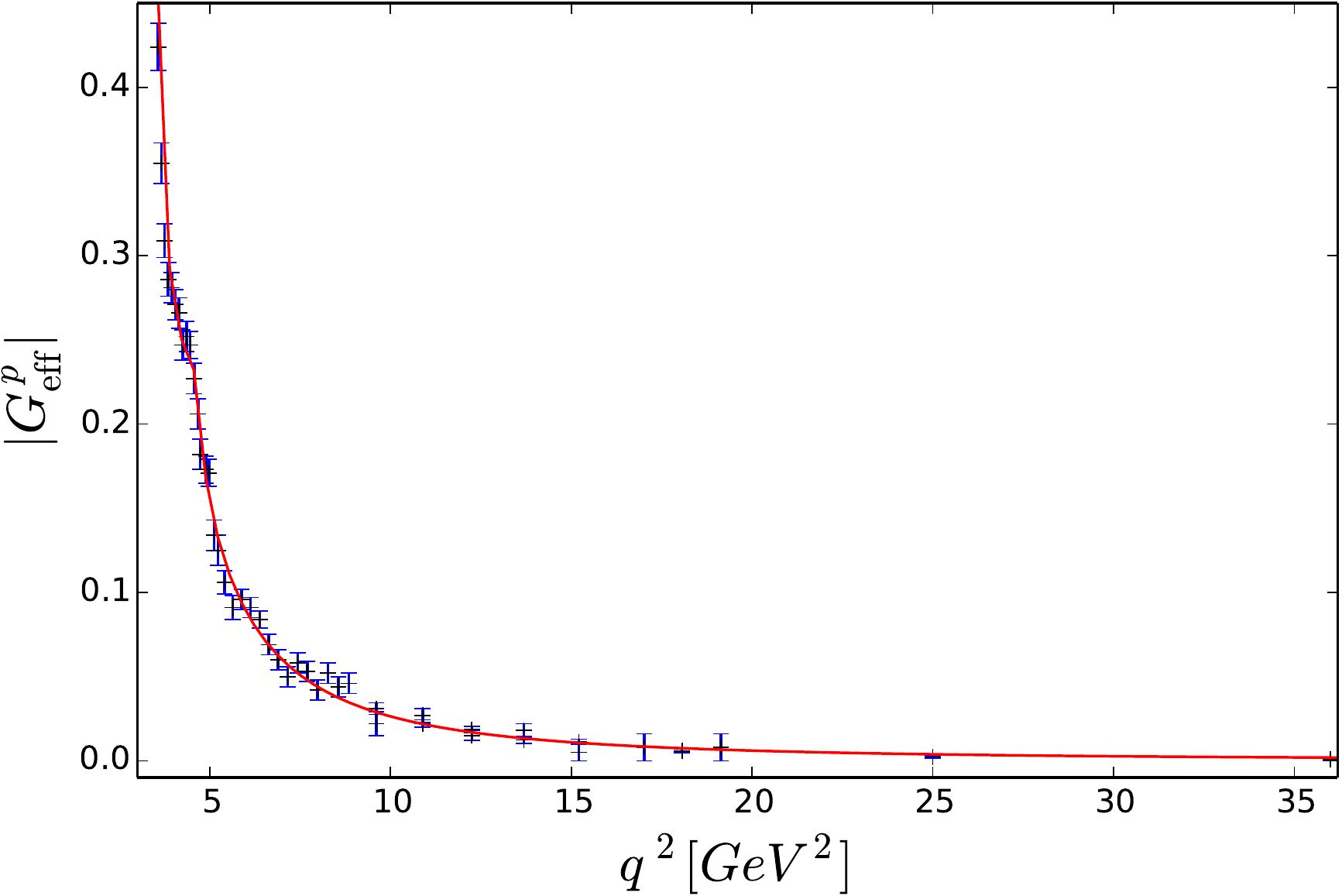}
\includegraphics[width=0.4\textwidth]{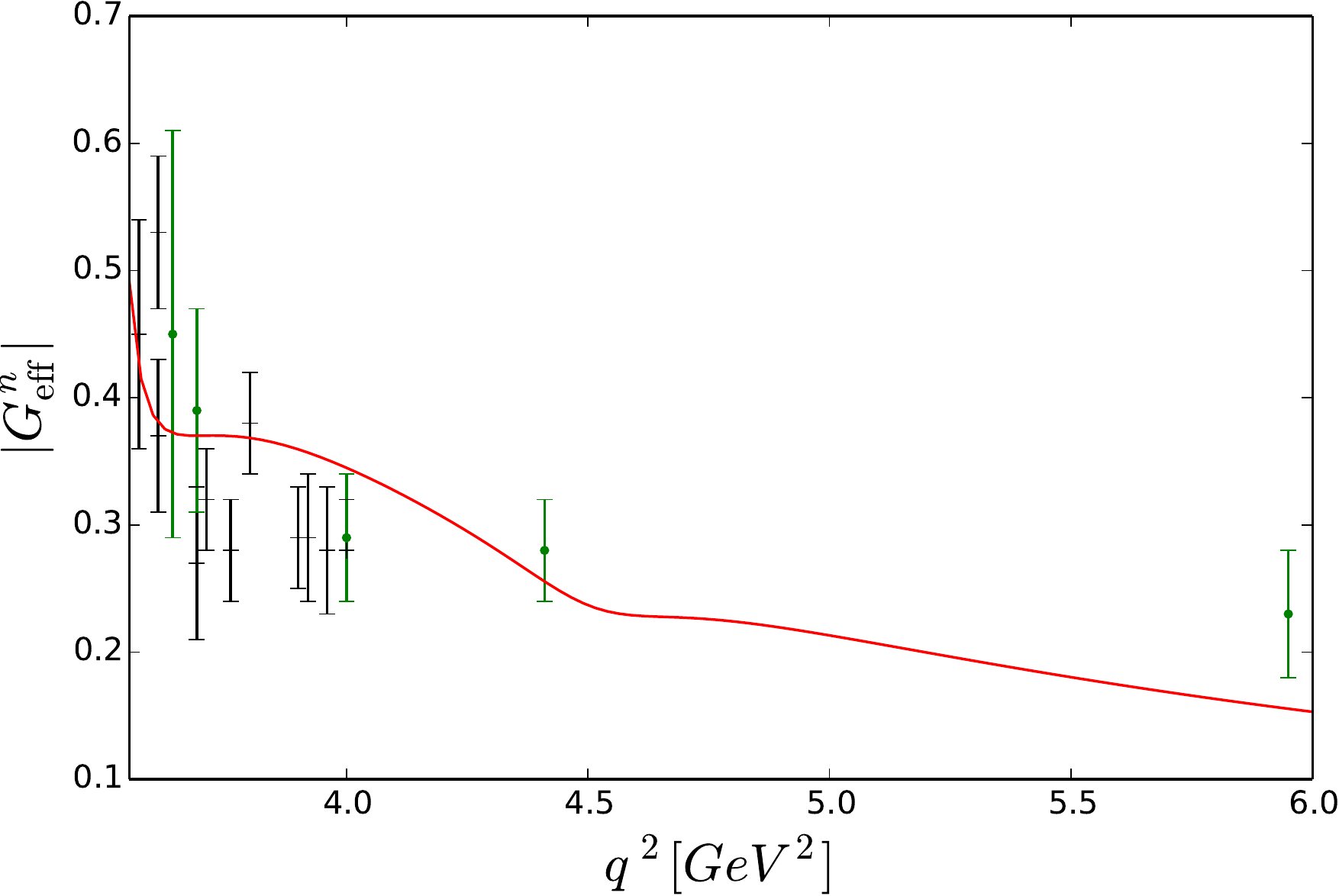}
\caption{The effective FF of the proton (upper panel) and the neutron (lower panel). Data for $G_{\rm eff}^p$
  are from BaBar~\cite{Lees:2013ebn,Lees:2013uta}. The neutron data are from 
  SND~\cite{Achasov:2014ncd} (crosses) and from FENICE~\cite{Antonelli:1998fv} (circles).}
\label{fig:eff}
\vspace{-3mm}
\end{figure}
\begin{figure}[t]
\centering
\includegraphics[width=0.4\textwidth]{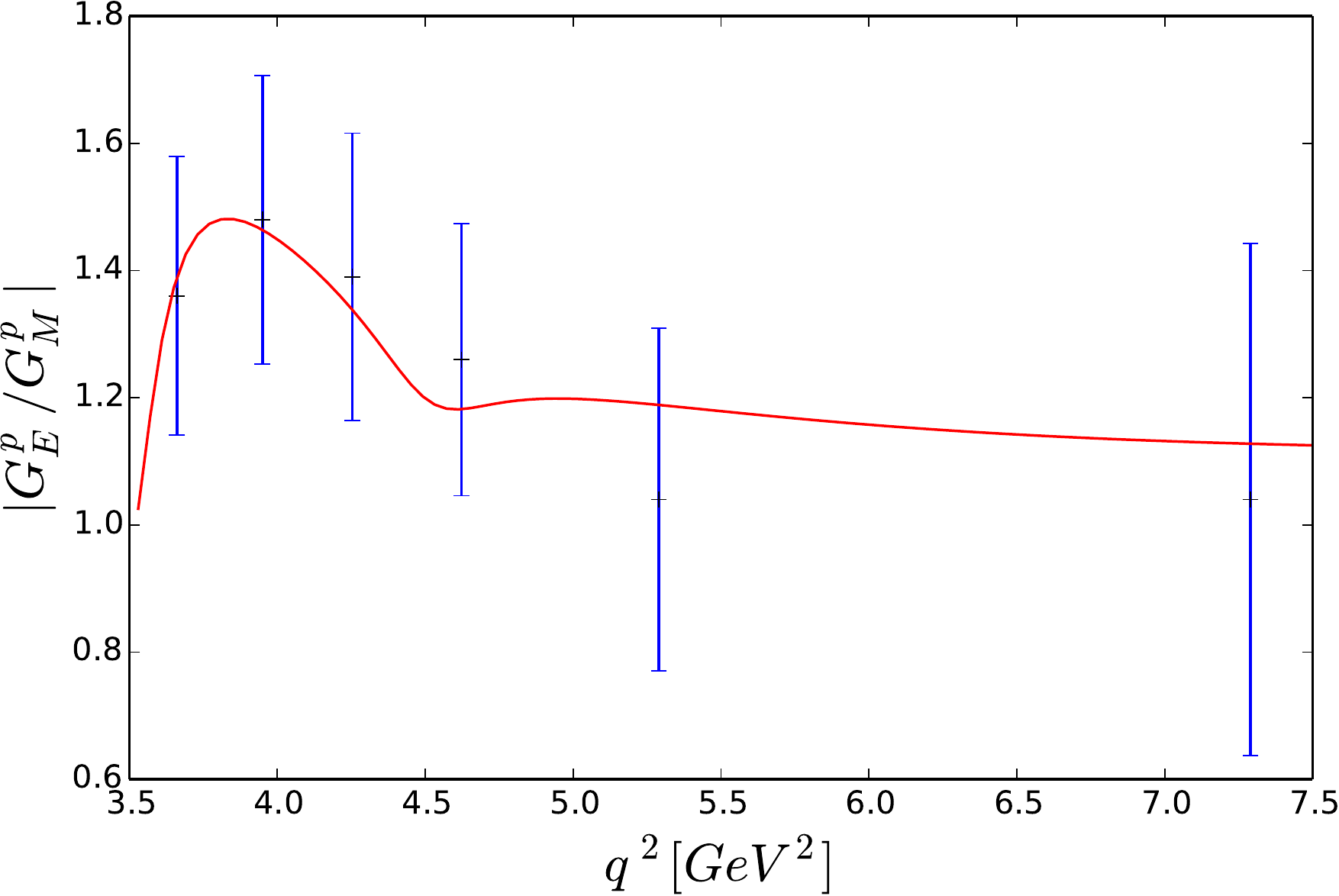}
\includegraphics[width=0.4\textwidth]{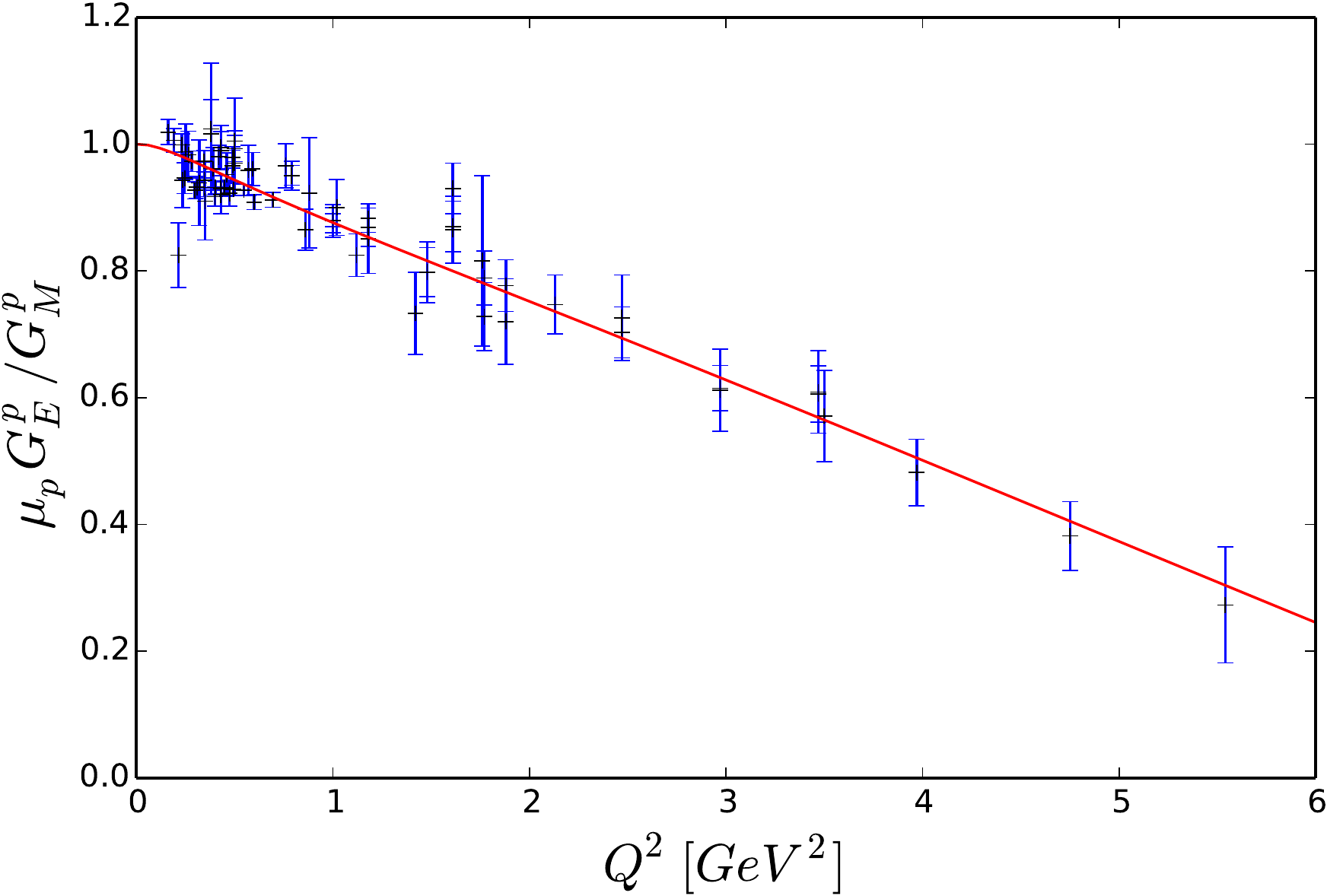}
\caption{The form factor ratio of the proton for space- (upper panel) and time-like (lower panel)
  momenta from the combined fit to space- and time-like data. The data are from
  BaBar~\cite{Lees:2013uta} (time-like region) and from the compilation in Ref.~\cite{Punjabi:2015bba}
  (space-like region)}
\label{fig:ratio}
\vspace{-3mm}
\end{figure}

Simultaneous fits to proton and neutron FFs for space- and time-like momenta
including the  $\phi(2170)$ were performed in Ref.~\cite{Lorenz:2015pba}. More
precisely, the differential cross sections and the ratio $G_E/G_M$ from polarization observables
on the scattering side in addition to the effective FF and $|G_E/G_M|$ on the production side were
included for the proton. In case of the neutron,  $G_E$ and $G_M$ from scattering data and again the
effective FF on the production side were considered. The spectral function included the
the $2\pi$, $K\bar{K}$ and $\rho\pi$-continuum, the $\omega$- and $\phi$-contribution and
three/five isoscalar/isovector poles restricted in the mass range $M_V = 1\ldots 1.8\,$GeV. In addition,
the new vector resonance was taken at a mass of 2.125~GeV and its width was determined in the
fit, which turned out to be $\Gamma = 0.088$~GeV. Good agreement with the existing data, as shown in
Fig.~\ref{fig:eff} for the proton and neutron effective form factors and in Fig.~\ref{fig:ratio}
for the proton form factor ratio in the space-like and the time-like region, was obtained.
In particular, the  SND data for the neutron effective FF show a very similar behavior to the
proton effective FF over a large range, which can be well described in this approach. However,
the range around the $\phi(2170)$ calls for further neutron measurements to allow for a
determination of the isospin of the structures in $G_{\rm eff}^p$.

Let us now discuss the various threshold effects in the time-like data, in particular the strong
enhancement at the $p\bar{p}$ threshold. This was first observed at LEAR in the inverse
reaction $p\bar{p}\to e^+e^-$ \cite{Bardin:1994am} and substantiated by the BaBar
collaboration~\cite{Aubert:2005cb}, which provided data for $ e^+e^-\to p\bar{p}$ down to the
threshold region. As noted before, this threshold enhancement was also observed in a number
of other production reactions such as $J/\psi$ and $B$ decays. Several explanations involving
unobserved  meson resonances or scenarios that involve $N\bar{N}$ bound states (baryonium)
have been put forward, see e.g. Ref.~\cite{Bai:2003sw}. More conventional but plausible
interpretations of this phenomenon were given in terms of the FSI between the proton and
the antiproton, employing either a Migdal-Watson approximation or meson-exchange models
of various levels of sophistication to describe the $p\bar{p}$ interaction, see e.g. the earlier works
Refs.~\cite{Zou:2003zn,Kerbikov:2004gs,Bugg:2004rk,Sibirtsev:2004id,Loiseau:2005cv,Haidenbauer:2006dm}.
The latest and arguably most sophisticated approach to this phenomenon employs simple
point-like form factors, whose energy dependence is entirely given by the proton-antiproton
FSI (or the $p\bar{p}$ initial state interactions in the annihilation process)~\cite{Haidenbauer:2014kja}.
The nucleon-antinucleon interaction is based on chiral effective field theory at NLO \cite{Kang:2013uia} and
NNLO~\cite{Dai:2017ont}. In this approach, the steep rise of the effective FF for energies
close to the $p\bar{p}$ threshold is explained solely in terms of the $p\bar{p}$ interactions,
cf. Fig.~\ref{fig:peff}, consistent with the findings in
Refs.~\cite{Haidenbauer:2006dm,Dmitriev:2007zz,Chen:2008ee,Chen:2010an,Dalkarov:2009yf,Bondar:2010zx,Haidenbauer:2012pu,Dmitriev:2013pfa,Dmitriev:2013xla,Entem:2007bb}.
Also existing experimental information (differential cross sections, form
factor ratio) is quantitatively described in this approach. In addition, predictions for various
spin-dependent observables, that can be tested in the future with PANDA at FAIR, are also given in that
work. Note, however, that this framework is only applicable to the threshold region, that is up excess
energies of about 100~MeV.

\begin{figure}[t!]
\begin{center}
\includegraphics[width=0.90\linewidth,clip]{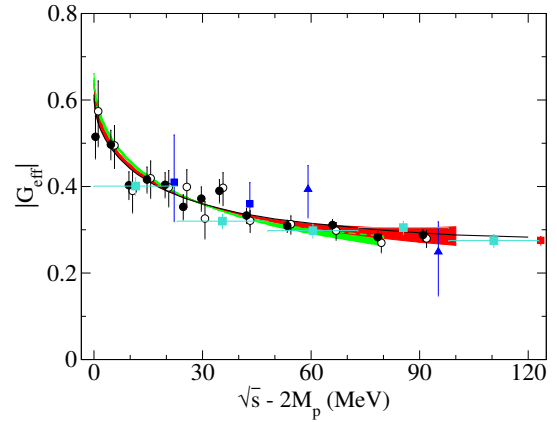}
\caption{Effective proton form factor as a function of the excess energy. 
The data are from the DM1~\cite{Delcourt:1979ed} (triangles), FENICE~\cite{Antonelli:1998fv} 
(squares), and BaBar~\cite{Aubert:2005cb} (empty circles), \cite{Lees:2013ebn} (filled squares)
collaborations. The green/red band refers to the calculation of the $p\bar{p}$ interaction
within chiral EFT at NLO/NNLO. Figure courtesy of Johann Haidenbauer.}
\label{fig:peff}
\end{center}
\vspace{-3mm}
\end{figure}

Triangle graphs with virtual $N\bar{\Delta}\pi$ and $\Delta\bar{\Delta}\pi$ states
were also considered in Ref.~\cite{Lorenz:2015pba}, picking up an idea
of Rosner~\cite{Rosner:2006vc}, in order to approximate possible cusp effects. However,
the vertices are not well known for these kinematics, so the calculation was reduced
to the scalar loop integral parameterized in terms of fitted strength parameter
$f_{N\Delta/\Delta\Delta}\sim {\cal O}(1)$ multiplying each loop structure. The explicit
momentum dependence of the vertices was accounted for in terms of overall form factors,
\begin{align}
 F(q^2) = \frac{1}{1 + q^2/\Lambda_{N\Delta/\Delta\Delta}^2}~,
\end{align}
with $\Lambda_{N\Delta/\Delta\Delta}$ the respective fitted cut-off parameter. Interestingly,
performing fits with these structures instead of the explicit poles discussed before leads
to an equally good description of the proton effective FF, very similar to what is shown in
Fig.~\ref{fig:eff} (upper panel). 

Furthermore, in Ref.~\cite{Lorenz:2015pba} the nucleon FFs were also discussed in
region of $t_0 = 4M_{\pi}^2 < t < t_{\rm thr} = 4m_p^2$ which is not accessible by direct measurements,
but by analytic continuation in $t = q^2 = -Q^2$. In fact, an additional particle emission from the
initial state proton can lower the energy of the (virtual proton) to reach below the threshold, as
discussed in Ref.~\cite{Guttmann:2012sq} for the process $p\bar{p}\to e^+e^-\pi^0$.
To get insight into the unphysical region, it is instructive to use a DR for the logarithm,
see e.g. Refs.~\cite{Gourdin:1974iq,Geshkenbein:1974gm,Baldini:1998qn,Pacetti:2007zz,Pacetti:2010nv}.
In principle, this also allows for a separation of the FF phase $\delta(t)$ and modulus in the
representation $G(t) = |G(t)|e^{i\delta(t)}$. A once subtracted DR for the function
$\ln [G(t)/G(0)]/(t\sqrt{t_0 - t})$ takes the form ($t<0$)
\bea
\ln G(t) &=& \ln G(0) + \frac{t\sqrt{t_0 - t}}{\pi}\int_{t_0}^{\infty}\frac{\ln|G(t')/G(0)|}{t'(t'-t)
\sqrt{t'-t_0}}dt'\nonumber\\
&\equiv& \int_{t_0}^{\infty} I(t,t_0,t')dt',\label{eq:ln}
\eea
where the first term vanishes due to the normalization $G_E(0) = G_M(0)/\mu_p = 1$. Experimental
information on this integral equation \eqref{eq:ln} is available in the space-like
region $t<0$ on $G(t)$ and in the time-like region for $t>t_{\rm thr}$ on the modulus $|G(t)|$. 
The solution of this integral equation is not straightforward, it requires  additional information
to be included (as the problem is ill-conditioned). One possible solution was proposed in  Refs.~\cite{Baldini:1998qn,Pacetti:2007zz},
namely to consider the integral contributions to the logarithm $\ln|G(t)|$ in the space-like region,
using definite values for the known part above $t_{\rm thr}$
\begin{align}
\ln G(t) - \int_{t_{\rm thr}}^{\infty} I(t,t_0,t')dt' = \int_{t_0}^{t_{\rm thr}} I(t,t_0,t')dt'~, ~~~t < 0.
\label{eq:minint}
\end{align}
In Ref.~\cite{Lorenz:2015pba}, as input for the left-hand-side of Eq.~\eqref{eq:minint}, the
discretized result of a simultaneous fit to data in all accessible regions were used. A typical
result for the modulus of $G_E^p(t)$ is shown in Fig.~\ref{fig:ln}. The enhancement just
below the production threshold could signal the appearance of a broad baryonium state, but
clearly more precise data on the time-like nucleon FFs are required to come to a definite
conclusion here.

\begin{figure}[t!]
\begin{center}
\includegraphics[width=0.4\textwidth]{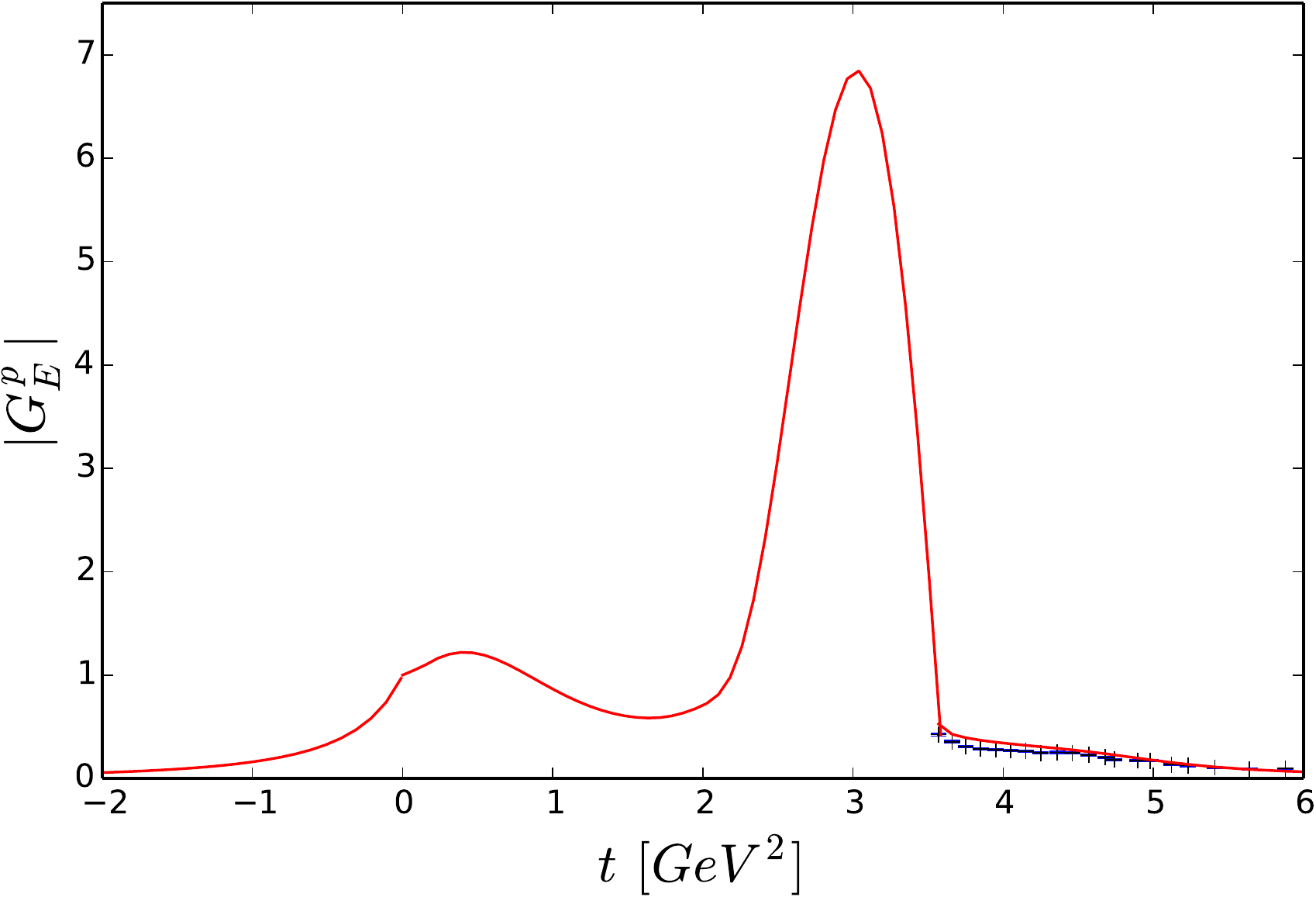}
\caption{An exemplary result for the modulus of electric form factor obtained from the
logarithmic integral Eq.~\eqref{eq:ln}. The form factors for $t<0$ are considered via
the differential cross sections, therefore they are not displayed here.}
\label{fig:ln}
\end{center}
\vspace{-3mm}
\end{figure}

\section{Summary and outlook}
\label{sec:summ}

This paper served two purposes. First, we have reviewed the dispersion-theoretical
approach to the electromagnetic form factors of the nucleon, with particular
emphasis on the constraints posed by unitarity and analyticity on the spectral functions.
Second, we have performed new fits including recent high-precision data on electron-proton
scattering for squared momentum transfers $Q^2<1\,$GeV$^2$, the proton form factor ratio in the
range $Q^2 \simeq 1\cdots 8.5\,$GeV$^2$ and the world data basis on the neutron electric and
magnetic form factors, including the recent accurate extraction of the neutron charge
radius squared from chiral nuclear EFT. We also have sharpened the toolbox to determine
the statistical and systematical uncertainties.  This led to a number of new results
concerning the various nucleon electromagnetic radii, the form factors and the
$\omega NN$ couplings. We would like to stress again that DRs have always found a
small proton charge radius, $r_E^p \simeq 0.84\,$fm, with a slightly larger proton magnetic
radius, $r_M^p \simeq 0.85\,$fm. As before, we find that the neutron magnetic radius
is the largest, $r_M^n \simeq 0.87\,$fm. Consistent with earlier analyses, the onset
of pQCD is not seen in the existing form factor data. We have also discussed our
present understanding of the physics in the time-like region, where a strong enhancement
of the cross section for $e^+e^-\to p \bar{p}$ (and in its reversed process $p\bar{p}\to e^+e^-$)
is observed, that can be understood in terms of proton-antiproton final-state interactions
(or initial-state interactions for the reserved process).  Furthermore, there are interesting
oscillating structures in the cross section that require additional poles and/or thresholds.

Clearly, there are a number of issues that require more data and/or further investigations:
\begin{itemize}
\item For the neutron data basis, a thorough analysis of the existing electron-deuteron
  and electron-$^3$He scattering data based on chiral effective field theory and
  including two-photon corrections should be performed. This would allow to consistently
  analyze the proton and neutron form factors based on the dispersive approach applied
  directly to cross section data.
\item A new combined analysis of the space-like data (as done here) with the time-like
  data should be performed, including the improved knowledge of the $p\bar{p}$ final-state
  interactions obtained in the last decade. This would also sharpen the predictions for
  future measurements with $\overline{\rm P}$ANDA at the FAIR facility.
\item Data on $ep$ scattering or the polarization transfer at $Q^2 \gg 10\,$GeV$^2$ are
  urgently needed to investigate the onset of perturbative QCD. It will also be interesting
  to find out whether the form factor ratio really crosses zero as the present data seem
  to indicate.
\item It would also be useful to improve our understanding of the spectral functions in the
  unphysical region, see Figs.~\ref{fig:FFgen} and \ref{fig:ln}. This requires more
  work based on logarithmic dispersion relations such as Eq.~\eqref{eq:ln}.
\end{itemize}

Finally, let us point out the the dispersion-theoretical approach to the nucleons electromagnetic
form factors has matured and become a precision tool to analyze electron scattering and
form factor ratio data. In the future, it will also be extended to analyze the upcoming
muon-proton scattering data from the MUSE~\cite{Downie:2014qna} and AMBER~\cite{Denisov:2018unj}.
experiments


\section*{Acknowledgments}

We are grateful to our (former) collaborators a for helping us to sharpen
our view on the issues discussed.
We acknowledge useful conversations with J\"urgen D\"olz on uncertainty
quantifications. We thank Johann Haidenbauer and Bernard Metsch with help to
prepare some of the figures.
This work of UGM and YHL is supported in
part by the Deutsche Forschungsgemeinschaft (DFG, German Research
Foundation) and the NSFC through the funds provided to the Sino-German Collaborative  
Research Center TRR~110 ``Symmetries and the Emergence of Structure in QCD''
(DFG Project-ID 196253076 - TRR 110, NSFC Grant No. 12070131001),
by the Chinese Aca\-de\-my of Sciences (CAS) through a President's
International Fellowship Initiative (PIFI) (Grant No. 2018DM0034), by the VolkswagenStiftung
(Grant No. 93562), and by the EU Horizon 2020 research and innovation programme,
STRONG-2020 project under grant agreement No. 824093. HWH was supported by the
Deut\-sche Forschungsgemeinschaft (DFG, German
Research Foundation) -- Projektnummer 279384907 -- CRC 1245
and by the German Federal Ministry of Education and Research (BMBF) (Grant
No. 05P18RDFN1).

\appendix
\section{Neutron form factors from light nuclei}
\label{app:nuclei}

As there are no free neutron targets, the neutron form factors must be
extracted from electron scattering off light nuclei. In this appendix,
we briefly outline how this can be achieved in case of the simplest
nucleus, the deuteron. The extension to systems with 3 or 4 nucleons
is similar but more complicated.

The deuteron is a spin-1 particle. Using Lorentz invariance, time-reversal invariance
as well as parity and current conservation, the matrix element of the deuteron em
currents takes the form, see e.g.~\cite{Arnold:1980zj,Garcon:2001sz} (as usual, in units
of the elementary charge $e$) 
\bea
\langle p',\lambda' | j_\mu^{\rm em}|p,\lambda\rangle
&=& -G_1(Q^2)\, \xi^*_{\nu} \xi^\nu \, (p'+p)_\mu \nonumber\\
&-& G_2(Q^2)\, (\xi_\mu \xi^{*,\nu}k_\nu - \xi^*_\mu \xi^\nu k_\nu)\nonumber\\
&+& G_3(Q^2)\, \frac{\xi^\nu k_\nu\,  \xi^{*,\nu'}k_{\nu'}\, (p'+p)_\mu}{2m_d^2}~,\nonumber\\
&&
\eea
where $p',p$ are the deuteron four-momenta, $\lambda',\lambda$ the corresponding helicities,
$m_d = 1.8756\,$GeV is the deuteron mass and  $Q^2 \equiv -k_\mu k^\mu = -(p'-p)^2\geq 0$.
Furthermore, the polarization four-vectors $\xi_\mu$ are subject to the constraints
$\xi_\mu(p,\lambda)p^\mu = 0$ and $\xi^*_\mu(p',\lambda')p'^\mu=0$. Instead of the scalar functions
$G_i(Q^2)\, (i=1,2,3)$, one often uses
the deuteron charge ($G_C$), magnetic ($G_M$) and quadrupole ($G_Q$) form factors given by
\bea
G_C(Q^2) &=& G_1(Q^2) + \frac{2}{3} \eta G_Q(Q^2)~, \nonumber\\
G_M(Q^2) &=& G_2(Q^2)~, \\
G_Q(Q^2) &=& G_1(Q^2) -G_2(Q^2) + (1+\eta) G_3(Q^2)~,\nonumber
\eea
with $\eta = Q^2/(4m_d^2)$. These form factors are subject to the normalizations:
\beq
G_C(0) = 1~,~~ 
G_M(0) = \frac{m_d}{m_p} \mu_d ~,~~
G_Q(0) = m_d^2 Q_d^{}~,
\eeq
in terms of the deuteron magnetic moment, $\mu_d = 0.857\, \mu_N$, and the deuteron quadrupole moment,
$Q_d = 0.286\,$fm$^2$.

In the one-photon exchange approximation, the unpolarized elastic
electron-deuteron ($ed$) scattering cross section in the laboratory (lab) frame is given by
\bea
\frac{d\sigma}{d\Omega} &=& \frac{\alpha^2}{4E^2}\frac{\cos^2(\theta/2)}{\sin^4(\theta/2)}
\left(1+\frac{2E}{m_d}\sin^2(\theta/2)\right)^{-1}\nonumber\\
&\times& \left[A(Q^2) + B(Q^2)\tan^2(\theta/2) \right]~,
\eea
with $E$ the energy of the incoming electron, $\theta$ the electron scattering angle
in the lab frame and $Q^2\geq 0$ is the squared momentum transfer. The structure functions 
$A(Q^2)$ and $B(Q^2)$ are related to the three form factors of the deuteron via
\bea
A(Q^2) &=& G_C^2(Q^2) + \frac{2}{3} \eta G_M^2(Q^2) + \frac{8}{9}\eta^2 G_Q^2(Q^2)~,\nonumber\\
B(Q^2) &=& \frac{4}{3} \eta (1+\eta)  G_M^2(Q^2)~.
\eea
As can be seen, from the unpolarized cross section one can not disentangle the charge and the
quadrupole form factors. This can be achieved by considering polarization data, e.g. the
tensor analyzing power $T_{20}(Q^2,\theta)$ is sensitive to a different combination of the
three deuteron FFs.

\begin{figure}[t]
\centering
\includegraphics[width=0.405\textwidth]{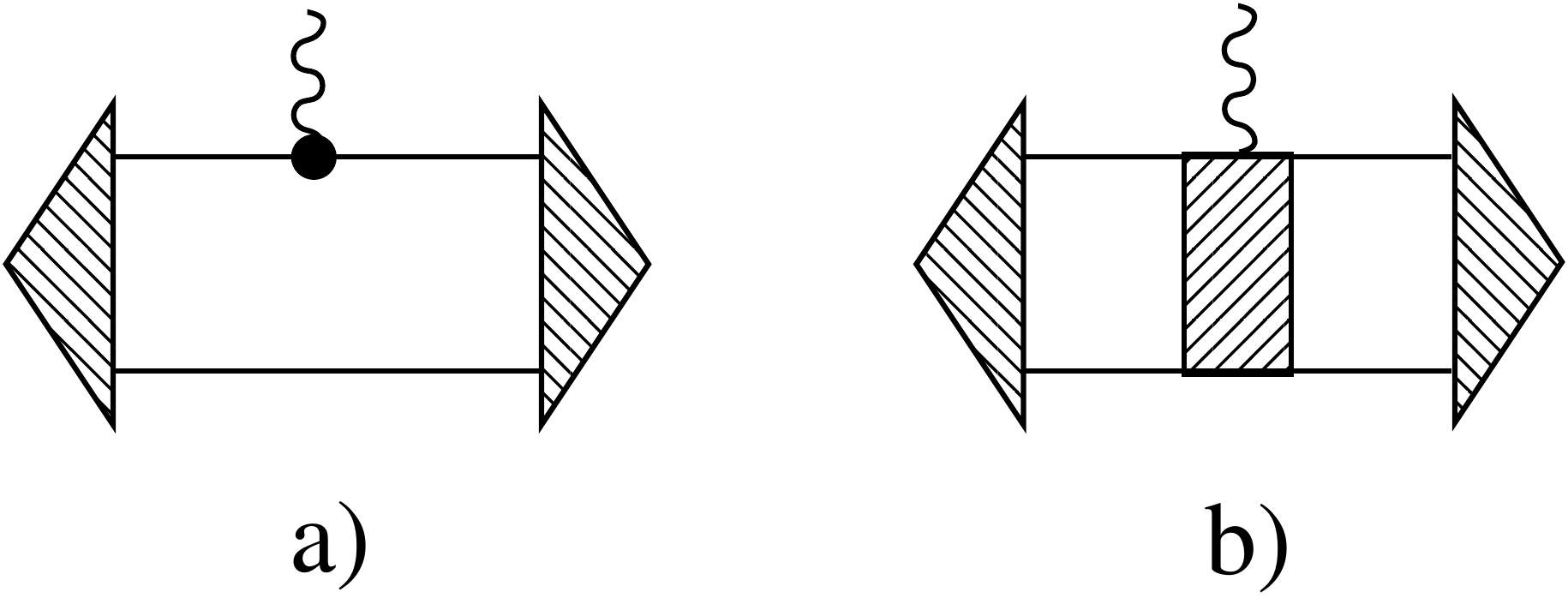}
\caption{Electromagnetic response of the deuteron. Diagram a) depicts the IA, that
  is sensitive to the nucleon em FFs (black circle) whereas diagram b) denotes the
  so-called two-body corrections (depicted by the shaded box) as explained in the
  text. The hatched triangles denote a deuteron wave function and solid (wiggly) lines
  represent nucleons (photons).
\label{fig:deutJ}}
\vspace{-3mm}
\end{figure}
Using a chiral expansion (or a meson-exchange model in older times), these deuteron FFs
can now be related to the single nucleon form factors as depicted in Fig.~\ref{fig:deutJ}.
The so-called impulse approximation (IA), where the photon couples to one nucleon,
is depicted in graph a). This contribution is evidently sensitive to the single nucleon
em FFs. To be precise, one takes the proton FFs as given and is then entirely sensitive
to $G_E^n$ and $G_M^n$, see below. There are, however, a number of corrections as displayed
by graph b) in Fig.~\ref{fig:deutJ}. This include one- and two-pion exchange currents as
well as photon--four-nucleon vertices (or heavy meson exchanges in the older language).
The status of the nuclear currents based on a systematic evaluation of the few-body
wave functions and the current operators is given in Ref.~\cite{Krebs:2020pii}. Consider for
example the charge density that generates the charge form factor. As the deuteron is an isoscalar,
its em response is entirely sensitive to the isoscalar form factors. Modulo higher order
corrections, at leading order in the chiral expansion, the charge density gives rise to the
form factors:
\bea
G_C^{\rm LO}(\vec{k}^2) &=& G_E^s(\vec{k}^2) \, \int_0^\infty (u^2(r)+w^2(r))\, j_0\left(\frac{kr}{2}\right) dr~,
\nonumber\\
G_Q^{\rm LO}(\vec{k}^2) &=& G_E^s(\vec{k}^2) \, \frac{6\sqrt{2}m_d^2}{\vec{k}^2}\\
&\times& \int_0^\infty \left(u(r)w(r)-\frac{w^2(r)}{\sqrt{8}}\right)\, j_2\left(\frac{kr}{2}\right) dr~,
\nonumber
\eea
where one works in the Breit frame with $\vec{k}^2 = Q^2$, the direction of the photon
momentum is taken along the positive $z$-axis and $k = |\vec{k}|$. Also, $u(r)$ and $w(r)$ are the 
deuteron S- and D-wave wave functions, normalized to one, see e.g.~\cite{Ericson:1988gk},
and $j_0(x),j_2(x)$ are the conventional Bessel functions.
Corrections to this leading order results can be worked out straightforwardly, these are
given for a chiral EFT approach in Ref.~\cite{Filin:2020tcs} (and references therein).

\section{Pion-nucleon scattering in the unphysical region}
\label{app:piN}

Here, we briefly discuss  the subthreshold expansion of the $\pi N$ amplitudes, which proceeds in terms of
the variables $\nu=(s-u)/(4m)$ and $t$ around $\nu=t=0$ ,
\begin{align}
\label{subthr}
\bar A^\pm(\nu,t)&=\begin{pmatrix}1\\\nu\end{pmatrix}
\sum_{n,m=0}^\infty a_{mn}^\pm\nu^{2m}t^n,\notag\\
\bar B^\pm(\nu,t)&=\begin{pmatrix}\nu\\1\end{pmatrix}
\sum_{n,m=0}^\infty b_{mn}^\pm\nu^{2m}t^n,
\end{align}
where the upper/lower entry corresponds to $I=\pm$, 
and the $a_{mn}^\pm, b_{mn}^\pm$ are the subthreshold
parameters. The $\bar{A}, \bar{B}$ are the
Born-term-subtracted amplitudes, defined as
\beq
\bar X^\pm(\nu,t)=X^\pm(\nu,t)-X^\pm_\text{pv}(\nu,t),\quad
X\in\{A,B\},
\eeq
with
\begin{align}
\label{Born_terms}
B^\pm_\text{pv}(\nu,t)&=g^2\bigg(\frac{1}{m^2-s}\mp\frac{1}{m^2-u}\bigg)
-\frac{g^2}{2m^2}\begin{pmatrix}0\\1\end{pmatrix},\notag\\
A^\pm_\text{pv}(\nu,t)&=\frac{g^2}{m}\begin{pmatrix}1\\ 0\end{pmatrix}.
\end{align}
Here, the subscript `pv' refers to the pseudovector $\pi N$ coupling as required from chiral symmetry and $g$ denotes the $\pi N$ coupling constant, to be identified later with $g_c$ for the charged-pion vertex. 
In the RS analysis of $\pi N$ scattering,  the value $g_c^2/(4\pi)=13.7(0.2)$~\cite{Baru:2011bw} has been used. 
This value is in line with the most recent determination from nucleon--nucleon scattering~\cite{Perez:2016aol,Reinert:2020mcu}.
More details on the subthreshold expansion of the $\pi N$ scattering amplitude is given in 
Refs.~\cite{Hoehler:1983,Hoferichter:2015hva}.

\section{Analysis of the three-pion contribution}
\label{app:3pi}

Here, we briefly review the arguments of Ref.~\cite{Bernard:1996cc}
that there is no strong enhancement on the left shoulder of the
$\omega$ resonance, the lowest vector meson in the three-pion channel
(for a recent update, see Ref.~\cite{Kaiser:2019irl}).
The imaginary parts of the isoscalar electromagnetic form factors open
at the three-pion threshold $t_0 = 9M_\pi^2$. The three-pion cut contribution is
given  by
\begin{equation}
{1\over 2}\int d\Gamma_3(A\, B) \,\, ,
\label{cut}
\end{equation} 
where the symbol 'A' refers to the $\gamma \to 3\pi$ and 'B' to the $3\pi
\to  \bar N N$ transition, respectively, and $d\Gamma_3$ is the measure on
the invariant three-body phase space.  This can be explicitly worked out
in baryon chiral perturbation theory~\cite{Bernard:1995dp}, where to leading order in
the small parameter $p$, namely at order ${\cal O}(p^7)$, the two-loop diagrams shown
in Fig.~\ref{fig:twoloop} can contribute to the isoscalar imaginary parts, however,
graph~(d) vanishes because of an isospin factor zero.
\begin{figure}[t!]
\centering
\includegraphics[width=0.305\textwidth]{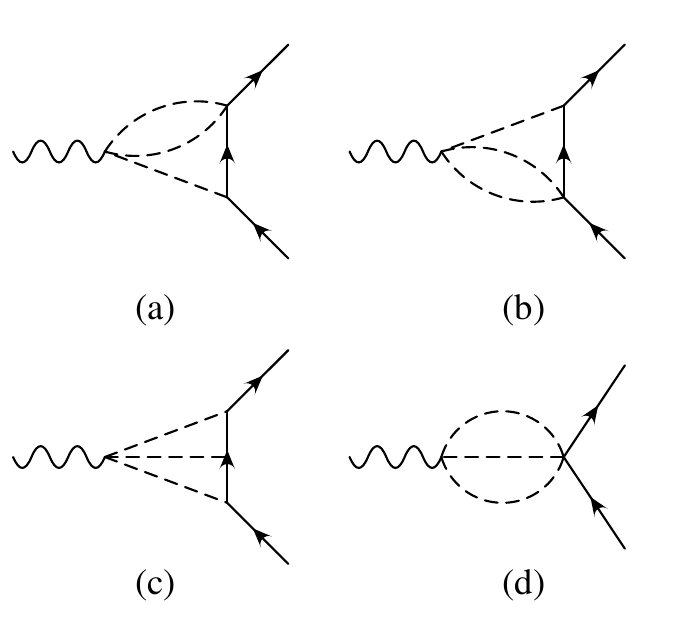}
\caption{ Two-loop diagrams contributing to the imaginary parts
of the isoscalar electromagnetic  nucleon form factors.  
 Wiggly, dashed and solid lines represent photons, pions and nucleons, in order.  
\label{fig:twoloop}}
\end{figure}
The  isoscalar imaginary parts in the heavy nucleon limit $m\to \infty$ can be given in compact
form (note that this corresponds to switching off all higher order corrections starting
at ${\cal O}(p^8)$)
\bea 
{\rm Im}\, G_E^S(t) &=& {3g_A^3 \,t\over (4 \pi)^5 F_\pi^6}
\int\int_{z^2<1} d \omega_1 d\omega_2\, |\vec{l}_1| \, |\vec{l}_2| \nonumber\\  
&\times& \sqrt{1-z^2} \, \arccos(-z) \,\,,
\label{imges}
\eea
\begin{eqnarray} 
&&{\rm Im}\, G_M^S(t) = {g_Am\over (8 \pi)^4 F_\pi^6}
\biggl\{L(t) \biggl[ 3t^2 -10 tM_\pi^2 + 2M_\pi^4 \nonumber\\
&& \qquad\qquad\qquad\qquad\qquad\qquad + g_A^2
\bigl(3t^2 -2 tM_\pi^2 - 2M_\pi^4 \bigr)\biggr]\nonumber\\ && \quad+W(t)\biggl[
t^3+2t^{5/2}M_\pi-39t^2 M_\pi^2 -12 t^{3/2}M_\pi^3\nonumber\\
&& \qquad\qquad\qquad  +65 t M_\pi^4 -50 
\sqrt{t} M_\pi^5 -27 M_\pi^6 \nonumber \\ && \quad +g_A^2 \bigl( 5t^3+10t^
{5/2}M_\pi- 147t^2 M_\pi^2 +36 t^{3/2}M_\pi^3 \nonumber\\
&& \qquad\qquad\qquad +277 t M_\pi^4 -58 \sqrt{t} M_\pi^5 -135 M_\pi^6 \bigr) \biggr] \biggr\}\,\,,
\label{imgms} 
\end{eqnarray}   
with
\bea 
L(t) &=& {M_\pi^4 \over 2t^{3/2}} \ln {\sqrt
t-M_\pi+\sqrt{t -2\sqrt t M_\pi-3M^2_\pi}\over2M_\pi} , \\
W(t) &=& {\sqrt t- M_\pi\over 96 t^{3/2}}\sqrt{t-2\sqrt t M_\pi-
3M^2_\pi}~, 
\eea
and in terms of the kinematical variables of the two independent pions
(the third one can be expressed in terms of these and the nucleon momenta),
\bea 
|\vec{l}_i| &=& \sqrt{\omega_i^2-M_\pi^2}\,\,, \, \, i=1,2~, \nonumber\\
 z &=&\hat{l}_1 \cdot \hat{l}_2=
 \frac{\omega_1 \omega_2 -\sqrt t (\omega_1+\omega_2)
+{1\over2} (t+M_\pi^2)}{|\vec{l}_1|\,|\vec{l}_2|}\, .
\eea

The behavior near threshold $t_0 = 9M_\pi^2$ of the imaginary parts
for finite pion mass, Eqs.~(\ref{imges},\ref{imgms}), is given by 
\bea
{\rm Im}\, G^S_E(t) &\sim& (\sqrt t-3M_\pi)^3\,, \nonumber\\
{\rm Im}\, G^S_M(t) &\sim& (\sqrt t-3M_\pi)^{5/2} 
\eea
which corresponds to a stronger growth than pure phase space  
\beq
 \int\int_{z^2<1} d\omega_1 d\omega_2 \,
|\vec{l}_1 \times \vec{l}_2|^2 \, \sim (\sqrt t-3M_\pi)^4 \,\, \, .
\eeq
This feature indicates (as in the isovector case) that in the heavy nucleon
mass limit $m \to \infty$ normal and anomalous thresholds coincide. In order to
find these singularities for finite nucleon mass $m$ an investigation of the Landau
equations is necessary \cite{polk}.  By using standard techniques
\cite{polk} one finds one anomalous threshold of diagrams (a) and (b) at  
\bea 
\sqrt{t_c} &=& M_\pi \biggl(\sqrt{4-M_\pi^2/m^2}+
\sqrt{1-M_\pi^2/m^2} \biggr)\,\,, \nonumber\\
t_c &=& 8.90\, M_\pi^2
\label{anthr}
\eea
which is very near to the (normal) threshold $t_0=9M_\pi^2$ and indeed
coalesces with $t_0$ in the infinite nucleon mass limit. Note that diagram
(d) does not possess this anomalous threshold, but only
the normal one.

\begin{figure}[t]
\centering
\includegraphics[width=0.405\textwidth]{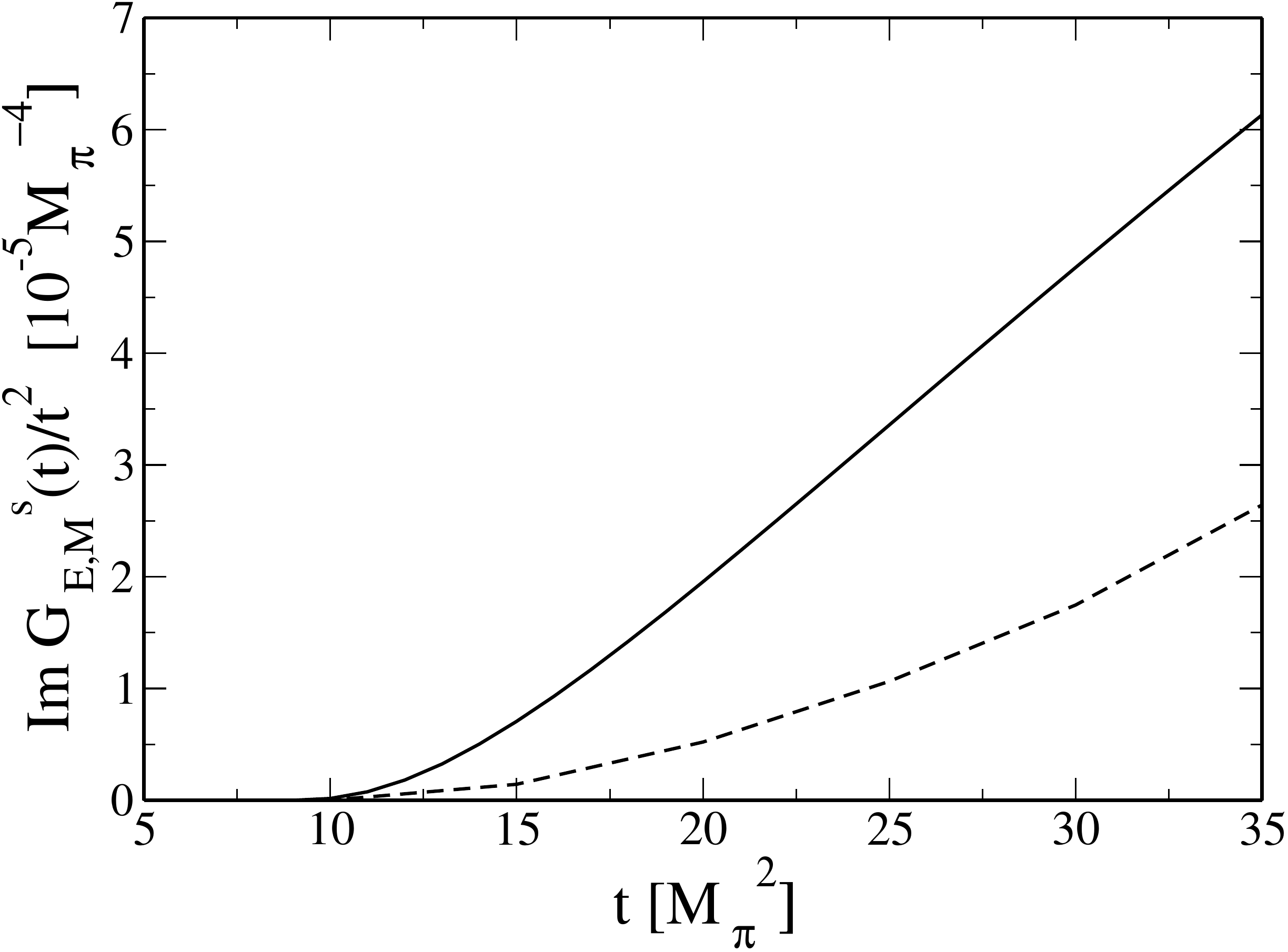}
\caption{Spectral distribution of the isoscalar electric and magnetic
nucleon form factors weighted with $1/t^2$ in the heavy nucleon limit.
Shown are Im\,$G_M^S (t)/t^2$ (solid line) and Im\,$G_E^S (t)/t^2$ (dashed
line). 
\label{fig:ImGtwoloop}}
\vspace{-3mm}
\end{figure}
The resulting spectral distributions  weighted with $1/t^2$ are
shown in Fig.~\ref{fig:ImGtwoloop}. Very different to the isovector
spectral functions discussed in Sec.~\ref{sec:2pi},
they show a smooth rise and are two orders of magnitude
smaller than the corresponding isovector ones, cf. Fig.~\ref{fig:ImG}.
There is indeed no enhancement  of the isoscalar electromagnetic spectral
function near threshold.  Even though the isoscalar and isovector
electromagnetic form factors behave formally very similar concerning the 
existence of anomalous thresholds $t_c$ very close to the normal thresholds
$t_0$, the influence of these on the physical spectral functions is
rather different for the two cases. Only in the isovector case a strong
enhancement is visible. This is presumably due to the different phase space
factors, which are $(t-t_0)^{3/2}$ and $(t-t_0)^4$ for the isovector
and isoscalar case, respectively. In the latter case,  the
anomalous threshold at $t_c = 8.9M_\pi^2$ is thus effectively masked. 
This justifies the standard DR approach of only taking the $\omega$-meson
as the lowest pole in isoscalar spectral function.

\section{{\boldmath$K\bar{K}\to N\bar{N}$} partial wave amplitudes}
\label{app:KN}

We seek the partial wave expansion of the $K\bar{K}\to N\bar{N}$ amplitude.
This allows one to impose the constraints of unitarity 
in a straightforward way. Here, we utilize  the helicity
amplitude formalism of Jacob and Wick \cite{Jacob:1959at}.
Denoting with $\lambda$ and $\bar{\lambda}$  the nucleon and antinucleon
helicities, the corresponding $S$-matrix element takes the form
\begin{eqnarray}
\label{kkbar:s_norm}
& &\bra{N(p,\lambda)\bar{N}(\pbar,\bar{\lambda})}{\hat S}\ket{K(k)\bar{K}
(\bar{k})}= \\ & &\phantom{N(p,\lambda)}
i(2\pi)^4\delta^4(p+\pbar-k-\bar{k})\nonumber\\
&& \times (2\pi)^2 \left[{64 t\over t-4 M_K^2}
\right]^{1/2} \bra{\theta,\phi,\lambda,\bar{\lambda}}{\hat S}(P)\ket{00}
\,,\nonumber
\end{eqnarray}
where $t=P^2=(p+\pbar)^2$ and $M_K$ is the kaon mass. The partial wave expansion
of the matrix element $\bra{\theta, \phi, \lambda, \bar{\lambda}}{\hat S}(P)\ket{00}$
is given by \cite{Jacob:1959at,Musolf:1996qt}
\begin{eqnarray}
\label{kkbar:par_d}
S_{\lambda, \bar{\lambda}} &\equiv&
\bra{\theta, \phi, \lambda, \bar{\lambda}}{\hat S}(P)\ket{00}\nonumber\\
&=&\sum_J \left({2J+1\over 4\pi}\right) b_J^{\lambda, \bar{\lambda}} \;
{\cal D}_{0\mu}^J(\phi, \theta, -\phi)^{\ast}\,,
\end{eqnarray}
where ${\cal D}_{\nu\, \nu'}^J(\alpha, \beta, \gamma)$ denotes a
Wigner rotation matrix with $\mu=\lambda-\bar{\lambda}$. Further,
the $b_J^{\lambda\, \bar{\lambda}}$ define partial waves of angular momentum $J$.
Because of the quantum numbers of the isoscalar em current, only
the $J=1$ partial waves contribute to the spectral functions. Moreover,
because of parity invariance only two of the four partial
waves are independent. In Ref.~\cite{Hammer:1998rz}, the $\bpp$ and $\bpm$
were chosen. These fulfill the threshold relation~  \cite{Musolf:1996qt}
\begin{equation}
\label{kkbar:th_rel}
\left.\bpm(t)\right|_{t=4\mns}=\sqrt{2}\,\left.\bpp(t)\right|_{t=4\mns}\,.
\end{equation}
Using the above definitions,  unitarity of the $S$-matrix, $S^{\dag} S = 1$, requires that
\begin{equation}
\label{kkbar:ubs}
\left(\frac{t/4-m^2}{t/4-M_K^2}\right)^{1/4}|b_J^{\lambda, \bar{\lambda}}(t)|\leq 1 \,,
\end{equation}
for $t\geq 4\mns$.
Thus, unitarity gives model-independent bounds on the
contribution of the physical region ($t\geq4\mns$) to the
imaginary part. In the unphysical region, that is for $4\mks \leq t \leq 4\mns$,
the partial waves are not bounded by unitarity.
Therefore, one must rely upon an analytic continuation of
$KN$ scattering amplitudes, which is not entirely straightforward.
This is discussed in detail in Ref.~\cite{Hammer:1998rz}.

\section{Fit parameters of the best fit}
\label{app:newresults}

We collect the various vector meson masses and couplings that appear in the
spectral functions and the normalization constants of the various
data sets in Table~\ref{tab:values}. 
\begin{table*}[ht!]
\centering
\begin{tabular}{|l|c|c|c||l|c|c|c|}
\hline
$V_{s}$ & $m_V$ & $a_1^V$ &$a_2^V$ & $V_{v}$ & $m_V$ & $a_1^V$ & $a_2^V$  \\
\hline
$\omega$ & $0.7830$	& $0.6893$  & $0.0431$	& $v_1$ & $1.1222$ & $1.0414$ & $-0.6239$\\
$\phi$   & $1.0190$	& $-0.0281$ & $-0.4705$	& $v_2$ & $1.5147$ & $-4.0062$ & $-3.0365$\\
$s_1$    & $1.8267$     & $0.3768$  & $0.5590$	& $v_3$ & $1.8062$ & $4.8533$ &  $2.1897$\\
$s_2$    & $4.0020$	& $-1.2786$ & $-4.882$	& $v_4$ & $2.2543$ & $-2.0208$ &$-0.0438$\\
$s_3$    & $4.0713$	& $1.8028$	& $4.0681$	&  & 	 & & 	\\
$s_4$    & $4.3075$	& $-0.6576$	& $0.4944$&  & 	 & & 	\\
		\hline
	\end{tabular}
	\begin{tabular}{|l|l|l|l|l|l|l|l|l|l|l|l|}
		\hline
		n1 & $0.9965$ & n2 & $1.0061$ &n3 & $1.0028$ & n4 & $1.0010$ &n5 & $1.0035$ & n6 & $0.9914$ \\
		n7 & $0.9982$ & n8 & $0.9929$ &n9 & $1.0076$ & n10 & $1.0000$ &n11 & $1.0000$ & n12 & $1.0037$ \\
		n13 & $1.0030$ & n14 & $1.0044$ &n15 & $1.0055$ & n16 & $1.0027$ &n17 & $1.0048$ & n18 & $1.0013$ \\
		n19 & $0.9995$ & n20 & $1.0029$ &n21 & $0.9977$ & n22 & $0.9905$ &n23 & $0.9985$ & n24 & $1.0100$ \\
		n25 & $1.0080$ & n26 & $1.0069$ &n27 & $0.9999$ & n28 & $1.0100$ &n29 & $1.0066$ & n30 & $0.9999$ \\
		n31 & $1.0100$ &$\mathrm{\tilde{\rm n}1}$ & $0.9989$ & $\mathrm{\tilde{\rm n}2}$ & $1.0059$ &  & &&&& \\
		\hline
	\end{tabular}
\caption{The parameters corresponding to our best fit described in Sec.~\ref{sec:newfits}.
  Vector meson (upper panel) and normalization (lower panel) parameters. The normalization
  constants n1$,\ldots ,$n31 refer to the MAMI data sets, whereas
  $\mathrm{\tilde{\rm n}1,\tilde{\rm n}2}$ normalize the PRad data. Masses $m_V$ are
  given in GeV and couplings $a_i^V$ in GeV$^{2}$.}
	\label{tab:values}
\end{table*}



\begin{thebibliography}{99}
\bibitem{Wilczek:2012ab}
  F.~Wilczek,
  Centr. Eur. J. Phys. \textbf{10}, 1021 (2012)
  [arXiv:1206.7114 [hep-ph]].


\bibitem{Denig:2012by}
A.~Denig and G.~Salme,
Prog. Part. Nucl. Phys. \textbf{68}, 113-157 (2013)
[arXiv:1210.4689 [hep-ex]].

\bibitem{Punjabi:2015bba}
V.~Punjabi, C.~F.~Perdrisat, M.~K.~Jones, E.~J.~Brash and C.~E.~Carlson,
Eur. Phys. J. A \textbf{51}, 79 (2015)
[arXiv:1503.01452 [nucl-ex]].

\bibitem{Armstrong:2012bi}
D.~S.~Armstrong and R.~D.~McKeown,
Ann. Rev. Nucl. Part. Sci. \textbf{62}, 337-359 (2012)
[arXiv:1207.5238 [nucl-ex]].

\bibitem{Maas:2017snj}
F.~E.~Maas and K.~D.~Paschke,
Prog. Part. Nucl. Phys. \textbf{95}, 209-244 (2017)

\bibitem{Hofstadter:1956} R.W.~McAllister and R.~Hofstadter,
  Phys.\ Rev.\ {\bf 102}, 851 (1956).

\bibitem{Hofstadter:1957wk} 
  R.~Hofstadter,
 Ann.\ Rev.\ Nucl.\ Part.\ Sci.\  {\bf 7}, 231 (1957).

 \bibitem{Karplus:1952}
 R.~Karplus, A.~Klein and J.~Schwinger,
 Phys.\ Rev.\ {\bf 86}, 288 (1952).
 
\bibitem{Mohr:2008fa} 
  P.~J.~Mohr, B.~N.~Taylor and D.~B.~Newell,
  Rev.\ Mod.\ Phys.\  {\bf 80}, 633 (2008)
  [arXiv:0801.0028 [physics.atom-ph]].

\bibitem{Pohl:2010zza} 
  R.~Pohl {\it et al.},
  Nature {\bf 466}, 213 (2010).

\bibitem{Bernauer:2010wm} 
  J.~C.~Bernauer {\it et al.} [A1 Collaboration],
  Phys.\ Rev.\ Lett.\  {\bf 105}, 242001 (2010)
  [arXiv:1007.5076 [nucl-ex]].

  

\bibitem{Beyer:2017gug} 
  A.~Beyer {\it et al.},
  Science {\bf 358}, 79 (2017).

\bibitem{Fleurbaey:2018fih} 
  H.~Fleurbaey {\it et al.},
  Phys.\ Rev.\ Lett.\  {\bf 120}, 183001 (2018)
  [arXiv:1801.08816 [physics.atom-ph]].


\bibitem{Bezginov:2019mdi} 
  N.~Bezginov, T.~Valdez, M.~Horbatsch, A.~Marsman, A.~C.~Vutha and E.~A.~Hessels,
  Science {\bf 365}, 1007 (2019).

\bibitem{Xiong:2019umf}
W.~Xiong, A.~Gasparian, H.~Gao, D.~Dutta, M.~Khandaker, N.~Liyanage, E.~Pasyuk, C.~Peng, X.~Bai and L.~Ye, \textit{et al.}
Nature \textbf{575}, no.7781, 147-150 (2019).

  
\bibitem{CODATAnew}  
  https://physics.nist.gov/cgi-bin/cuu/Value?rp  
  
\bibitem{Hammer:2019uab}
H.-W.~Hammer and U.-G.~Mei{\ss}ner,
Sci. Bull. \textbf{65}, 257-258 (2020)
[arXiv:1912.03881 [hep-ph]].



\bibitem{Frazer:1959gy}
W.~R.~Frazer and J.~R.~Fulco,
Phys. Rev. Lett. \textbf{2}, 365 (1959).

\bibitem{Frazer:1960zza}
W.~R.~Frazer and J.~R.~Fulco,
Phys. Rev. \textbf{117}, 1603-1609 (1960).

\bibitem{Frazer:1960zzb}
W.~R.~Frazer and J.~R.~Fulco,
Phys. Rev. \textbf{117}, 1609-1614 (1960).

\bibitem{Pacetti:2015iqa}
S.~Pacetti, R.~Baldini Ferroli and E.~Tomasi-Gustafsson,
Phys. Rept. \textbf{550-551}, 1-103 (2015).



\bibitem{Hohler:1976ax}
G.~H\"ohler, E.~Pietarinen, I.~Sabba Stefanescu, F.~Borkowski, G.~G.~Simon, V.~H.~Walther
and R.~D.~Wendling,
Nucl. Phys. B \textbf{114}, 505-534 (1976).


\bibitem{Okubo:1963fa}
S.~Okubo,
Phys. Lett. \textbf{5}, 165-168 (1963).

\bibitem{Zweig:1964jf}
G.~Zweig,
CERN-TH-412 (1964).

\bibitem{Iizuka:1966fk}
J.~Iizuka,
Prog. Theor. Phys. Suppl. \textbf{37}, 21-34 (1966).

\bibitem{Mergell:1995bf}
P.~Mergell, U.-G.~Mei{\ss}ner and D.~Drechsel,
Nucl. Phys. A \textbf{596}, 367-396 (1996)
[arXiv:hep-ph/9506375 [hep-ph]].


\bibitem{Gari:1984ia}
M.~Gari and W.~Kr\"umpelmann,
Z. Phys. A \textbf{322}, 689-693 (1985).

\bibitem{Gari:1986rj}
M.~Gari and W.~Kr\"umpelmann,
Phys. Lett. B \textbf{173}, 10-14 (1986).


\bibitem{Simon:1980hu}
G.~G.~Simon, C.~Schmitt, F.~Borkowski and V.~H.~Walther,
Nucl. Phys. A \textbf{333}, 381-391.

\bibitem{Hammer:1996kx}
H.-W.~Hammer, U.-G.~Mei{\ss}ner and D.~Drechsel,
Phys. Lett. B \textbf{385}, 343-347 (1996)
[arXiv:hep-ph/9604294 [hep-ph]].

\bibitem{Jones:1999rz}
M.~K.~Jones \textit{et al.} [Jefferson Lab Hall A],
Phys. Rev. Lett. \textbf{84}, 1398-1402 (2000)
[arXiv:nucl-ex/9910005 [nucl-ex]].

\bibitem{Gayou:2001qd}
O.~Gayou \textit{et al.} [Jefferson Lab Hall A],
Phys. Rev. Lett. \textbf{88}, 092301 (2002)
[arXiv:nucl-ex/0111010 [nucl-ex]].

\bibitem{Hammer:2003ai}
H.-W.~Hammer and U.-G.~Mei{\ss}ner,
Eur. Phys. J. A \textbf{20}, no.3, 469-473 (2004)
[arXiv:hep-ph/0312081 [hep-ph]].

\bibitem{Belushkin:2006qa}
M.~A.~Belushkin, H.-W.~Hammer and U.-G.~Mei{\ss}ner,
Phys. Rev. C \textbf{75}, 035202 (2007)
[arXiv:hep-ph/0608337 [hep-ph]].

\bibitem{Hammer:1998rz}
H.-W.~Hammer and M.J.~Ramsey-Musolf,
Phys.\ Rev.\ C {\bf 60}, 045205 (1999)
[Erratum-ibid.\ C {\bf 62}, 049903 (2000)]
[arXiv:hep-ph/9812261].


\bibitem{Hammer:1999uf}
H.-W.~Hammer and M.J.~Ramsey-Musolf,
Phys.\ Rev.\ C {\bf 60}, 045204 (1999)
[Erratum-ibid.\ C {\bf 62}, 049902 (2000)]
[arXiv:hep-ph/9903367].


\bibitem{Meissner:1997qt}
U.-G.~Mei{\ss}ner, V.~Mull, J.~Speth and J.~W.~van Orden,
Phys.\ Lett.\ B {\bf 408}, 381 (1997)
[arXiv:hep-ph/9701296].

\bibitem{Belushkin:2005ds}
M.~A.~Belushkin, H.-W.~Hammer and U.-G.~Mei{\ss}ner,
Phys. Lett. B \textbf{633}, 507-511 (2006)
[arXiv:hep-ph/0510382 [hep-ph]].

\bibitem{Melnikov:1999xp}
K.~Melnikov and T.~van Ritbergen,
Phys. Rev. Lett. \textbf{84}, 1673-1676 (2000)
[arXiv:hep-ph/9911277 [hep-ph]].


\bibitem{Belushkin:2007zv}
M.~A.~Belushkin, H.-W.~Hammer and U.-G.~Mei{\ss}ner,
Phys. Lett. B \textbf{658}, 138-142 (2008)
[arXiv:0705.3385 [hep-ph]].

\bibitem{Blunden:2003sp}
P.~G.~Blunden, W.~Melnitchouk and J.~A.~Tjon,
Phys. Rev. Lett. \textbf{91}, 142304 (2003)
[arXiv:nucl-th/0306076 [nucl-th]].

\bibitem{Blunden:2005ew}
P.~G.~Blunden, W.~Melnitchouk and J.~A.~Tjon,
Phys. Rev. C \textbf{72}, 034612 (2005)
[arXiv:nucl-th/0506039 [nucl-th]].

\bibitem{Kondratyuk:2005kk}
S.~Kondratyuk, P.~G.~Blunden, W.~Melnitchouk and J.~A.~Tjon,
Phys. Rev. Lett. \textbf{95}, 172503 (2005)
doi:10.1103/PhysRevLett.95.172503
[arXiv:nucl-th/0506026 [nucl-th]].

\bibitem{Bernauer:2013tpr}
J.~C.~Bernauer \textit{et al.} [A1],
Phys. Rev. C \textbf{90}, no.1, 015206 (2014)
doi:10.1103/PhysRevC.90.015206
[arXiv:1307.6227 [nucl-ex]].


\bibitem{Lorenz:2012tm}
I.~T.~Lorenz, H.-W.~Hammer and U.-G.~Mei{\ss}ner,
Eur. Phys. J. A \textbf{48}, 151 (2012)
[arXiv:1205.6628 [hep-ph]].

\bibitem{Lorenz:2014yda}
I.~T.~Lorenz, U.-G.~Mei{\ss}ner, H.-W.~Hammer and Y.~B.~Dong,
Phys. Rev. D \textbf{91}, no.1, 014023 (2015)
[arXiv:1411.1704 [hep-ph]].

\bibitem{Ron:2011rd}
G.~Ron \textit{et al.} [Jefferson Lab Hall A],
Phys. Rev. C \textbf{84}, 055204 (2011)
[arXiv:1103.5784 [nucl-ex]].

\bibitem{Zhan:2011ji}
X.~Zhan, K.~Allada, D.~S.~Armstrong, J.~Arrington, W.~Bertozzi, W.~Boeglin, J.~P.~Chen, K.~Chirapatpimol, S.~Choi and E.~Chudakov, \textit{et al.}
Phys. Lett. B \textbf{705}, 59-64 (2011)
[arXiv:1102.0318 [nucl-ex]].

\bibitem{Lorenz:2015pba}
I.~T.~Lorenz, H.-W.~Hammer and U.-G.~Mei{\ss}ner,
Phys. Rev. D \textbf{92} (2015) no.3, 034018
[arXiv:1506.02282 [hep-ph]].

\bibitem{Lees:2013ebn}
J.~P.~Lees \textit{et al.} [BaBar],
Phys. Rev. D \textbf{87} (2013) no.9, 092005
[arXiv:1302.0055 [hep-ex]].






\bibitem{Hoferichter:2016duk}
M.~Hoferichter, B.~Kubis, J.~Ruiz de Elvira, H.-W.~Hammer and U.-G.~Mei{\ss}ner,
Eur. Phys. J. A \textbf{52}, no.11, 331 (2016)
[arXiv:1609.06722 [hep-ph]].


\bibitem{Hoferichter:2015hva}
M.~Hoferichter, J.~Ruiz de Elvira, B.~Kubis and U.-G.~Mei{\ss}ner,
Phys. Rept. \textbf{625}, 1-88 (2016)
[arXiv:1510.06039 [hep-ph]].


\bibitem{Lin:2021umk}
Y.~H.~Lin, H.-W.~Hammer and U.-G.~Mei{\ss}ner,
Phys. Lett. B \textbf{816}, 136254 (2021)
[arXiv:2102.11642 [hep-ph]].


\bibitem{Paz:2020prs}
G.~Paz,
[arXiv:2004.03077 [hep-ph]].

\bibitem{Alarcon:2017lhg}
J.~M.~Alarc\'on and C.~Weiss,
Phys. Rev. C \textbf{97} (2018) no.5, 055203
[arXiv:1710.06430 [hep-ph]].

\bibitem{Alarcon:2018irp}
J.~M.~Alarc\'on and C.~Weiss,
Phys. Lett. B \textbf{784} (2018), 373-377
[arXiv:1803.09748 [hep-ph]].

\bibitem{Alarcon:2018zbz}
J.~M.~Alarc\'on, D.~W.~Higinbotham, C.~Weiss and Z.~Ye,
Phys. Rev. C \textbf{99} (2019) no.4, 044303
[arXiv:1809.06373 [hep-ph]].

\bibitem{Alarcon:2020kcz}
J.~M.~Alarc\'on, D.~W.~Higinbotham and C.~Weiss,
Phys. Rev. C \textbf{102} (2020) no.3, 035203
[arXiv:2002.05167 [hep-ph]].

\bibitem{Leupold:2017ngs}
S.~Leupold,
Eur. Phys. J. A \textbf{54} (2018) no.1, 1
[arXiv:1707.09210 [hep-ph]].


\bibitem{Jaffe:1989mj}
R.~L.~Jaffe,
Phys. Lett. B \textbf{229}, 275-279 (1989).

\bibitem{Hammer:1995de}
H.-W.~Hammer, U.-G.~Mei{\ss}ner and D.~Drechsel,
Phys. Lett. B \textbf{367}, 323-328 (1996)
[arXiv:hep-ph/9509393 [hep-ph]].

\bibitem{Forkel:1995ff}
H.~Forkel,
Phys. Rev. C \textbf{56}, 510-525 (1997)
[arXiv:hep-ph/9607452 [hep-ph]].

\bibitem{Pohl:2013yb}
R.~Pohl, R.~Gilman, G.~A.~Miller and K.~Pachucki,
Ann. Rev. Nucl. Part. Sci. \textbf{63}, 175-204 (2013)
[arXiv:1301.0905 [physics.atom-ph]].

\bibitem{Karr:2020wgh}
J.~P.~Karr, D.~Marchand and E.~Voutier,
Nature Rev. Phys. \textbf{2}, no.11, 601-614 (2020).

\bibitem{Rosenbluth:1950yq}
M.~N.~Rosenbluth,
Phys. Rev. \textbf{79}, 615-619 (1950).

\bibitem{Akhiezer:1968ek}
A.~I.~Akhiezer and M.~P.~Rekalo,
Sov. Phys. Dokl. \textbf{13}, 572 (1968).

\bibitem{Arnold:1980zj}
R.~G.~Arnold, C.~E.~Carlson and F.~Gross,
Phys. Rev. C \textbf{23}, 363 (1981).


\bibitem{TomasiGustafsson:2001za}
E.~Tomasi-Gustafsson and M.~P.~Rekalo,
Phys. Lett. B \textbf{504}, 291-295 (2001).

\bibitem{DZbook}
S.~D.~Drell and F.~Zachariasen,
``Electromagnetic Structure of Nucleons,''
Oxford University Press, London, 1965.
  
\bibitem{Chew:1958zjr}
G.~F.~Chew, R.~Karplus, S.~Gasiorowicz and F.~Zachariasen,
Phys. Rev. \textbf{110}, no.1, 265 (1958).

\bibitem{Federbush:1958zz}
P.~Federbush, M.~L.~Goldberger and S.~B.~Treiman,
Phys. Rev. \textbf{112}, 642-665 (1958).

\bibitem{Hohler:1974eq}
G.~H\"ohler and E.~Pietarinen,
Phys. Lett. B \textbf{53}, 471-475 (1975).


\bibitem{Eidelman:2003uh}
S.~Eidelman and L.~Lukaszuk,
Phys.\ Lett.\ B {\bf 582}, 27 (2004)
[hep-ph/0311366].

\bibitem{Gasser:1990bv}
J.~Gasser and U.-G.~Mei{\ss}ner,
Nucl. Phys. B \textbf{357}, 90-128 (1991).

\bibitem{Watson:1954uc}
K.~M.~Watson,
Phys. Rev. \textbf{95}, 228-236 (1954).

\bibitem{Omnes:1958hv}
R.~Omnes,
Nuovo Cim. \textbf{8}, 316-326 (1958).

\bibitem{Aubert:2009ad}
B.~Aubert \textit{et al.} [BaBar],
Phys. Rev. Lett. \textbf{103}, 231801 (2009)
[arXiv:0908.3589 [hep-ex]].

\bibitem{Babusci:2012rp}
D.~Babusci \textit{et al.} [KLOE],
Phys. Lett. B \textbf{720}, 336-343 (2013)
[arXiv:1212.4524 [hep-ex]].

\bibitem{Ablikim:2015orh}
M.~Ablikim \textit{et al.} [BESIII],
Phys. Lett. B \textbf{753}, 629-638 (2016)
[erratum: Phys. Lett. B \textbf{812}, 135982 (2021)]
[arXiv:1507.08188 [hep-ex]].

\bibitem{Caprini:2011ky}
I.~Caprini, G.~Colangelo and H.~Leutwyler,
Eur. Phys. J. C \textbf{72}, 1860 (2012)
[arXiv:1111.7160 [hep-ph]].

\bibitem{GarciaMartin:2011cn}
R.~Garcia-Martin, R.~Kaminski, J.~R.~Pelaez, J.~Ruiz de Elvira and F.~J.~Yndurain,
Phys. Rev. D \textbf{83}, 074004 (2011)
[arXiv:1102.2183 [hep-ph]].

\bibitem{Schneider:2012ez}
S.~P.~Schneider, B.~Kubis and F.~Niecknig,
Phys. Rev. D \textbf{86}, 054013 (2012)
[arXiv:1206.3098 [hep-ph]].

\bibitem{Ditsche:2012fv}
C.~Ditsche, M.~Hoferichter, B.~Kubis and U.-G.~Mei{\ss}ner,
JHEP \textbf{06}, 043 (2012)
[arXiv:1203.4758 [hep-ph]].

\bibitem{Hoferichter:2012wf}
M.~Hoferichter, C.~Ditsche, B.~Kubis and U.-G.~Mei{\ss}ner,
JHEP \textbf{06}, 063 (2012)
[arXiv:1204.6251 [hep-ph]].

\bibitem{Hoferichter:2015dsa}
M.~Hoferichter, J.~Ruiz de Elvira, B.~Kubis and U.~G.~Mei{\ss}ner,
Phys. Rev. Lett. \textbf{115}, 092301 (2015)
[arXiv:1506.04142 [hep-ph]].

\bibitem{RuizdeElvira:2017stg}
J.~Ruiz de Elvira, M.~Hoferichter, B.~Kubis and U.-G.~Mei{\ss}ner,
J. Phys. G \textbf{45}, no.2, 024001 (2018)
[arXiv:1706.01465 [hep-ph]].

\bibitem{Koch:1980ay}
R.~Koch and E.~Pietarinen,
Nucl. Phys. A \textbf{336}, 331-346 (1980).

\bibitem{Hoehler:1983}
G.~H\"{o}hler,
{\it Pion--Nukleon-Streuung: Methoden und Ergebnisse},
in Landolt-B\"ornstein, {\bf 9b2}, ed.\ H.~Schopper,
Springer Verlag, Berlin, 1983.  

\bibitem{Steiner:1968}
F.~Steiner, PhD thesis, University of Karlsruhe, 1968.
  
\bibitem{Hite:1973pm}
G.~E.~Hite and F.~Steiner,
Nuovo Cim. A \textbf{18}, 237-270 (1973).

\bibitem{Baru:2011bw}
V.~Baru, C.~Hanhart, M.~Hoferichter, B.~Kubis, A.~Nogga and D.~R.~Phillips,
Nucl. Phys. A \textbf{872}, 69-116 (2011)
[arXiv:1107.5509 [nucl-th]].

\bibitem{Hirtl:2021zqf}
A.~Hirtl, D.~F.~Anagnostopoulos, D.~S.~Covita, H.~Fuhrmann, H.~Gorke, D.~Gotta, A.~Gruber, M.~Hennebach, P.~Indelicato and T.~S.~Jensen, \textit{et al.}
Eur. Phys. J. A \textbf{57}, no.2, 70 (2021).

\bibitem{Gounaris:1968mw}
G.~J.~Gounaris and J.~J.~Sakurai,
Phys. Rev. Lett. \textbf{21}, 244-247 (1968),

\bibitem{Wess:1971yu}
J.~Wess and B.~Zumino,
Phys. Lett. B \textbf{37}, 95-97 (1971).

\bibitem{Witten:1983tw}
E.~Witten,
Nucl. Phys. B \textbf{223}, 422-432 (1983).


\bibitem{Bernard:1996cc}
V.~Bernard, N.~Kaiser and U.-G.~Mei{\ss}ner,
Nucl. Phys. A \textbf{611}, 429-441 (1996)
[arXiv:hep-ph/9607428 [hep-ph]].

\bibitem{Kaiser:2019irl}
N.~Kaiser and E.~Passemar,
Eur. Phys. J. A \textbf{55}, no.2, 16 (2019)
[arXiv:1901.02865 [nucl-th]].

\bibitem{Janssen:1996kx}
G.~Janssen, K.~Holinde and J.~Speth,
Phys. Rev. C \textbf{54}, 2218-2234 (1996).


\bibitem{Sakurai:1960ju}
J.~J.~Sakurai,
Annals Phys. \textbf{11}, 1-48 (1960).

\bibitem{Kroll:1967it}
N.~M.~Kroll, T.~D.~Lee and B.~Zumino,
Phys. Rev. \textbf{157}, 1376-1399 (1967)

\bibitem{Bando:1987br}
M.~Bando, T.~Kugo and K.~Yamawaki,
Phys. Rept. \textbf{164}, 217-314 (1988).

\bibitem{Meissner:1987ge}
U.-G.~Mei{\ss}ner,
Phys. Rept. \textbf{161}, 213 (1988).


\bibitem{sabba}
S.~Ciulli, C.~Pomponiu, I.~Sabba-Stefanescu,
Phys. Rept. \textbf{17}, 133 (1975).

\bibitem{SabbaStefanescu:1978hvt}
I.~Sabba Stefanescu,
J. Math. Phys. \textbf{21}, 175 (1980).

\bibitem{Grein:1977mn}
W.~Grein and P.~Kroll,
Nucl. Phys. B \textbf{137}, 173-188 (1978).

\bibitem{Kopecky:1995zz}
S.~Kopecky, P.~Riehs, J.~A.~Harvey and N.~W.~Hill,
Phys. Rev. Lett. \textbf{74}, 2427-2430 (1995).

\bibitem{Kopecky:1997rw}
S.~Kopecky, M.~Krenn, P.~Riehs, S.~Steiner, J.~A.~Harvey, N.~W.~Hill and M.~Pernicka,
Phys. Rev. C \textbf{56}, 2229-2237 (1997).

\bibitem{Filin:2019eoe}
A.~A.~Filin, V.~Baru, E.~Epelbaum, H.~Krebs, D.~M\"oller and P.~Reinert,
Phys. Rev. Lett. \textbf{124}, no.8, 082501 (2020)
[arXiv:1911.04877 [nucl-th]].

\bibitem{Filin:2020tcs}
A.~A.~Filin, D.~M\"oller, V.~Baru, E.~Epelbaum, H.~Krebs and P.~Reinert,
Phys. Rev. C \textbf{103} (2021) no.2, 024313
[arXiv:2009.08911 [nucl-th]].

\bibitem{Lepage:1980fj}
G.~P.~Lepage and S.~J.~Brodsky,
Phys. Rev. D \textbf{22}, 2157 (1980).


\bibitem{Bianconi:2015owa}
A.~Bianconi and E.~Tomasi-Gustafsson,
Phys. Rev. Lett. \textbf{114} (2015) no.23, 232301
[arXiv:1503.02140 [nucl-th]].




\bibitem{Arrington:2003df}
J.~Arrington,
Phys. Rev. C \textbf{68}, 034325 (2003)
[arXiv:nucl-ex/0305009 [nucl-ex]].

\bibitem{Meister:1963zz}
N.~Meister and D.~R.~Yennie,
Phys. Rev. \textbf{130}, 1210-1229 (1963).

\bibitem{Mo:1968cg}
L.~W.~Mo and Y.~S.~Tsai,
Rev. Mod. Phys. \textbf{41}, 205-235 (1969).

\bibitem{Maximon:2000hm}
L.~C.~Maximon and J.~A.~Tjon,
Phys. Rev. C \textbf{62}, 054320 (2000)
[arXiv:nucl-th/0002058 [nucl-th]].

\bibitem{Arrington:2011dn}
J.~Arrington, P.~G.~Blunden and W.~Melnitchouk,
Prog. Part. Nucl. Phys. \textbf{66}, 782-833 (2011)
[arXiv:1105.0951 [nucl-th]].

\bibitem{Afanasev:2017gsk}
A.~Afanasev, P.~G.~Blunden, D.~Hasell and B.~A.~Raue,
Prog. Part. Nucl. Phys. \textbf{95}, 245-278 (2017)
[arXiv:1703.03874 [nucl-ex]].

\bibitem{Ahmed:2020uso}
J.~Ahmed, P.~G.~Blunden and W.~Melnitchouk,
Phys. Rev. C \textbf{102}, no.4, 045205 (2020)
[arXiv:2006.12543 [nucl-th]].

\bibitem{Rarita:1941mf}
W.~Rarita and J.~Schwinger,
Phys. Rev. \textbf{60}, 61 (1941).


\bibitem{Jones:1972ky}
H.~F.~Jones and M.~D.~Scadron,
Annals Phys. \textbf{81}, 1-14 (1973).

\bibitem{Kondratyuk:2001qu}
S.~Kondratyuk and O.~Scholten,
Phys. Rev. C \textbf{64}, 024005 (2001)
[arXiv:nucl-th/0103006 [nucl-th]].

\bibitem{Lalakulich:2006sw}
O.~Lalakulich, E.~A.~Paschos and G.~Piranishvili,
Phys. Rev. D \textbf{74}, 014009 (2006)
[arXiv:hep-ph/0602210 [hep-ph]].

\bibitem{Tiator:2003uu}
L.~Tiator, D.~Drechsel, S.~Kamalov, M.~M.~Giannini, E.~Santopinto and A.~Vassallo,
Eur. Phys. J. A \textbf{19}, 55-60 (2004)
[arXiv:nucl-th/0310041 [nucl-th]].

\bibitem{Graczyk:2013pca}
K.~M.~Graczyk,
Phys. Rev. C \textbf{88}, 065205 (2013)
[arXiv:1306.5991 [hep-ph]].

\bibitem{Zhou:2014xka}
H.~Q.~Zhou and S.~N.~Yang,
Eur. Phys. J. A \textbf{51}, no.8, 105 (2015)
[arXiv:1407.2711 [nucl-th]].

\bibitem{McKinley:1948zz}
W.~A.~McKinley and H.~Feshbach,
Phys. Rev. \textbf{74}, 1759-1763 (1948)


\bibitem{Arrington:2011kv}
J.~Arrington,
Phys. Rev. Lett. \textbf{107}, 119101 (2011)
[arXiv:1108.3058 [nucl-ex]].


\bibitem{Hammer:2006mw}
H.-W.~Hammer,
Eur. Phys. J. A \textbf{28}, 49-57 (2006)
[arXiv:hep-ph/0602121 [hep-ph]].


\bibitem{Efron:1993qfh}
B.~Efron and R.~Tibshirani,
Statist. Sci. \textbf{57}, no.1, 54-75 (1986).


\bibitem{Schindler:2008fh}
M.~R.~Schindler and D.~R.~Phillips,
Annals Phys. \textbf{324} (2009), 682-708
[erratum: Annals Phys. \textbf{324} (2009), 2051-2055]

\bibitem{Wesolowski:2015fqa}
S.~Wesolowski, N.~Klco, R.~J.~Furnstahl, D.~R.~Phillips and A.~Thapaliya,
J. Phys. G \textbf{43} (2016) no.7, 074001
[arXiv:1511.03618 [nucl-th]].

\bibitem{ParaMonte}
A.~Shahmoradi and F.~Bagheri, 
arXiv:2009.14229 [cs.MS]



\bibitem{Punjabi:2005wq}
V.~Punjabi, C.~F.~Perdrisat, K.~A.~Aniol, F.~T.~Baker, J.~Berthot, P.~Y.~Bertin, W.~Bertozzi, A.~Besson, L.~Bimbot and W.~U.~Boeglin, \textit{et al.}
Phys. Rev. C \textbf{71} (2005), 055202
[erratum: Phys. Rev. C \textbf{71} (2005), 069902]
[arXiv:nucl-ex/0501018 [nucl-ex]].

\bibitem{Puckett:2010ac}
A.~J.~R.~Puckett, E.~J.~Brash, M.~K.~Jones, W.~Luo, M.~Meziane, L.~Pentchev, C.~F.~Perdrisat, V.~Punjabi, F.~R.~Wesselmann and A.~Ahmidouch, \textit{et al.}
Phys. Rev. Lett. \textbf{104} (2010), 242301
[arXiv:1005.3419 [nucl-ex]].

\bibitem{Meziane:2010xc}
M.~Meziane \textit{et al.} [GEp2gamma],
Phys. Rev. Lett. \textbf{106} (2011), 132501
[arXiv:1012.0339 [nucl-ex]].


\bibitem{Puckett:2011xg}
A.~J.~R.~Puckett, E.~J.~Brash, O.~Gayou, M.~K.~Jones, L.~Pentchev, C.~F.~Perdrisat, V.~Punjabi, K.~A.~Aniol, T.~Averett and F.~Benmokhtar, \textit{et al.}
Phys. Rev. C \textbf{85} (2012), 045203
[arXiv:1102.5737 [nucl-ex]].

\bibitem{Djukanovic:2021cgp}
D.~Djukanovic, T.~Harris, G.~von Hippel, P.~M.~Junnarkar, H.~B.~Meyer, D.~Mohler, K.~Ottnad, T.~Schulz, J.~Wilhelm and H.~Wittig,
Phys. Rev. D \textbf{103} (2021), 094522
[arXiv:2102.07460 [hep-lat]].

\bibitem{Hammer:2003qv}
H.-W.~Hammer, D.~Drechsel and U.-G.~Mei{\ss}ner,
Phys. Lett. B \textbf{586}, 291-296 (2004)
[arXiv:hep-ph/0310240 [hep-ph]].

\bibitem{Meissner:2007tp}
U.-G.~Mei{\ss}ner,
AIP Conf. Proc. \textbf{904} (2007) no.1, 142-150
[arXiv:nucl-th/0701094 [nucl-th]].

\bibitem{Hohler:1994rt}
G.~H\"ohler,
PiN Newslett. \textbf{1993}, no.9, 1-36 (1993).


\bibitem{Ericson:1988gk}
T.~E.~O.~Ericson and W.~Weise,
Int. Ser. Monogr. Phys. {\bf 74} (1988).


\bibitem{Ablikim:2021kjh}
M.~Ablikim \textit{et al.} [BESIII],
Phys. Lett. B \textbf{817} (2021), 136328
[arXiv:2102.10337 [hep-ex]].

\bibitem{CMD-3:2018kql}
R.~R.~Akhmetshin \textit{et al.} [CMD-3],
Phys. Lett. B \textbf{794} (2019), 64-68
[arXiv:1808.00145 [hep-ex]].



\bibitem{Delcourt:1979ed}
B.~Delcourt, I.~Derado, J.~L.~Bertrand, D.~Bisello, J.~C.~Bizot, J.~Buon, A.~Cordier, P.~Eschstruth, L.~Fayard and J.~Jeanjean, \textit{et al.}
Phys. Lett. B \textbf{86} (1979), 395-398.

\bibitem{Antonelli:1998fv}
A.~Antonelli, R.~Baldini, P.~Benasi, M.~Bertani, M.~E.~Biagini, V.~Bidoli, C.~Bini, T.~Bressani, R.~Calabrese and R.~Cardarelli, \textit{et al.}
Nucl. Phys. B \textbf{517} (1998), 3-35.

\bibitem{Bai:2003sw}
J.~Z.~Bai \textit{et al.} [BES],
Phys. Rev. Lett. \textbf{91} (2003), 022001
[arXiv:hep-ex/0303006 [hep-ex]].

\bibitem{Aubert:2005gw}
B.~Aubert \textit{et al.} [BaBar],
Phys. Rev. D \textbf{72} (2005), 051101
[arXiv:hep-ex/0507012 [hep-ex]].

\bibitem{BESIII:2011aa}
M.~Ablikim \textit{et al.} [BESIII],
Phys. Rev. Lett. \textbf{108} (2012), 112003
[arXiv:1112.0942 [hep-ex]].



\bibitem{Achasov:2014ncd}
M.~N.~Achasov, A.~Y.~Barnyakov, K.~I.~Beloborodov, A.~V.~Berdyugin, D.~E.~Berkaev, A.~G.~Bogdanchikov, A.~A.~Botov, T.~V.~Dimova, V.~P.~Druzhinin and V.~B.~Golubev, \textit{et al.}
Phys. Rev. D \textbf{90} (2014) no.11, 112007
[arXiv:1410.3188 [hep-ex]].


\bibitem{Tomasi-Gustafsson:2020vae}
E.~Tomasi-Gustafsson, A.~Bianconi and S.~Pacetti,
Phys. Rev. C \textbf{103} (2021) no.3, 035203
[arXiv:2012.14656 [hep-ph]].

\bibitem{Lees:2013uta}
J.~P.~Lees \textit{et al.} [BaBar],
Phys. Rev. D \textbf{88} (2013) no.7, 072009
[arXiv:1308.1795 [hep-ex]].

\bibitem{Bardin:1994am}
G.~Bardin, G.~Burgun, R.~Calabrese, G.~Capon, R.~Carlin, P.~Dalpiaz, P.~F.~Dalpiaz, J.~Derr\'e, U.~Dosselli and J.~Duclos, \textit{et al.}
Nucl. Phys. B \textbf{411} (1994), 3-32.



\bibitem{Aubert:2005cb}
B.~Aubert \textit{et al.} [BaBar],
Phys. Rev. D \textbf{73} (2006), 012005
[arXiv:hep-ex/0512023 [hep-ex]].


\bibitem{Zou:2003zn}
B.~S.~Zou and H.~C.~Chiang,
Phys. Rev. D \textbf{69} (2004), 034004
[arXiv:hep-ph/0309273 [hep-ph]].
  
\bibitem{Kerbikov:2004gs}
B.~Kerbikov, A.~Stavinsky and V.~Fedotov,
Phys. Rev. C \textbf{69} (2004), 055205
[arXiv:hep-ph/0402054 [hep-ph]].

\bibitem{Bugg:2004rk}
D.~V.~Bugg,
Phys. Lett. B \textbf{598} (2004), 8-14
[arXiv:hep-ph/0406293 [hep-ph]].
  
\bibitem{Sibirtsev:2004id}
A.~Sibirtsev, J.~Haidenbauer, S.~Krewald, U.-G.~Mei{\ss}ner and A.~W.~Thomas,
Phys. Rev. D \textbf{71} (2005), 054010
[arXiv:hep-ph/0411386 [hep-ph]].


\bibitem{Loiseau:2005cv}
B.~Loiseau and S.~Wycech,
Phys. Rev. C \textbf{72} (2005), 011001
[arXiv:hep-ph/0501112 [hep-ph]].

\bibitem{Haidenbauer:2006dm}
J.~Haidenbauer, H.-W.~Hammer, U.-G.~Mei{\ss}ner and A.~Sibirtsev,
Phys. Lett. B \textbf{643} (2006), 29-32
[arXiv:hep-ph/0606064 [hep-ph]].

\bibitem{Haidenbauer:2014kja}
J.~Haidenbauer, X.~W.~Kang and U.-G.~Mei\ss{}ner,
Nucl. Phys. A \textbf{929} (2014), 102-118
[arXiv:1405.1628 [nucl-th]].

\bibitem{Kang:2013uia}
X.~W.~Kang, J.~Haidenbauer and U.-G.~Mei\ss{}ner,
JHEP \textbf{02} (2014), 113
[arXiv:1311.1658 [hep-ph]].

\bibitem{Dai:2017ont}
L.~Y.~Dai, J.~Haidenbauer and U.-G.~Mei\ss{}ner,
JHEP \textbf{07} (2017), 078
[arXiv:1702.02065 [nucl-th]].


\bibitem{Dmitriev:2007zz}
V.~F.~Dmitriev and A.~I.~Milstein,
Phys. Lett. B \textbf{658} (2007), 13-16.


\bibitem{Chen:2008ee}
G.~Y.~Chen, H.~R.~Dong and J.~P.~Ma,
Phys. Rev. D \textbf{78} (2008), 054022
[arXiv:0806.4661 [hep-ph]].

\bibitem{Chen:2010an}
G.~Y.~Chen, H.~R.~Dong and J.~P.~Ma,
Phys. Lett. B \textbf{692} (2010), 136-142
[arXiv:1004.5174 [hep-ph]].

\bibitem{Dalkarov:2009yf}
O.~D.~Dalkarov, P.~A.~Khakhulin and A.~Y.~Voronin,
Nucl. Phys. A \textbf{833} (2010), 104-118
[arXiv:0906.0266 [nucl-th]].

\bibitem{Bondar:2010zx}
A.~E.~Bondar, V.~F.~Dmitriev, A.~I.~Milstein and V.~M.~Strakhovenko,
Phys. Lett. B \textbf{697} (2011), 159-163
[arXiv:1012.4638 [hep-ph]].

\bibitem{Haidenbauer:2012pu}
J.~Haidenbauer and U.-G.~Mei{\ss}ner,
Phys. Rev. D \textbf{86} (2012), 077503
[arXiv:1208.3343 [hep-ph]].

\bibitem{Dmitriev:2013pfa}
V.~F.~Dmitriev and A.~I.~Milstein,
Phys. Lett. B \textbf{722} (2013), 83-85
[arXiv:1303.0653 [hep-ph]].

\bibitem{Dmitriev:2013xla}
V.~F.~Dmitriev, A.~I.~Milstein and S.~G.~Salnikov,
Phys. Atom. Nucl. \textbf{77} (2014) no.9, 1173-1177
[arXiv:1307.0936 [hep-ph]].

\bibitem{Entem:2007bb}
D.~R.~Entem and F.~Fernandez,
Phys. Rev. D \textbf{75} (2007), 014004.

\bibitem{Rosner:2006vc}
J.~L.~Rosner,
Phys. Rev. D \textbf{74} (2006), 076006
[arXiv:hep-ph/0608102 [hep-ph]].

\bibitem{Guttmann:2012sq}
J.~Guttmann and M.~Vanderhaeghen,
Phys. Lett. B \textbf{719} (2013), 136-142
[arXiv:1210.3290 [hep-ph]].

\bibitem{Gourdin:1974iq}
M.~Gourdin,
Phys. Rept. \textbf{11} (1974), 29.

\bibitem{Geshkenbein:1974gm}
B.~V.~Geshkenbein, B.~L.~Ioffe and M.~A.~Shifman,
Yad. Fiz. \textbf{20} (1974), 128-136.

\bibitem{Baldini:1998qn}
R.~Baldini, S.~Dubnicka, P.~Gauzzi, S.~Pacetti, E.~Pasqualucci and Y.~Srivastava,
Eur. Phys. J. C \textbf{11} (1999), 709-715.

\bibitem{Pacetti:2007zz}
S.~Pacetti,
Eur. Phys. J. A \textbf{32} (2007) no.4, 421-427.

\bibitem{Pacetti:2010nv}
S.~Pacetti,
Chin. Phys. C \textbf{34} (2010), 874-876
[arXiv:1012.1232 [hep-ph]].


\bibitem{Downie:2014qna}
E.~J.~Downie [MUSE],
EPJ Web Conf. \textbf{73} (2014), 07005.

\bibitem{Denisov:2018unj}
B.~Adams, C.~A.~Aidala, R.~Akhunzyanov, G.~D.~Alexeev, M.~G.~Alexeev, A.~Amoroso, V.~Andrieux, N.~V.~Anfimov, V.~Anosov and A.~Antoshkin, \textit{et al.}
[arXiv:1808.00848 [hep-ex]].



\bibitem{Garcon:2001sz}
M.~Garcon and J.~W.~Van Orden,
Adv. Nucl. Phys. \textbf{26} (2001), 293
[arXiv:nucl-th/0102049 [nucl-th]].

\bibitem{Krebs:2020pii}
H.~Krebs,
Eur. Phys. J. A \textbf{56} (2020) no.9, 234
[arXiv:2008.00974 [nucl-th]].



\bibitem{Perez:2016aol}
R.~Navarro P\'erez, J.~E.~Amaro and E.~Ruiz Arriola,
Phys. Rev. C \textbf{95}, no.6, 064001 (2017)
[arXiv:1606.00592 [nucl-th]].

\bibitem{Reinert:2020mcu}
P.~Reinert, H.~Krebs and E.~Epelbaum,
Phys. Rev. Lett. \textbf{126}, no.9, 092501 (2021)
[arXiv:2006.15360 [nucl-th]].


\bibitem{Bernard:1995dp}
V.~Bernard, N.~Kaiser and U.-G.~Mei{\ss}ner,
Int. J. Mod. Phys. E \textbf{4}, 193-346 (1995)
[arXiv:hep-ph/9501384 [hep-ph]].

\bibitem{polk}
R.J. Eden, P.V. Landshoff, D.I. Olive and J.C. Polkinghorne,  {\it The
Analytic S--Matrix} (Cambridge University Press, Cambridge, 1966).

\bibitem{Jacob:1959at}
M.~Jacob and G.~C.~Wick,
Annals Phys. \textbf{7} (1959), 404-428.

\bibitem{Musolf:1996qt}
M.~J.~Musolf, H.-W.~Hammer and D.~Drechsel,
Phys. Rev. D \textbf{55} (1997), 2741
[erratum: Phys. Rev. D \textbf{62} (2000), 079901]
[arXiv:hep-ph/9610402 [hep-ph]].



\end{thebibliography}
\end{document}